\documentclass[rmp,aps,reprint,amsmath,amssymb]{revtex4-1}

\usepackage[pdftex]{graphicx}
\pdfoutput=1
\usepackage{overcirc}

\newcommand{\bra}[1]{\langle {#1} |}
\newcommand{\ket}[1]{| {#1} \rangle}
\newcommand{\inproduct}[2]{\langle #1 | #2 \rangle}
\def \cJ{{ \cal J}}

\def\<{\langle}
\def\>{\rangle}
\newcommand{\del}{\partial}
\newcommand{\Hhat}{\hat{H}}
\newcommand{\Nhat}{\hat{N}}
\newcommand{\Qhat}{\hat{Q}}
\newcommand{\Phat}{\hat{P}}

\newcommand{\That}{\hat{\Theta}}
\newcommand{\Nt}{\tilde{N}}
\newcommand{\Hc}{{\cal H}}

\newcommand{\dbeta}{\frac{\partial}{\partial\beta}}
\newcommand{\dgamma}{\frac{\partial}{\partial\gamma}}

\newcommand{\bg}{\beta,\gamma}
\newcommand{\Dhatp}{\hat{D}^{(+)}}
\newcommand{\Dp}{D^{(+)}}
\newcommand{\Ddotp}{\dot{D}^{(+)}}

\newcommand{\comma}{,}

\begin{document}
\DeclareGraphicsExtensions{.pdf}

\title{Time-dependent density-functional description of nuclear dynamics}

\author{Takashi Nakatsukasa}
\affiliation{Center for Computational Sciences\comma
University of Tsukuba\comma
Tsukuba 305-8577\comma
Japan}
\affiliation{RIKEN Nishina Center\comma
2-1 Hirosawa\comma
Wako 351-0198\comma
Japan}
\author{Kenichi Matsuyanagi}
\affiliation{RIKEN Nishina Center\comma
2-1 Hirosawa\comma
Wako 351-0198\comma
Japan}
\affiliation{Yukawa Institute for Theoretical Physics\comma
Kyoto University\comma
Kyoto 606-8502\comma
Japan}
\author{Masayuki Matsuo}
\affiliation{Department of Physics\comma
Faculty of Science\comma
Niigata University\comma
Niigata 950-2181\comma
Japan}
\author{Kazuhiro Yabana}
\affiliation{Center for Computational Sciences\comma
University of Tsukuba\comma
Tsukuba 305-8577\comma
Japan}

\begin{abstract}  
We present the basic concepts and recent developments
in the time-dependent density functional theory (TDDFT)
for describing nuclear dynamics at low energy.
The symmetry breaking is inherent in nuclear energy density functionals
(EDFs),
which provides a practical description of important correlations at the
ground state.
Properties of elementary modes of excitation are strongly influenced
by the symmetry breaking and can be studied with TDDFT.
In particular, a number of recent developments in the linear response
calculation have demonstrated their usefulness
in description of collective modes of excitation in nuclei.
Unrestricted real-time calculations have also become available
in recent years, with new developments for quantitative
description of nuclear collision phenomena.
There are, however, limitations in the real-time approach;
for instance, it cannot describe the many-body quantum tunneling.
Thus, we treat the quantum fluctuations associated with slow collective motions
assuming that time evolution of densities are determined 
by a few collective coordinates and momenta. 
The concept of collective submanifold is introduced 
in the phase space associated with the TDDFT and 
used to quantize the collective dynamics. 
Selected applications are presented to demonstrate the usefulness and
quality of the new approaches.
Finally, conceptual differences
between nuclear and electronic TDDFT are discussed, with
some recent applications to studies of electron dynamics in
the linear response and under a strong laser field.
\end{abstract}

\maketitle
\tableofcontents

\section{INTRODUCTION}
\label{sec:intro}

\subsection{Scope of the present review}

In the study of strongly correlated many-particle systems,
a fundamental challenge is to find basic properties of a variety of
elementary modes of excitation, and to identify the degrees
of freedom that are suitable for describing the collective phenomena.
The collective motion in such complex systems, with an ample supply
of experimental data and theoretical study, may often lead to
deeper insight
into the basic concepts of quantum many-body physics.

Among a variety of many-particle systems in the universe,
the nucleus provides a unique opportunity to investigate fundamental
aspects of the quantum many-body problems.
The nucleus is a self-bound system with finite number of fermionic
particles, called nucleons, which have the isospin degrees of freedom,
($t=1/2$), in addition to the intrinsic spin ($s=1/2$).
The strong interplay between the collective and single-particle
degrees of freedom plays important roles in nuclei,
which produces a rich variety of unique phenomena.
A prominent example of the consequence of this
coupling
is given by
the manifestation of nuclear deformation and rotational spectra.
Is is also closely related with
the damping and particle decay of the collective motion,
the particle transfer in the heavy-ion collision,
and the dissipation process in the nuclear fission.
In fact, the coupling between the single-particle motion and
collective motion is a key issue in nuclear structure.
It is the basic idea behind the unified model of nuclei \cite{BM69,BM75},
in which the collective motion is described by a shape change
of the average one-body nuclear potential.
It is easy to see that the basic concept of the unified model is
similar to that of the mean-field theory.
We therefore expect that
the mean-field theory 
may provide
a microscopic description of
the phenomena described above,
although it is limited to the one-body dynamics.

The self-consistent mean-field models for nuclei
are currently a leading theory for describing
properties of heavy nuclei \cite{BHR03,LPT03}.
They self-consistently determine the nuclear one-body mean-field potential,
starting from effective energy density functionals (EDF).
They are
capable of
describing almost all nuclei, including infinite nuclear matter,
with a single universal EDF.
The concept is very similar to the density functional theory (DFT)
in electronic systems, utilized in atomic, molecular, and solid state
physics.
Major conceptual difference is that, for the isolated finite nucleus,
all the currently available nuclear DFT models are designed to reproduce
the {\it intrinsic} ground state.
The self-consistent solution produces a density distribution
which spontaneously violates symmetries of the system,
such as translational, rotational, and gauge symmetries.
This feature has advantages and disadvantages.
The spontaneous breaking of symmetries (SSB) provides us with an intuitive
explanation of a variety of nuclear phenomena.
A typical example is the observed rotational spectra as a
consequence of the intrinsic density deformation.
On the other hand, when the symmetries are restored in finite nuclei,
an additional correlation energy is
generated.
A question arises then, concerning whether all the correlation energy should
be included in the EDF or not.
We do not think this issue is completely settled yet.
(Nevertheless, there are also attempts to justify the use of
symmetry-violating (wave-packet) densities in a rigorous sense,
which we will present in Sec.~\ref{sec:basic_formalism}.)
Perhaps, because of this unsettled problem,
it is common to use terminologies of the mean-field theory,
such as the time-dependent Hartree-Fock (TDHF) equations,
instead of the time-dependent Kohn-Sham (TDKS) equations.
In this article, we mainly use the DFT terminologies, since the naive
mean-field theory is not applicable to nuclear systems,
due to a strong two-body correlations (Sec.~\ref{sec:conceptual}).
Moreover, the mean-field calculation with a density-independent
(state-independent) interaction
cannot account for the nuclear saturation properties (Sec.~\ref{sec:IPM}).

An extension of the DFT to the time-dependent DFT (TDDFT) provides
a feasible description of many-body dynamics, which contains information
on excited states in addition to the ground state.
The TDDFT and its KS scheme are formally justified by
the one-to-one correspondence between the
time-dependent density and time-dependent external potential,
assuming the $v$-representability \cite{RG84}.
The TDDFT has vast applications to quantum phenomena in many-body systems.
In nuclear physics applications, there exist extensive studies in
simulation of the heavy-ion collision dynamics, especially
of nuclear fusion and deep inelastic scattering \cite{Neg82}.
Ultimately, the nuclear TDDFT aims at describing nuclear excitations
with different characters, such as vibration, rotation, and clustering,
nuclear reactions of many kinds, such as fusion/fission, particle transfer,
fragmentation,
and even collective excitations
in the crust and the interior of neutron stars.

One of the
most extensively studied area of the nuclear TDDFT is small
amplitude vibrations or linear response to external perturbations.
This is a perturbative regime of the TDDFT, but it provides a powerful
method to explore a variety of modes of excitation in nuclei.
Many kinds of approaches to the linear response calculations
have been developed and will be presented in
Sec.~\ref{sec:linear_response}.
In addition to the conventional matrix formalism, we present
some recent developments, such as the finite amplitude method
and the Green's function method for the quasiparticle formalism
with finite pair densities.

It is of significant interest and 
challenge
to go beyond the
perturbative regime.
Nuclei show numerous phenomena related to the large amplitude collective
motion.
In particular, nuclear
reactions
involving
collective and non-collective dynamics of many nucleons are
extensively studied using the real-time calculations in the past.
In Sec.~\ref{sec:real_time}, we show some recent developments
and applications.
A recent review articles on the real-time approaches in normal
\cite{Sim12} and superfluid systems \cite{Bul13},
may be supplementary to the present review.
It is also of great interest to study the strong quantum nature of
large amplitude 
collective motion, such as spontaneous fission, shape transition,
shape coexistence, anharmonic vibrations, and so on.
For these phenomena, the real-time simulation of the TDDFT
is not directly applicable to the problems.
In most cases, we need requantization of the TDDFT dynamics.
The requantization of TDHF and the imaginary-time TDHF for
classically forbidden dynamics were previously discussed in another
review paper in great details \cite{Neg82}.
Unfortunately,
the method has not been 
applied
to realistic problems,
due to number of difficulties, such as
finding suitable periodic orbits to quantize
\cite{BSW03}.
We present, in Sec.~\ref{sec:collective_submanifold},
an alternative theory to identify an optimal collective submanifold
in the TDDFT phase space.
Consequently,
with a small number of canonical variables, it is much more
practical to quantize the collective dynamics.

Since the DFT and TDDFT are commonly adopted in many domains of
quantum many-body systems,
current problems and new ideas in other fields are
of significant interest.
Similarly to nuclear physics,
there are linear response TDDFT calculations
and TDDFT for large amplitude motion as an initial value problem.
However, it should be noted that
there are conceptual and qualitative differences of EDFs
between nuclear and electronic DFT/TDDFT.
These issues will be discussed in Sec.~\ref{sec:electronic}.

We try to make the present review somewhat pedagogical
and tractable for non-practitioners,
to explain essential elements of the theories.
For more details, readers should be referred to literature.

\subsection{Saturation and the mean-field picture}
\label{sec:IPM}

The saturation is a fundamental property of the nuclear system,
that is analogous to the liquid.
The volume and total binding energy of
observed nuclei in nature are approximately
proportional to the mass number $A$.
Extrapolating the observed property to the infinite nuclear matter
with neglect of the Coulomb interaction,
the nuclear matter should have an equilibrium state with
$\rho_0\approx 0.17$ fm$^{-3}$ and $B/A\approx 16$ MeV,
at zero pressure and zero temperature.
The empirical mass formula of Bethe and Weizs\"acker \cite{Wei35,BB36},
which is based on this liquid drop picture of nuclei,
well accounts of the bulk part of the nuclear binding.
 
In contrast, there are many evidences 
pointing to
the fact that the mean-free path of nucleons
is larger than the size of nucleus.
The great success of the nuclear shell model \cite{MJ55},
in which nucleons are assumed to move independently inside an
average one-body potential, gives one of them.
The scattering experiments with incident neutrons and protons
provide more quantified information on the mean-free path.
In fact, the mean free path depends on the nucleon's energy,
and becomes larger for lower energy \cite{BM69}.
Therefore, it is natural to assume that
the nucleus can be primarily approximated by the
independent-particle model with an average one-body potential.
For the symmetric nuclear matter,
this approximation leads to the degenerate Fermi gas of the same number of
protons and neutrons ($Z=N=A/2$).
The observed saturation density of $\rho_0\approx 0.17$ fm$^{-3}$
gives the Fermi momentum, $k_F\approx 1.36$ fm$^{-1}$,
which corresponds to the Fermi energy (the maximum kinetic energy),
$T_F=k_F^2/2m\approx 40$ MeV.

The justification
of the independent-particle motion
encourage us to investigate the mean-field models of nuclei.
However, the naive mean-field models
cannot properly describe
the nuclear saturation property.
Here, the ``naive'' mean-field models mean those using
any kind of state-independent two-body interactions.
This has been known for many years \cite{BM69}.
Since it contains useful insights and relations to the nuclear
DFT, let us explain the essential point.
It is easy to consider the uniform nuclear matter
with a constant attractive ``mean-field'' potential $V<0$.
The constancy of $B/A$ means that it is equal to the
separation energy of nucleons, $S$.
In the Fermi-gas model, it is estimated as
\begin{equation}
\label{B1}
S \approx B/A \approx -(T_F + V) .
\end{equation}
Since the binding energy is $B/A\approx 16$ MeV,
the potential $V$ is about $-56$ MeV.
It should be noted that the relatively small separation energy is
the consequence of the significant cancellation between
the kinetic and the potential energies.
This indicates that the nucleus has a strong quantum nature.
In the mean-field theory,
the total (binding) energy is given by
\begin{equation}
\label{B2}
-B = \sum_{i=1}^A \left( T_i + \frac{V}{2} \right)
        = A \left( \frac{3}{5}T_F + \frac{V}{2} \right) ,
\end{equation}
where we assume that the average potential results from a two-body
interaction.
The two kinds of expressions for $B/A$, Eqs. (\ref{B1}) and (\ref{B2}),
lead to $T_F\approx -5V/4\approx 70$ MeV,
which is different from the previously estimated value ($\sim 40$ MeV).
Moreover,  the negative separation energy ($T_F +V >0$)
contradicts the fact that the nucleus is bound!
The presence of a three-body interaction may change this argument.
However, solving the present contradiction would require
an unrealistically strong three-body repulsive
effect whose magnitude is comparable to that of the attractive two-body
interaction.

To reconcile
the independent-particle motion with the saturation property of the nucleus,
the nuclear average potential must be state dependent.
Allowing the potential $V_i$ to depend on the state $i$,
the potential $V$ should be replaced by that for the highest occupied
orbital $V_F$ in Eq. (\ref{B1}),
and by its average value $\langle V \rangle$ in
the right-hand side of Eq. (\ref{B2}).
Then, we obtain the following relation:
\begin{equation}
\label{V_F}
V_F \approx \langle V \rangle + T_F/5 + B/A .
\end{equation}
The potential $V_F$ is shallower
than its average value $\langle V \rangle$.
\textcite{Wei57} suggested the momentum-dependent potential $V$, which can be
expressed in terms of an effective mass $m^*$:
\begin{equation}
\label{mom_dep_pot}
V_i=U_0+U_1\frac{k_i^2}{k_F^2} .
\end{equation}
In fact, the non-local mean-field potential can be
expressed by the momentum dependence \cite{RS80}.
Equation (\ref{mom_dep_pot})
leads to the effective mass, $m^*/m = (1+U_1/T_F)^{-1}$.
Using Eqs. (\ref{B1}), (\ref{V_F}), and (\ref{mom_dep_pot}),
we obtain the effective mass given by
\begin{equation}
\label{m*/m}
\frac{m^*}{m} = \left\{ \frac{3}{2} + \frac{5}{2}\frac{B}{A}\frac{1}{T_F}
 \right\}^{-1} \approx 0.4 .
\end{equation}
Quantitatively, this value disagrees with the experimental data.
Although the empirical values of the effective mass
vary according to the energy of nucleons,
$0.7 \lesssim m^*/m \lesssim 1$,
they are almost twice larger than
the value of Eq. (\ref{m*/m}).
Furthermore, the total energy, Eq.~(\ref{B2}), is written as
\begin{equation}
\label{B3}
-B=\frac{1}{2}\sum_{i=1}^A \left( T_i + \epsilon_i \right) ,
\end{equation}
where $\epsilon_i$ are single-particle energies.
Within the constraint of Eq. (\ref{B3}), 
it is impossible to reproduce both the total binding energy and
the single-particle spectra observed in experiments.
As far as we use a normal two-body interaction,
these discrepancies should be present in the mean-field calculation
with any interaction,
because Eqs. (\ref{m*/m}) and (\ref{B3}) are valid in general
for a saturated self-bound system.
Therefore, the naive mean-field models
have a fundamental difficulty to describe the nuclear saturation.

The DFT provides a practical solution to this problem,
in which we start from an EDF, $E[\rho]$, instead of the interaction.
The KS field is calculated as $h[\rho]=\partial E/\partial\rho$,
which may contain the non-trivial
density dependence different from that of the mean-field theory
starting from the interaction.
In nuclear physics, this additional density dependence was introduced
by the density-dependent effective interaction, thus, it was
called ``density-dependent Hartree-Fock'' (DDHF) method \cite{Neg70}.
In this terminology,
the variation of the total energy with respect to the density
contains re-arrangement potential, $\partial V_{\rm eff}[\rho]/\partial\rho$,
which comes from the density dependence of the effective force
$V_{\rm eff}[\rho]$.
These terms are crucial to obtain the saturation
and to provide a consistent
independent-particle description of nuclei.

\subsection{Symmetry breaking and restoration
by the Anderson-Nambu-Goldstone (ANG) modes}
\label{sec:ANG}

\begin{figure}[t]
\includegraphics[width=5cm]{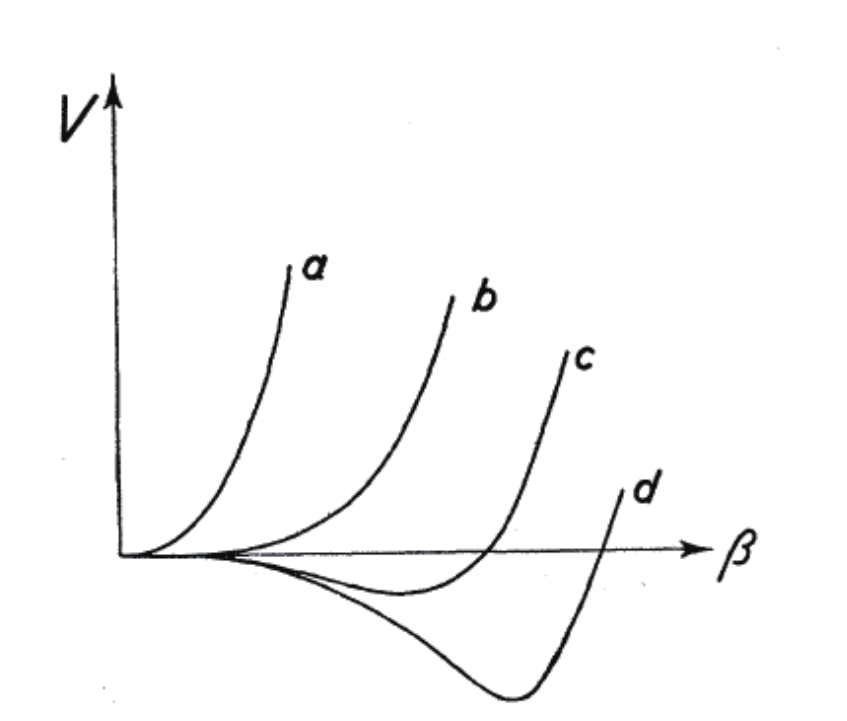}
\caption{
Sketch of potential energy curves as functions of deformation parameter
$\beta$.
The curve a corresponds to spherical nuclei,
b and c correspond to transitional nuclei,
and d to well deformed nuclei.
From \textcite{Ald56}.
}
\label{fig:SSB}
\end{figure}

One of the prominent features in the nuclear EDF approaches is
the appearance of the SSB.
For the system of a small number of particles, such as nuclei,
the SSB is hidden.
The experimental measurements probe the states which preserve
the symmetries of the original Hamiltonian.
In nuclear physics, the state with a broken symmetry is often called
``intrinsic'' state.
Nevertheless, we clearly observe a number of nuclear phenomena associated with
effects of the SSB, both in the ground-state properties and
in excitation spectra.
In nuclear physics, this was realized in 1950's, soon after the experimental
identification of the characteristic patterns of rotational spectra.
Figure~\ref{fig:SSB} is taken from 
a seminal review paper on the Coulomb excitation \cite{Ald56}.
The nuclear potential energy function
clearly indicates the nuclear
deformation as the phase transition involving the SSB.
The SSB in small finite-size systems has been an important concept
in nuclear physics and chemistry for many years, and has become so
in fields of quantum dots and ultracold atoms \cite{YL07}.

\begin{figure}[t]
\includegraphics[width=0.3\textwidth]{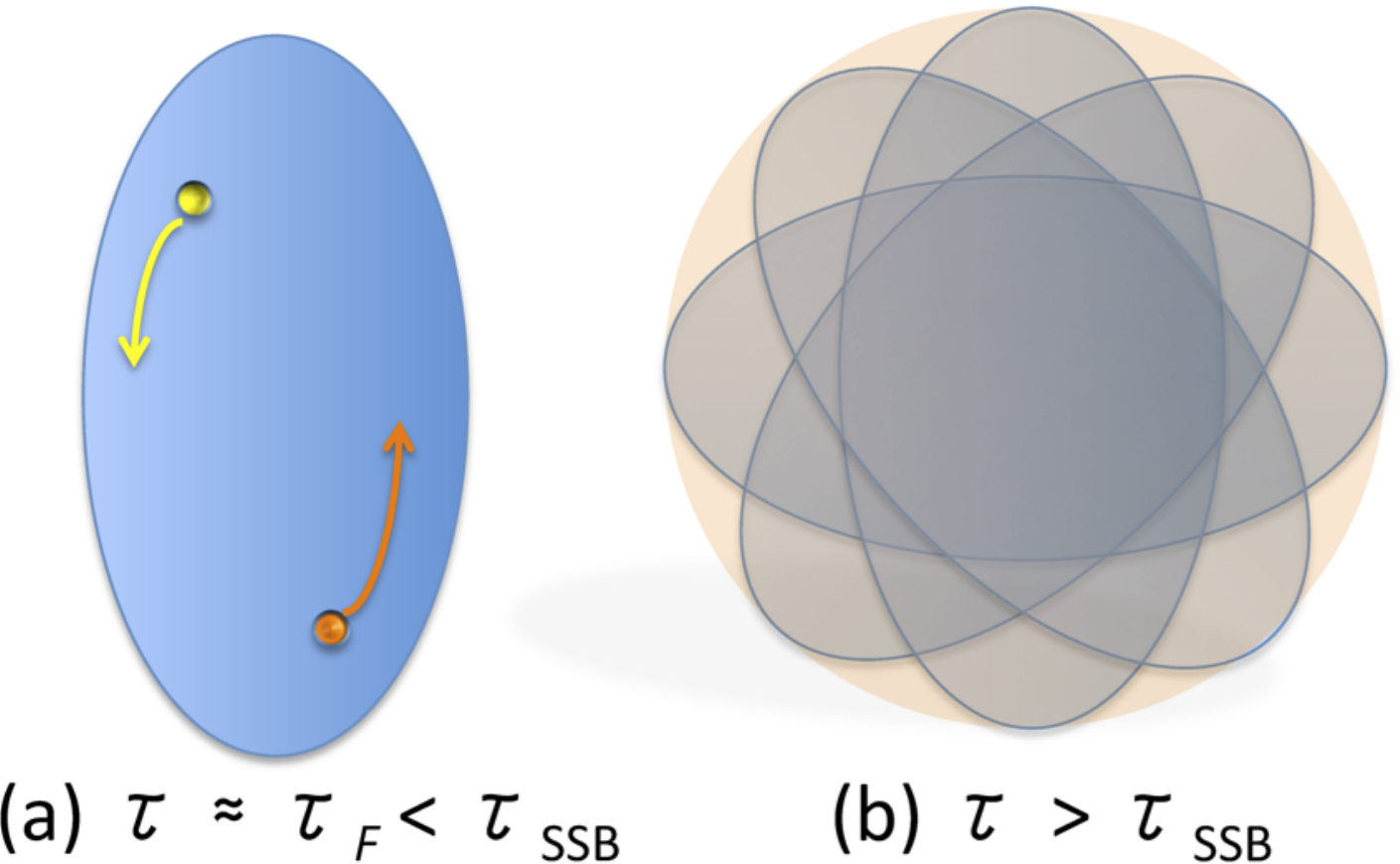}
\caption{
(Color online)
Illustration of deformed nuclei in different time scales,
(a) shorter than the symmetry restoring time $\tau_\textrm{\sc ssb}$,
and (b) larger than $\tau_\textrm{\sc ssb}$.
}
\label{fig:tau_SSB}
\end{figure}

The symmetry restoration is a quantum fluctuation effect.
When the spontaneous breaking of the continuous symmetry occurs,
there exists the Anderson-Nambu-Goldstone (ANG) modes,
to restore the broken symmetry \cite{And58-1,Nam60,Gol61,And63}.
This symmetry restoration process is extremely slow for macroscopic objects,
thus, the SSB is realized in a rigorous sense.
In other words, the quantum fluctuation associated with the ANG mode
is negligibly small in those cases.
If the deformed nucleus with extremely heavy mass ($A\rightarrow\infty$)
existed,
the moment of inertia ${\cal J}$ should be macroscopically large.
Then, the excitation spectra of this heavy rotor would be nearly degenerate
with the ground state, $E_I=I(I+1)/2{\cal J}$ for the state with the total
angular momentum $I$, leading to a stable deformed wave packet.
In reality, the restoration of the rotational symmetry even in heaviest nuclei
takes place much faster than the shortest time resolution we can achieve with
the present experimental technologies.
In this sense, the SSB in nuclei is hidden.
Nevertheless, the nucleonic motion is strongly influenced by the SSB,
since the time scale of the symmetry restoration, $\tau_\textrm{\sc ssb}$,
is much longer than
the periodic time of single-particle motion in the nucleus of
radius $R$,
$\tau_F=R/v_F\sim 10^{-22}$~s.
This is schematically illustrated in Fig.~\ref{fig:tau_SSB}.
We believe that it is important to distinguish
these two types of correlations in nuclei,
those of relatively short time scales $\tau\sim \tau_F$ ({\it ``fast'' motion}),
and of long time scales $\tau\sim \tau_\textrm{\sc ssb}$
({\it ``slow'' motion}).

The nuclear superfluidity can be understood exactly in an
analogous way, as the SSB leading to the deformation in the gauge space
\cite{BB05}.
The condensate of the nucleonic Cooper pairs is expressed as an intrinsic
deformation in the magnitude of the pair field.
The pair field creates and annihilates the pairs of nucleons
giving rise to the quasiparticles that are superpositions of
particles and holes, expressed by the Bogoliubov transformation.
The ANG mode, called pair rotation,
corresponds to the addition and removal of the nucleon pairs
from the pair condensate.
In this case, the ``angular momentum'' in the gauge space
corresponds to the particle number, and
the ``moment of inertia'' is defined by
the second derivative of the ground-state energy with respect to the
particle number, $\cJ_{\rm pair}=(d^2 E_N/d N^2)^{-1}$.
See also Sec.~\ref{sec:normal_modes}.

Since the broken symmetry is restored by the quantum fluctuation,
its time scale $\tau_\textrm{\sc ssb}$ can be estimated by
the uncertainty principle.
The time is proportional to the moment of inertia $\mathcal{J}$ as
$\tau_\textrm{\sc ssb}\sim \mathcal{J}/\hbar$,
which amounts to $10^{-20}-10^{-19}$ s for typical deformed nuclei
in the rare-earth and the actinide regions.
Thus, the symmetry restoration is a slow motion, compared to
the nucleonic Fermi motion.
Here, it is important to distinguish this time scale of the ``quantum''
fluctuation from that of the ``classical'' rotation,
$\omega_{\rm rot}^{-1}\approx \mathcal{J}/I$.
The latter could be comparable to $\tau_F$ at very high spin (large $I$),
however, the concept of the deformation (symmetry breaking) still holds.
For the pair rotation,
using an observed value of the moment of inertia for the pair rotation
in Sn isotopes \cite{BB05},
$\tau_\textrm{\sc ssb}$ for the symmetry breaking in the gauge space
can be given by
$\tau_\textrm{\sc ssb}= 10^{-21}-10^{-20}$ s.

These concepts of SSB are 
invoked
in the nuclear DFT and TDDFT.
The symmetry restoration can be treated either by the projection
method or by the (time-dependent) large-amplitude collective motion
of the ANG modes \cite{RS80}.
In the present review, we mainly discuss the latter treatment with
the time-dependent description.

\section{Basic formalism: DFT and TDDFT}
\label{sec:basic_formalism}

The DFT describes a many-particle system
exactly in terms of its
local one-body density $\rho(\vec{r})$ alone.
The DFT is based on the original theorem of Hohenberg and Kohn (HK)
\cite{HK64}
which was proved for the ground-state of the many-particle system.
Every observable can be written, in principle, as
a functional of density.

In nuclear physics, as is discussed in Sec.~\ref{sec:ANG},
many kinds of SSB takes place without an external potential.
In fact, the minimization of the nuclear EDF for finite nuclei always
produces a localized density profile, which
spontaneously violates the translational symmetry.
Furthermore, it often violates the rotational symmetry
in the real space and the gauge space.
The SSB enables us to introduce 
an intrinsic (wave-packet) state.
A possible justification of the DFT for the intrinsic state
is presented in Sec.~\ref{sec:DFT_wave_packet}.

For finite many-fermion systems, the shell effects associated with
the quantum nature of the Fermi motion
play a major role to determine the ground state.
The KS scheme \cite{KS65} gives a practical treatment of
the shell effects in the density functional.
This is presented in Sec.~\ref{sec:Kohn-Sham}

The DFT is designed for calculating the ground-state properties.
For excited-state properties and reactions,
the TDDFT
is a powerful and useful tool.
The basic theorem for the TDDFT has been developed as an exact
theorem \cite{RG84}, similar to the HK theorem in the static case.
This is reviewed in Sec.~\ref{sec:TDDFT}.

Both the DFT and TDDFT have been extended to the superconductors,
introducing an external pair-removal and pair-addition potential
\cite{OGK88,WKG94}.
These extensions are relevant to nuclear physics as well,
to account for various properties of heavy open-shell nuclei.
In this article, we call them ``superconducting nuclei'' or
``nuclear superfluidity''.
Properties of the (time-dependent) Bogoliubov-de-Gennes equations will be
presented in Secs.~\ref{sec:Kohn-Sham} and \ref{sec:TDDFT}.

\subsection{Nuclear EDF models}
\label{sec:NEDF_models}

Before presenting the theorem of DFT,
we recapitulate basic equations of
nuclear EDF models
and their properties
\cite{RS80,BR86,BHR03}.

\subsubsection{Basic equations}
\label{sec:NEDF_basic_eq}

To simplify the discussion, we assume that the EDF
$F[\rho]$, which represents the total energy of the nucleus,
is a functional of local density $\rho(\vec{r})$
without the spin-orbit coupling.
The KS equations read with the spin index
$\sigma=(\uparrow,\downarrow)$,
\begin{equation}
\label{KS_eq}
\left(-\frac{1}{2m}\nabla^2 + v_s(\vec{r}) \right)
 \varphi_i (\vec{r}\sigma)= \epsilon_i \varphi_i(\vec{r}\sigma) .
\end{equation}
Hereafter, we use the unit $\hbar=1$.
We decompose $F[\rho]$ into two parts, $F[\rho]=T_s[\rho] + E_c[\rho]$,
where
$
T_s[\rho]=\sum_{i=1}^N \bra{\varphi_i} \left(-\frac{1}{2m} \nabla^2 \right)
\ket{\varphi_i}
$
and the rest $E_c[\rho]$.
The KS potential is defined by
$v_s(\vec{r})=\delta E_c/\delta\rho(\vec{r})$.
The density is given by summing up the KS orbitals,
\begin{equation}
\label{KS_density}
\rho(\vec{r})=
\sum_\sigma \sum_{i=1}^N |\varphi_i(\vec{r}\sigma)|^2
.
\end{equation}

When we take into account the nuclear superfluidity,
we adopt an EDF which is a functional of $\rho$ and $(\kappa,\kappa^*)$,
including the pair tensor $\kappa(\vec{r},\vec{r'})$
whose definition (\ref{kappa})
requires a symmetry-broken wave-packet state in
Sec.~\ref{sec:DFT_wave_packet}.
If the EDF depends only on their diagonal parts (pair density
$\kappa(\vec{r})$), 
$F[\rho,\kappa]=T_s[\rho,\kappa]+E_c[\rho,\kappa]$,
Eq. (\ref{KS_eq}) should be extended to
the Bogoliubov-de-Gennes-KS (BdGKS) equations:
\begin{equation}
\label{BdGKS_eq}
\begin{split}
\sum_{\sigma'}
\begin{pmatrix}
(h_s(\vec{r})-\mu)\delta_{\sigma\sigma'} & \Delta_s(\vec{r})\gamma_{\sigma\sigma'} \\
-\Delta_s^*(\vec{r})\gamma_{\sigma\sigma'} & -(h_s(\vec{r})-\mu)^*\delta_{\sigma\sigma'}
\end{pmatrix}
\\
\times
\begin{pmatrix}
U_i(\vec{r}\sigma') \\
V_i(\vec{r}\sigma')
\end{pmatrix}
= E_i
\begin{pmatrix}
U_i(\vec{r}\sigma) \\
V_i(\vec{r}\sigma)
\end{pmatrix}
,
\end{split}
\end{equation}
where $h_s(\vec{r})\equiv -\nabla^2/(2m)+v_s(\vec{r})$,
$\gamma_{\uparrow\downarrow}=-\gamma_{\downarrow\uparrow}=1$,
and $\gamma_{\uparrow\uparrow}=\gamma_{\downarrow\downarrow}=0$.
The chemical potential $\mu$ is introduced to control the
total particle number.
The potentials, $v_s(\vec{r})$ and $\Delta_s(\vec{r})$,
are respectively defined by
\begin{equation}
\label{BdGKS_potentials}
v_s(\vec{r})=\frac{\delta E_c}{\delta\rho(\vec{r})} ,
\quad\quad
\Delta_s(\vec{r})=\frac{\delta E_c}{\delta\kappa^*(\vec{r})} .
\end{equation}
The normal and pair densities are given by
$
\rho(\vec{r})=\sum_\sigma \sum_i |V_i(\vec{r}\sigma)|^2$  and
$
\kappa(\vec{r})= \sum_i V_i^*(\vec{r}\uparrow) U_i(\vec{r}\downarrow)
$,
where the summation with respect to $i$ is taken over all the states with
positive quasiparticle energies, $E_i>0$.
The same convention is assumed in this article.

In nuclear physics, Eqs. (\ref{KS_eq}) and (\ref{BdGKS_eq}) are
often called Hartree-Fock (HF) and Hartree-Fock-Bogoliubov (HFB) equations,
respectively\footnote{
The time-dependent equations are also called
TDHF (TDHFB) in nuclear physics,
instead of TDKS (TDBdGKS).
}.
Accordingly, the quasiparticle vacuum, $\ket{\phi_0}$, is introduced 
and called HFB ground state, where the Bogoliubov transformation
\begin{equation}
\label{psi}
\hat{\psi}(\vec{r}\sigma)=\sum_i \left\{
U_i(\vec{r}\sigma) a_i + V_i^*(\vec{r}\sigma) a_i^\dagger 
\right\} ,
\end{equation}
defines the quasiparticle annihilation and creation operators,
$(a_i,a_i^\dagger)$,
with the vacuum condition
\begin{equation}
\label{vacc}
a_i\ket{\phi_0}=0 .
\end{equation}

These mean-field terminologies are due to practical usage of effective 
density-dependent interaction.
The EDF is provided by the expectation value of the effective
Hamiltonian at $\ket{\phi_0}$,
$F[\rho,\kappa]=\bra{\phi_0}\hat{H}-\mu\hat{N}\ket{\phi_0}$.
The variation,
$\delta(F[\rho,\kappa]-\mu\int \rho(\vec{r})d\vec{r})=0$, leads to
Eqs. (\ref{KS_eq}) and (\ref{BdGKS_eq}).

\subsubsection{Properties of BdGKS equations and useful notations}
\label{sec:useful_notations}

Solutions of the BdGKS (HFB) equations (\ref{BdGKS_eq})
are {\it paired} in the following sense:
For each quasiparticle eigenstate with $\Psi^0_i$
with a positive eigenvalue $E_i$,
there exists a partner eigenstate,
$\tilde{\Psi}^0_i$ with the negative energy $-E_i$.
\begin{equation*}
\Psi^0_i=\begin{pmatrix}U_i \\ V_i \end{pmatrix},
\quad
\tilde{\Psi}^0_i=\begin{pmatrix}V_i^* \\ U_i^* \end{pmatrix} .
\end{equation*}
Introducing the collective notation of the quasiparticles with
positive [negative] energies
at the ground state $\Psi^0=(\Psi^0_1,\Psi^0_2,\cdots)$
[$\tilde\Psi^0=(\tilde\Psi^0_1,\tilde\Psi^0_2,\cdots)$],
the orthonormal and completeness relations are equivalent to
the unitarity condition, ${\cal W W^\dagger = W^\dagger W} = 1$,
of the matrix
\begin{equation*}
{\cal W}\equiv
\begin{pmatrix}
\Psi^0 & \tilde\Psi^0
\end{pmatrix}
=
\begin{pmatrix}
U_1 & U_2 \cdots & V_1^* & V_2^* & \cdots \\
V_1 & V_2 \cdots & U_1^* & U_2^* & \cdots
\end{pmatrix}
.
\end{equation*}

The generalized density matrix $R^0$ is defined as
\begin{equation}
\label{R}
R^0=\begin{pmatrix}
\rho & \kappa \\
-\kappa^* & 1-\rho^* 
\end{pmatrix}
= 1-\Psi^0 \Psi^{0\dagger}
= \tilde\Psi^0 \tilde\Psi^{0\dagger}
,
\end{equation}
which is Hermitian and idempotent ($(R^0)^2=R^0$).
The orthonormal property immediately gives
\begin{equation}
\label{orthonormal}
R^0 \Psi^0_i = 0 ,
\quad
R^0 \tilde\Psi^0_i = \tilde\Psi^0_i .
\end{equation}
The BdGKS equations can be rewritten in terms of $R$ as
\begin{equation}
\label{BdGKS_R_eq}
\left[ H_s[R^0], R^0 \right] = 0 ,
\end{equation}
where $H_s[R^0]$ is the BdGKS (or HFB) Hamiltonian
in the left hand side of Eq. (\ref{BdGKS_eq}).

Any unitary transformation among the quasiparticles,
$a'_i=\sum U_{ij} a_j$, keeps $R^0$ ($\rho$,$\kappa$) invariant.
The quasiparticles defined by Eq. (\ref{BdGKS_eq}) gives
one choice of the gauge (``quasiparticle representation'').
Another common choice is called ``canonical representation'',
in which the density matrix $\rho$ is diagonal.
Note that Eqs. (\ref{R}), (\ref{orthonormal}),
and (\ref{BdGKS_R_eq}) are all independent of the choice of
the gauge.

In Sec.~\ref{sec:NEDF_basic_eq},
we used the coordinate-space representation $(\vec{r},\sigma)$,
and have assumed that the density functional $E_c[\rho,\kappa]$ depends
only on the diagonal densities ($\rho(\vec{r}),\kappa(\vec{r})$)
without spin dependence.
It can be easily generalized to other representation
$(\alpha, \beta, \cdots)$ and to functionals of density matrices in general;
Hermitian $\rho_{\alpha\beta}$ and anti-symmetric $\kappa_{\alpha\beta}$.
The potentials are given by
\begin{equation*}
v_s(\alpha\beta)= \left.
\frac{\delta E_c}{\delta \rho_{\beta\alpha}}\right|_{R^0} , \quad
\Delta_s(\alpha\beta)=\left.
-\frac{\delta E_c}{\delta \kappa^*_{\beta\alpha}}\right|_{R^0} .
\end{equation*}
Then, the BdGKS Hamiltonian can be written as
\begin{equation}
\label{H_s}
H_s[R^0](\alpha'\beta')
=\left. \frac{\delta F}{\delta R_{\beta'\alpha'}}\right|_{R^0},
\end{equation}
where $F[R]=T_s[R]+E_c[R]$.
Here, we have introduced the {\it primed} indices,
which are double the dimension.
Let the dimension of the single-particle space be $d$, then
the {\it unprimed} index runs over
$
\alpha = 1,\cdots,d,
$
while the {\it primed} one
$\alpha' = 1,\cdots,2d$.

It is not necessary, but often useful to introduce the
generalized Slater determinant (quasiparticle vacuum) $\ket{\phi}$,
defined by $a_i\ket{\phi}=0$.
Then, we may denote $R^0$ as
\begin{equation*}
R^0_{\alpha'\beta'}=\bra{\phi}
\begin{pmatrix}
\hat{\psi}^\dagger_\beta \hat{\psi}_\alpha &
\hat{\psi}_\beta \hat{\psi}_\alpha \\
\hat{\psi}^\dagger_\beta \hat{\psi}^\dagger_\alpha &
\hat{\psi}_\beta \hat{\psi}^\dagger_\alpha
\end{pmatrix}
\ket{\phi} .
\end{equation*}
A one-body operator $\hat{O}$
can be written in a form
\begin{eqnarray}
\hat{O}&=&\sum_{\alpha\beta} \left[
f_{\alpha\beta} \hat{\psi}^\dagger_\alpha\hat{\psi}_\beta
+\frac{1}{2} \left\{
g_{\alpha\beta} \hat{\psi}^\dagger_\alpha\hat{\psi}^\dagger_\beta
+g'_{\alpha\beta} \hat{\psi}_\alpha\hat{\psi}_\beta
\right\} \right] \nonumber\\
&=& \textrm{const.} +
\frac{1}{2}
\begin{pmatrix}
\hat{\psi}^\dagger & \hat{\psi}
\end{pmatrix}
\begin{pmatrix}
f & g \\
g' & -f^T
\end{pmatrix}
\begin{pmatrix}
\hat{\psi} \\
\hat{\psi}^\dagger
\end{pmatrix}
,
\label{O_operator}
\end{eqnarray}
where $g$ and $g'$ are anti-symmetric.
If $\hat{O}$ is Hermitian, we have
$f^T=f^*$ and $g^*=-g'$ ($g^\dagger=g'$).

The Bogoliubov transformation (\ref{psi}) is written with
the unitary matrix ${\cal W}$ as
\begin{equation*}
\begin{pmatrix}
a \\
a^\dagger
\end{pmatrix}
= {\cal W}^\dagger
\begin{pmatrix}
\hat{\psi} \\
\hat{\psi}^\dagger
\end{pmatrix}
,\quad
\begin{pmatrix}
\hat{\psi} \\
\hat{\psi}^\dagger
\end{pmatrix}
= {\cal W}
\begin{pmatrix}
a \\
a^\dagger
\end{pmatrix}
.
\end{equation*}
This transforms Eq. (\ref{O_operator}) into
\begin{eqnarray}
\hat{O}&=& 
 \sum_{ij}\left[
O_{ij}^{(++)} a_i^\dagger a_j
+\frac{1}{2}\left\{
O^{(+-)}_{ij} a_i^\dagger a_j^\dagger
+O^{(-+)}_{ij} a_i a_j
\right\} \right] \nonumber\\
&=&  
\frac{1}{2}
\begin{pmatrix}
a^\dagger & a
\end{pmatrix}
\begin{pmatrix}
O^{(++)} & O^{(+-)} \\
O^{(-+)} & O^{(--)}
\end{pmatrix}
\begin{pmatrix}
a \\
a^\dagger
\end{pmatrix}
\label{O_matrix}
,
\end{eqnarray}
where $O^{(--)}=-O^{(++)T}$ and the constant shift is ignored.
The matrices appearing in Eqs. (\ref{O_operator}) and (\ref{O_matrix}) are
essentially identical, but in different representation.
We symbolically denote this as $O$.
The superscript indices ``$+$'' and ``$-$''
 indicate the positive- and negative-energy
states; $O^{(++)}_{ij}\equiv\Psi_i^{0\dagger} O \Psi_j^0$,
 $O^{(+-)}_{ij}\equiv\Psi_i^{0\dagger} O \tilde{\Psi}_j^0$,
 $O^{(-+)}_{ij}\equiv\tilde{\Psi}_i^{0\dagger} O \Psi_j^0$, and
 $O^{(--)}_{ij}\equiv\tilde{\Psi}_i^{0\dagger} O \tilde{\Psi}_j^0$.
The matrix elements, $O_{ij}^{(+-)}$ and $O_{ij}^{(-+)}$,
correspond to the two-quasiparticle
creation and annihilation parts, respectively,
which are occasionally denoted as $O^{20}_{ij}$ and $O^{02}_{ji}$
in literature \cite{RS80,AN11}.
The block elements of the density is also written as
$R=
\begin{pmatrix}
R^{(++)} & R^{(+-)} \\
R^{(-+)} & R^{(--)}
\end{pmatrix}
$.
The expectation value of $\hat{O}$ is given by
$\bra{\phi} \hat{O} \ket{\phi} = (1/2)\textrm{tr}[OR^0]$.
These matrix notations are frequently used in Sec.~\ref{sec:linear_response}.

\subsection{DFT theorem for a wave-packet state}
\label{sec:DFT_wave_packet}

The DFT is based on the HK theorem which 
guarantees a one-to-one mapping between a one-body
density $\rho(\vec{r})$ for the ground state
and an external potential $v_0(\vec{r})$.
According to the recent progress \cite{Eng07,Bar07,Gir08,MBS09},
the theorem is extended to functionals of the localized intrinsic density
of self-bound systems.
Thus, it is a functional of density $\rho(\vec{r}-\vec{R})$ where
$\vec{R}$ is the center of mass.
In contrast to the center-of-mass motion,
a strict definition of the intrinsic state is not trivial for the
rotational motion of a deformed nucleus.
In this section, we show a possible justification of the functional
of the wave-packet density produced by the SSB in finite systems.
The arguments presented here were given by \textcite{GJB08}.
The argument is exact for the SSB in the translational symmetry,
while it is approximate for the SSB in the rotational symmetry.

\subsubsection{Principles}
\label{sec:principles}

A useful fact is that the SSB of the continuous symmetries produces
ANG modes which are decoupled
from the other degrees of freedom.
It is exactly true in the case of translational symmetry.
Consequently, there appear the collective variables associated
with the ANG modes, which are symbolically denoted as $(q,p)$.
Here, $p$ are conserved and $q$ are cyclic variables.
The decoupling allows us to define the collective subspace
$\Sigma_\textrm{\sc ang}$
in the whole Hilbert space of many-particle systems.
$\Sigma_\textrm{\sc ang}$
is the space spanned by the collective wave functions, $\{ \chi(q) \}$.
The subspace orthogonal to $\Sigma_\textrm{\sc ang}$, which
is denoted as $\Sigma_\textrm{intr}$, describes the intrinsic motion.

In the ideal case of the SSB in the translational symmetry,
the center-of-mass variables, $(q,p)=(\vec{R},\vec{P})$, and
the intrinsic variables, $(\xi,\pi)$, are exactly decoupled.
The state $\ket{\Phi}$ is rigorously given by a product wave function of
$\phi(\xi)$ and $\chi(\vec{R})$.
\begin{equation*}
\ket{\Phi} = \ket{\phi} \otimes \ket{\chi} , \quad
\hat{H} = \hat{H}_{\rm intr}(\xi,\pi) + \frac{\vec{P}^2}{2M} ,
\end{equation*}
where $M$ is the total mass.
In this case, the intrinsic subspace $\Sigma_\textrm{intr}$ is
defined by the space spanned by $\{ \phi(\xi) \}$.
The intrinsic ground state is obtained by
the minimization of the intrinsic energy
$\bra{\phi}\hat{H}_\textrm{intr}\ket{\phi}$
 in the subspace $\Sigma_\textrm{intr}$.
The choice of the center-of-mass motion $\chi(\vec{R})$ is arbitrary for
the determination of $\phi(\xi)$.
Thus, we can adopt a localized  form of $\chi(\vec{R})$,
such as a Gaussian form
$\chi(\vec{R})\propto \exp[-(\vec{R}-\vec{R}_0)^2/2b^2]$.
This leads to the wave-packet state $\ket{\Phi}$.
Using the operator $\hat{P}$ which projects $\chi(q)$ onto
the $\vec{P}=0$ state,
the ground-state energy can be obtained 
by the variation after the projection:
\begin{equation}
\label{chi-fixed}
E_0\equiv
\min
\left[ \frac{\bra{\Phi}\hat{H} \hat{P}\ket{\Phi}}{\bra{\Phi}\hat{P}\ket{\Phi}}
\right]_{\Sigma_\textrm{intr}}
\end{equation}
where the variation is performed only in $\Sigma_\textrm{intr}$
with a fixed $\chi(\vec{R})$.

In general, the wave-packet state is constructed in an analogous way.
Choosing a localized form of $\chi(q)$,
e.g., $\chi(q)\propto \exp[-(q-q_0)^2/2b^2]$,
the variation after projection is performed in
a restricted space $\Sigma_\textrm{intr}$.
The projection operator $\hat{P}$ 
makes the state an eigenstate of the collective momentum
(symmetry operator) $p$; $p \hat{P}\ket{\Phi}=p_0 \hat{P}\ket{\Phi}$.
Then, Eq. (\ref{chi-fixed}) produces the ground-state energy with $p=p_0$.
In nuclear physics interests, in addition to the total momentum,
$p$ may stand for either
the total angular momentum $J$, or the neutron (proton) number $N$ ($Z$).
The wave-packet density profile is simply given by
\begin{equation*}
\rho(\vec{r})\equiv \sum_\sigma \bra{\Phi} \hat{\psi}^\dagger(\vec{r}\sigma)
                                   \hat{\psi}(\vec{r}\sigma)\ket{\Phi},
\end{equation*}
that depends on the choice of $\chi(q)$.
In this article, we omit the isospin index $\tau=(n,p)$ for simplicity.
Since we adopt a localized $\chi(q)$ which violates the symmetry,
the density $\rho(\vec{r})$ is also localized, or ``deformed''.

In order to find the (wave-packet) density functional,
we use the constrained search \cite{Lev79}.
The minimization in Eq. (\ref{chi-fixed}) is divided into two steps;
one first considers only states that produce a given wave-packet density
$\rho(\vec{r})$,
and next takes the variation with respect to the density.
\begin{equation*}
E_0=\min_\rho \left\{
\min_{\Phi\rightarrow\rho}
\left[ \frac{\bra{\Phi}\hat{H} \hat{P}\ket{\Phi}}{\bra{\Phi}\hat{P}\ket{\Phi}}
\right]_{\Sigma_\textrm{intr}} \right\}
.
\end{equation*}
This leads to the universal density functional
\begin{equation*}
F[\rho]\equiv
\min_{\Phi\rightarrow\rho}
\left[ \frac{\bra{\Phi}\hat{H} \hat{P}\ket{\Phi}}{\bra{\Phi}\hat{P}\ket{\Phi}}
\right]_{\Sigma_\textrm{intr}} .
\end{equation*}
Thus, the energy of the ground state with $p=p_0$ 
may be obtained by the minimization,
$E_0=\min F[\rho]$.

The SSB of the gauge symmetry in nuclear superfluidity
is caused by the pairing correlations
among nucleons.
Thus, in practice, it is convenient to introduce the pair tensors
for the wave-packet state as
\begin{equation}
\label{kappa}
\kappa(\vec{r}\sigma;\vec{r}'\sigma')\equiv
\bra{\Phi} \hat{\psi}(\vec{r}'\sigma') \hat{\psi}(\vec{r}\sigma)\ket{\Phi} .
\end{equation}
In other words, it is easier to construct the density functional
$F[\rho,\kappa,\kappa^*]$ than $F[\rho]$,
which takes account of essential aspects of the pairing correlations.
Hereafter, we denote $F[\rho,\kappa]$ omitting $\kappa^*$
for simplicity.
Following the above idea of the constrained search,
it is easy to define the functional of $\rho$ and $\kappa$,
\begin{equation}
\label{F_rho_kappa}
F[\rho,\kappa]\equiv
\min_{\Phi\rightarrow(\rho,\kappa)}
\left[ \frac{\bra{\Phi}\hat{H} \hat{P}\ket{\Phi}}{\bra{\Phi}\hat{P}\ket{\Phi}}
\right]_{\Sigma_\textrm{intr}} .
\end{equation}
Instead of adopting the full pair tensors of (\ref{kappa}),
one can restrict them to their ``diagonal'' parts,
$\kappa(\vec{r})\equiv \kappa(\vec{r}\uparrow,\vec{r}\downarrow)
=-\kappa(\vec{r}\downarrow,\vec{r}\uparrow)$,
in the functional.
The inclusion of other form of densities, in addition to $\rho$ and $\kappa$,
can also be achieved exactly in the same manner.

Let us make a few remarks here.
First, in general, $\rho(\vec{r})$ and $\kappa(\vec{r})$ are not
the exact densities in the laboratory frame
\cite{SR91}.
Thus, $F[\rho,\kappa]$ is the functional of ``localized''
wave-packet densities $\rho(\vec{r})$ and $\kappa(\vec{r})$.
Second, when $\Sigma_\textrm{\sc ang}$ describes the center-of-mass motion,
the decoupling is exact, and in principle, $E_0=\min F[\rho]$
gives the exact ground state energy.
On the other hand, when the decoupling is approximate, such as
the SSB in the rotational symmetry,
$E_0=\min F[\rho,\kappa]$ provides an approximate ground state energy with
the deformed densities.
In the strict sense, an ``infinite'' system has an exact deformed ground state
with an arbitrary fixed orientation, $\chi(q)\sim \delta(q-q_0)$.
This limit is not realized in finite nuclei.
Nevertheless,
we expect that the approximation becomes better for heavier nuclei.
Third, in the approximate decoupling,
the subspace $\Sigma_\textrm{intr}$ should be chosen to be optimal
for a certain eigenvalue of $p_0$.
Therefore, $F[\rho,\kappa]$ may depend on $p_0$.

\subsubsection{Practices}
\label{sec:practices}

We have discussed the principles of the DFT for the wave-packet
state.
The universal density functional $F[\rho,\kappa]$ can be a functional
of ``local'' densities, $\rho(\vec{r})$ and $\kappa(\vec{r})$,
in principle.
However, 
even if the existence is guaranteed,
it is another issue in practice whether we can construct
the accurate density functional in terms of
$\rho(\vec{r})$ and $\kappa(\vec{r})$ only.

For a proper account of the shell effects,
the inclusion of the kinetic density is the only practical solution
at present (Sec.~\ref{sec:Kohn-Sham}).
Furthermore, for the shell structure in finite nuclei with
correct magic numbers, it is indispensable to take into account the
spin-orbit splitting \cite{MJ55}.
Currently, we need to adopt the spin-current densities for this
purpose.
In the end, the currently available EDFs for realistic applications 
contains several kinds of densities.
The Skyrme and point-coupling covariant EDFs
consist of local densities,
while the Gogny and covariant (relativistic) EDFs contain
non-local ones.
Actual forms of the EDFs can be found in \textcite{BHR03}.

In Sec.~\ref{sec:principles}, we have presented an argument that
the approximate decoupling for the rotational degrees of freedom
justifies the use of the EDF of {\it deformed} densities.
It is reliable for describing the correlations at $\tau\sim\tau_F$
in Fig.~\ref{fig:tau_SSB} (a).
These include the shell effects and the saturation properties.
Conversely, the existing EDFs have difficulties to simultaneously
reproduce binding energies of spherical and deformed nuclei.
This may be due to a missing correlation associated with
the quantum rotation of deformed intrinsic shapes
shown in Fig.~\ref{fig:tau_SSB} (b).
In Sec.~\ref{sec:linear_response_appli}, we show that
the giant resonances (fast collective motion) are well reproduced in the
linear-response calculations, while the low-energy vibrations
(slow collective motion) are not as good as those.
In
our opinions,
correlations associated with large-amplitude shape fluctuations at
low energy, which are in a time scale $\tau \gg \tau_F$,
are missing in the available EDF.
In practice, these correlations should be treated in addition to
the conventional DFT and TDDFT calculations
\cite{BBH06}.
We shall address this issue later,
in Sec.~\ref{sec:collective_submanifold}.

The non-universality ($p_0$ dependence) of $F[\rho,\kappa]$
is treated by enlarging the space $\Sigma_\textrm{intr}$ to include
all the $p_0$ states,
and adding an additional condition to the constrained search
of Eq. (\ref{F_rho_kappa}), for
the average value of $p_0$ ($J$ and $N$) of the wave packet.
This also limits the strictness of the nuclear DFT.

\subsection{Kohn-Sham (KS) scheme}
\label{sec:Kohn-Sham}

For many-fermion systems, the Fermi motion plays an important
role in various quantum phenomena, such as the shell effects.
This is a main source of difficulties in the local density
approximation (LDA) \cite{RS80}.
At present, a scheme given by Kohn and Sham \cite{KS65}
only provides a practical solution for this problem.
Eventually, this leads to the self-consistent equations similar to
those in the mean-field approximation.

\subsubsection{Normal systems}
Now, we derive the KS equations (\ref{KS_eq})
according to the argument by Kohn and Sham.
Let us assume that
the EDF is a functional of $\rho(\vec{r})$ only, $F[\rho]$.
We introduce a {\it reference system} which is a
``virtual'' non-interacting system with
an external potential $v_s(\vec{r})$.
The ground state of the reference system is obviously given as
a Slater determinant constructed by the solution of Eq.~(\ref{KS_eq}).
Alternatively, it is obtained by
the minimization of the total energy of the reference system,
$
E_s[\rho] = T_s[\rho] + \int v_s(\vec{r})\rho(\vec{r}) d^3\vec{r}
$.
Since $E_s[\rho]$ is a functional of density, the minimization can be
performed in terms of density variation
with the particle number constraint,
$\delta(E_s[\rho]-\mu \int \rho(\vec{r})d^3\vec{r})=0$.
This leads to
\begin{equation}
\label{Euler_eq}
\mu = \frac{\delta T_s[\rho]}{\delta\rho(\vec{r})} + v_s(\vec{r}) .
\end{equation}
The state determined by Eq. (\ref{Euler_eq}) should be identical to that
of Eq. (\ref{KS_eq}).

The success of the KS scheme comes from a simple idea to
decompose the kinetic energy in the physical interacting system 
into two parts;
$T_s[\rho]$ and the rest.
The former is a major origin of the shell effects, and the latter
is treated as a part of ``correlation energy'' $E_c[\rho]$.
$E_c[\rho]$ corresponds to the ``exchange-correlation energy''
$E_\textrm{xc}[\rho]$ in electronic DFT.
The EDF is given by the sum, $F[\rho] = T_s[\rho] + E_c[\rho]$.
Then, the variation, $\delta (F[\rho]-\mu \int \rho(\vec{r})d\vec{r})=0$,
leads to Eq.~(\ref{Euler_eq})
where the potential $v_s(\vec{r})$ is 
defined by
\begin{equation*}
v_s(\vec{r})\equiv \frac{\delta E_c[\rho]}{\delta\rho(\vec{r})} .
\end{equation*}
Therefore, the solution of Eq.~(\ref{KS_eq}) provides the ground-state
density of $F[\rho]$.
The only practical difference between the reference system and
the interacting system
is that, since $v_s(\vec{r})$
is a functional of density in the latter,
Eqs. (\ref{KS_eq}) and (\ref{KS_density}) with
$v_s$ must be self-consistently solved.
The success of the KS scheme is attributed to the goodness of
the LDA for $E_c[\rho]$.

\subsubsection{Superconducting systems}
\label{sec:superconducting_systems}

Next, with the density functional of Eq.~(\ref{F_rho_kappa}),
we introduce a non-interacting reference system
under an external pair potential $\Delta_s(\vec{r})$
in addition to $v_s(\vec{r})$.
The Hamiltonian with a constraint on the particle number,
\begin{eqnarray*}
&&\hat{H}_s  - \mu \hat{N}
= \int \left[ \Delta_s^*(\vec{r})
\hat{\psi} (\vec{r}\downarrow) 
\hat{\psi} (\vec{r}\uparrow) 
+ \textrm{h.c.}
\right] d^3\vec{r}
 \nonumber \\
&&+
\sum_{\sigma=\uparrow,\downarrow} \int
\hat{\psi}^\dagger (\vec{r}\sigma) 
\left\{
-\frac{\nabla^2}{2m} +v_s(\vec{r}) - \mu
\right\}
\hat{\psi}(\vec{r}\sigma) d^3\vec{r} \quad
\end{eqnarray*}
can be diagonalized by the Bogoliubov transformation (\ref{psi}),
in which $(U_i,V_i)$ are the solutions of Eq. (\ref{BdGKS_eq}).
Alternatively, Eq. (\ref{BdGKS_eq}) can be derived by minimizing
\begin{eqnarray*}
F_s[\rho,\kappa] &-&\mu N_\textrm{av}
= T_s[\rho,\kappa]
+\int \left\{ v_s(\vec{r})-\mu \right\}\rho(\vec{r})d^3\vec{r}
\nonumber\\
 &&+ \int \left\{\Delta_s^*(\vec{r}) \kappa(\vec{r})
 + \Delta_s(\vec{r}) \kappa^*(\vec{r})\right\} d^3\vec{r} .
\end{eqnarray*}
where
$
T_s[\rho,\kappa]= \sum_\sigma \sum_i \int V_i(\vec{r}\sigma)
\frac{-\nabla^2}{2m} V_i^*(\vec{r}\sigma) d^3\vec{r}
$.
The same minimization can be done
with respect to $(\rho,\kappa,\kappa^*)$.
\begin{equation}
\label{Euler_eq_2}
\mu = \frac{\delta T_s}{\delta\rho(\vec{r})} +v_s(\vec{r}) , \quad\quad
0 = \frac{\delta T_s}{\delta\kappa^*(\vec{r})} + \Delta_s(\vec{r}) .
\end{equation}
Equations (\ref{BdGKS_eq}) and (\ref{Euler_eq_2}) should
provide the identical
state.

According to the Kohn-Sham's idea,
we express the energy density functional of the interacting system
in the form
$F[\rho,\kappa]\equiv T_s[\rho,\kappa]+E_c[\rho,\kappa]$.
Then, the variation, $\delta(F[\rho,\kappa]-\mu\int\rho(\vec{r})d^3\vec{r})=0$,
leads to Eq. (\ref{Euler_eq_2})
but the potentials, $v_s(\vec{r})$ and $\Delta_s(\vec{r})$, are given by
Eq. (\ref{BdGKS_potentials}).
Equations (\ref{BdGKS_eq}) with potentials
(\ref{BdGKS_potentials}) constitutes the BdGKS scheme \cite{OGK88}.

\subsection{Time-dependent density functional theory}
\label{sec:TDDFT}

\subsubsection{Foundation: Runge-Gross theorem}
\label{sec:TDDFT_basic_theorem}

The basic theorem of the TDDFT
tells us that, starting from a common initial state $\ket{\Phi_0}$ at $t=t_0$,
there is one-to-one correspondence between a pair of time-dependent densities
$(\rho(\vec{r},t),\kappa(\vec{r},t))$ and
a pair of time-dependent external potentials $(v(\vec{r},t),\Delta(\vec{r},t))$
\cite{RG84,WKG94}.
Here, we recapitulate the proof.
The external potential is required to be expandable in a Taylor series about
the initial time $t_0$.
\begin{eqnarray}
\label{Taylor_expansion_v}
v(\vec{r},t)&=&\sum_{k=0}^\infty  \frac{1}{k!} v_k(\vec{r}) (t-t_0)^k , \\
\Delta(\vec{r},t)&=&\sum_{k=0}^\infty  \frac{1}{k!} \Delta_k(\vec{r}) (t-t_0)^k .
\end{eqnarray}
If two potentials $(v,v')$ differ merely by a time-dependent function,
$c(t) = v(\vec{r},t)-v'(\vec{r},t)$,
they should be regarded as {\it identical} potentials.
For the different potentials,
there should exist some non-negative integer $n$ such that
$\vec{\nabla} w_n(\vec{r})\neq 0$ where
$w_n(\vec{r})\equiv v_n(\vec{r})-v'_n(\vec{r})$.
Similarly, 
the pair potentials are different
if $D_n(\vec{r},t)\neq 0$ at a certain $n$ where
$D_n(\vec{r},t)\equiv\Delta_n(\vec{r},t)-\Delta'_n(\vec{r},t)$.

Let us first assume that
two different external potentials, $v(\vec{r},t)$
and $v'(\vec{r},t)$,
produce current densities
$\vec{j}(\vec{r},t)$ and
$\vec{j'}(\vec{r},t)$,
respectively.
The pair potential is assumed to be equal,
$\Delta(\vec{r},t)=\Delta'(\vec{r},t)$.
In the following, we assume the Heisenberg picture, and
the quantities associated with the potentials
$v'(\vec{r},t)$ are denoted with primes,
while those with $v(\vec{r},t)$ are without primes.
The equation of motion for the current,
$\vec{j}(\vec{r},t)\equiv \bra{\Phi_0}\hat{\vec{j}}(\vec{r},t)\ket{\Phi_0}$,
is written as
\begin{equation}
\label{dj_dt}
i\frac{\partial}{\partial t} \vec{j}(\vec{r},t)
= \bra{\Phi_0} [ \hat{\vec{j}}(\vec{r},t), \hat{H}(t) ] \ket{\Phi_0} .
\end{equation}
We have the same equation for $\vec{j'}(\vec{r},t)$,
with $\hat{H}(t)$ replaced by $\hat{H}'(t)$.
Since the field operators at $t=t_0$ are identical to each other,
$\psi(\vec{r}\sigma,t_0)=\psi'(\vec{r}\sigma,t_0)$, they lead to
\begin{eqnarray*}
&&\left. i\frac{\partial}{\partial t}
\left\{\vec{j}(\vec{r},t)-\vec{j'}(\vec{r},t)\right\} \right|_{t=t_0}
\nonumber \\
&&\quad\quad\quad\quad = \bra{\Phi_0} [ \hat{\vec{j}}(\vec{r},t_0),
                    \hat{H}(t_0)-\hat{H}'(t_0) ] \ket{\Phi_0}
  \nonumber\\
&&\quad\quad\quad\quad = - \frac{i}{m}\rho(\vec{r},t_0)
 \vec{\nabla} w_0(\vec{r}) .
\end{eqnarray*}
If $\vec{\nabla} w_0(\vec{r})\neq 0$, it is easy to see that
$\vec{j}(\vec{r},t)$ and $\vec{j'}(\vec{r},t)$ are different at $t>t_0$.
In case that $\vec{\nabla} w_0(\vec{r})= 0$ and
$\vec{\nabla} w_1(\vec{r})\neq 0$,
we need to further calculate derivative of Eq. (\ref{dj_dt})
with respect to $t$.
\begin{eqnarray}
\label{d2j_dt2}
&&\left. \left(i\frac{\partial}{\partial t}\right)^2
   \vec{j}(\vec{r},t) \right|_{t=t_0}
= \bra{\Phi_0} \left[ \hat{\vec{j}}(\vec{r},t),
        i \bar\partial \hat{H}/\bar\partial t \right]_{t=t_0} \ket{\Phi_0} \nonumber \\
&& \quad\quad\quad\quad
+ \bra{\Phi_0} [ [ \hat{\vec{j}}(\vec{r},t_0), \hat{H}(t_0) ], \hat{H}(t_0) ]
  \ket{\Phi_0} ,
\end{eqnarray}
where $\bar\partial/\bar\partial t$ indicates the time derivative
of the potentials only, not of the field operators.
The second term of Eq. (\ref{d2j_dt2}) vanishes for the difference,
$\partial^2/\partial t^2\{\vec{j}(\vec{r},t) - \vec{j'}(\vec{r},t)\}|_{t=t_0}$,
because $\hat{H}'(t_0)=\hat{H}(t_0)+\mbox{const}$.
Thus,
\begin{equation*}
\left. \frac{\partial^2}{\partial t^2}
\left\{\vec{j}(\vec{r},t_0)-\vec{j'}(\vec{r},t)\right\} \right|_{t=t_0}
= - \frac{1}{m}\rho(\vec{r},t_0) \vec{\nabla} w_1(\vec{r}) \neq 0.
\end{equation*}
Again, we conclude that $\vec{j}(\vec{r},t)\neq \vec{j'}(\vec{r},t)$
at $t>t_0$.
In general, if $\vec{\nabla} w_k(\vec{r})= 0$ for $k<n$ and
$\vec{\nabla} w_n(\vec{r})\neq 0$, we repeat the same argument to reach
\begin{eqnarray*}
&& \frac{\partial^{n+1}}{\partial t^{n+1}}
\left\{\vec{j}(\vec{r},t)-\vec{j'}(\vec{r},t)\right\}_{t=t_0}
\\
&&\quad\quad
 = -i\bra{\Phi_0} \left[ \hat{\vec{j}}(\vec{r},t_0),
    \left.
     {\bar\partial^n}
     \{ \hat{H}(t) - \hat{H}'(t) \}/{\bar\partial t^{n}}
    \right|_{t=t_0} \right]
     \ket{\Phi_0} \nonumber \\
&&\quad\quad
 = - \frac{1}{m}\rho(\vec{r},t_0) \vec{\nabla} w_n(\vec{r}) \neq 0.
\end{eqnarray*}
Therefore, there exists a mapping from the expandable potential $v(\vec{r},t)$
and the current density $\vec{j}(\vec{r},t)$.
The continuity equation relates the current density $\vec{j}(\vec{r},t)$
with the density $\rho(\vec{r},t)$.
Therefore, we can conclude that the densities
$\rho(\vec{r},t)$ and $\rho'(\vec{r},t)$ are different at $t>t_0$.

Next, let us assume the different pair potentials,
$\Delta(\vec{r},t)$ and $\Delta'(\vec{r},t)$.
The same argument above leads to
\begin{eqnarray*}
\left. \frac{\partial^{n+1}}{\partial t^{n+1}}
\left\{\kappa(\vec{r},t)-\kappa'(\vec{r},t)\right\} \right|_{t=t_0}
 = f(\vec{r})  D_n(\vec{r}) 
\neq 0 ,
\end{eqnarray*}
where $f(\vec{r})
=i \{ \rho(\vec{r},t_0) -\delta^3(0) \}$.
The appearance of the delta function $\delta^3(0)$ is a consequence of
the local nature of the pair potential $\Delta(\vec{r,t})$.
Anyway, $f(\vec{r})$ is nonzero, and the pair densities,
$\kappa(\vec{r},t)$ and $\kappa'(\vec{r},t)$,
become different immediately after $t=t_0$.
This completes the proof of the one-to-one correspondence between
the potentials $(v(\vec{r},t),\Delta(\vec{r},t))$ and
the densities $(\rho(\vec{r},t),\kappa(\vec{r},t))$.
As is obvious in the proof here,
the one-to-one correspondence also holds when
the density $\rho(\vec{r},t)$ is replaced by the current density
$\vec{j}(\vec{r},t)$.

\subsubsection{TDKS scheme: van Leeuwen theorem}

In practice, the KS scheme is indispensable for quantum systems.
According to the basic theorem in Sec.~\ref{sec:TDDFT_basic_theorem}
there is a one-to-one correspondence
between given time-dependent densities
and external potentials
for any system.
Let us introduce
a {\it virtual} reference system of non-interacting particles
by choosing the potentials, $v_s(\vec{r},t)$ and $\Delta_s(\vec{r},t)$,
in such a way that
it exactly produces the densities, $\rho(\vec{r},t)$ and $\kappa(\vec{r},t)$,
of a {\it real} interacting system.
This results in
the time-dependent Bogoliubov-de-Gennes-Kohn-Sham (TDBdGKS) equations:
\begin{equation}
\label{TDBdGKS_eq}
\begin{split}
&i\frac{\partial}{\partial t}
\begin{pmatrix}
U_i(\vec{r}\sigma;t) \\
V_i(\vec{r}\sigma;t)
\end{pmatrix}
= 
\\
&\sum_{\sigma'}
\begin{pmatrix}
h_s(\vec{r};t)\delta_{\sigma\sigma'} & \Delta_s(\vec{r};t)\gamma_{\sigma\sigma'} \\
-\Delta_s^*(\vec{r};t)\gamma_{\sigma\sigma'} & -h_s^*(\vec{r};t)\delta_{\sigma\sigma'}
\end{pmatrix}
\begin{pmatrix}
U_i(\vec{r}\sigma';t) \\
V_i(\vec{r}\sigma';t)
\end{pmatrix}
.
\end{split}
\end{equation}
Here,
$h_s(\vec{r},t)\equiv -\nabla^2/(2m) + v_s(\vec{r},t)$.
With $\Delta_s=0$,
they reduce to
the TDKS equations ($i=1,\cdots,N$);
\begin{equation}
\label{TDKS_eq}
i\frac{\partial}{\partial t}\varphi_i(\vec{r}, t)
= h_s(\vec{r},t) \varphi_i(\vec{r},t) .
\end{equation}

The next obvious question is the following:
Do such potentials in non-interacting systems
exist to reproduce the densities in real systems?
This question was answered affirmatively by \textcite{van99} as follows.
For simplicity, let us consider the TDKS equations without pairing.
Hereafter, quantities associated with the reference system
are denoted with a subscript ``$s$''.
First, calculating the right-hand side of Eq. (\ref{dj_dt}) gives
$i{\partial\vec{j}(\vec{r},t)}/{\partial t} =
-\rho(\vec{r},t)\vec{\nabla} v(\vec{r},t) - \vec{f}(\vec{r},t)$.
Here, $\vec{f}(\vec{r},t)$ is given by the momentum-stress tensor
and the interaction parts, but these details are not important in the
proof.
Taking the divergence of this equation and using the continuity equation,
we find
\begin{equation}
\label{d2rho_dt2}
\frac{\partial^2\rho}{\partial t^2} = \vec{\nabla}\cdot
(\rho(\vec{r},t)\vec{\nabla} v(\vec{r},t)) + q(\vec{r},t) ,
\end{equation}
where $q(\vec{r},t)=\vec{\nabla}\cdot\vec{f}(\vec{r},t)$.
Assuming the density is identical in two systems all the times,
the difference of Eq. (\ref{d2rho_dt2}) between the two leads to
\begin{equation}
\label{nabla_n_nabla_w}
\vec{\nabla}\cdot (\rho(\vec{r},t)\vec{\nabla} w(\vec{r},t)) = \zeta(\vec{r},t) ,
\end{equation}
where $w=v-v_s$ and $\zeta=q_s-q$.
This equation plays a key role in the proof.
Now, the question is whether we can uniquely determine $w(\vec{r},t)$
if $\rho(\vec{r},t)$ is given.

Necessary conditions for the initial state $\ket{\Phi_s}$ of the
reference system are only two:
(i) The two initial states, $\ket{\Phi_0}$ and $\ket{\Phi_s}$,
yield the same density, $\rho(\vec{r},t_0)=\rho_s(\vec{r},t_0)$.
(ii) Their time derivatives are identical,
$\dot{\rho}(\vec{r},t_0)=\dot{\rho}_s(\vec{r},t_0)$.
With $\ket{\Phi_s}$ satisfying these initial conditions, 
we determine the solution $w$ of
Eq. (\ref{nabla_n_nabla_w}).
We should first notice that
Eq. (\ref{nabla_n_nabla_w}) does not contain time derivatives,
which means that $t$ can be regarded as a parameter.
Furthermore, Eq. (\ref{nabla_n_nabla_w}) is of the Sturm-Liouville type,
thus it has a unique solution with the boundary condition,
$w(\vec{r},t)=0$ at infinity.
It is now obvious that we can uniquely determine $w(\vec{r},t_0)$
at $t=t_0$ because $\zeta(\vec{r},t_0)$ is calculable with
the initial states $\ket{\Phi_0}$ and $\ket{\Phi_s}$.
This means in the Taylor-series expansion, Eq. (\ref{Taylor_expansion_v}),
$w_0(\vec{r})=v_0(\vec{r})-{v_s}_0(\vec{r})$ is solved.
Taking the time derivative of Eq. (\ref{nabla_n_nabla_w}) at
$t=t_0$, we can determine $w_n(\vec{r})$ for higher-order terms
in a recursive manner ($n=1,2,\cdots$).
This procedure completely determines $v_s(\vec{r},t)$.

\subsubsection{TDBdGKS equation and its properties}

The key quantity in TDDFT is the time-dependent potentials
 $(v_s(\vec{r},t),\Delta_s(\vec{r},t))$.
So far, we simply adopt the {\it adiabatic approximation}:
We take the BdGKS potentials in Eq. (\ref{BdGKS_eq})
from static DFT and
use it in the TDBdGKS equations (\ref{TDBdGKS_eq}),
by replacing ground-state densities 
with the time-dependent ones.
\begin{equation}
\label{adiabatic_potential}
v_s(t) = \left. v_s[\rho_0]\right|_{\rho_0 \rightarrow \rho(t)} ,
\quad
v_s[\rho]\equiv \delta E_c/\delta \rho
\end{equation}
and the same prescription is applied to $\Delta_s$.
This obviously lacks the memory effect.

Properties of the static BdGKS equations shown
in Sec.~\ref{sec:useful_notations}
also hold for the time-dependent case, except for
BdGKS equation (\ref{BdGKS_R_eq}) which should be replaced by
\begin{equation}
\label{TDBdGKS_R_eq}
i\frac{\partial}{\partial t}R(t)=\left[ H_s[R](t), R(t) \right] .
\end{equation}
Here, $H_s[R]$ is given by Eq. (\ref{H_s}) with $R_0\rightarrow R(t)$
in the adiabatic approximation.
We use the notations same as those in Sec.~\ref{sec:useful_notations}
with obvious changes introducing the time dependence, such as
$\Psi^0\rightarrow\Psi(t)$, $R^0\rightarrow R(t)$, etc.

With respect to a time-dependent unitary transformation
$\Psi(t)\rightarrow \Psi(t) U(t)$,
$R(t)$ and Eq. (\ref{TDBdGKS_R_eq}) are invariant,
while Eq. (\ref{TDBdGKS_eq}) is not.
Including this gauge freedom, 
the TDBdGKS equations (\ref{TDBdGKS_eq}) should be generalized to
\begin{equation}
\label{gTDBdGKS_eq}
i\frac{\partial}{\partial t} \Psi(t)
= H_s[R](t) \Psi(t) - \Psi(t) \Xi(t) ,
\end{equation}
with a Hermitian matrix $\Xi=i(dU/dt) U^\dagger
=-iU dU^\dagger/dt$.
The choice of $\Xi(t)$ is arbitrary and does not affect the physical
contents of the calculation.
The TDBdGKS equations (\ref{TDBdGKS_eq}) correspond to
a special gauge $\Xi(t)=0$.

In case that the pair density and potential are absent,
$\kappa(t)=\Delta(t)=0$,
Eq. (\ref{TDBdGKS_R_eq}) reduces to
\begin{equation}
\label{TDKS_rho_eq}
i\frac{\partial}{\partial t}\rho(t)=
\left[ h_s[\rho](t), \rho(t) \right] .
\end{equation}
This is equivalent to the TDKS equations,
\begin{equation*}
i\frac{\partial}{\partial t}\varphi(t)=
h_s[\rho](t)\varphi(t)-\varphi(t)\xi(t) ,
\quad \rho(t)=\varphi(t)\varphi^\dagger(t) ,
\end{equation*}
where $\varphi(t)$ is a collective notation of the non-vanishing
vectors $V_i^*(t)$, $\varphi(t)=(V_1^*(t),V_2^*(t),\cdots,V_N^*(t))$,
that are upper components of $\tilde{\Psi}(t)$.
The quantity $\xi(t)$ is an arbitrary $N\times N$ Hermitian matrix.
A choice of $\xi(t)=0$ leads to Eq.~(\ref{TDKS_eq}).

\subsubsection{Local gauge invariance}

The practical success of the TDBdGKS equations relies on
the availability of a good correlation functional $E_c[\rho,\kappa]$.
Many applications, so far employ the
functional of the static potentials (\ref{BdGKS_potentials})
with time-dependent densities
({\it adiabatic approximation}).
Although the memory effect is missing,
this simple choice guarantees exact properties of the functional,
such as the harmonic potential theorem (HPT) \cite{Dob94,Vig95}.
In nuclear physics applications, it is also customary to adopt a
functional in the same form as the static one.
Since a local density form (Skyrme-type) of the nuclear correlation energy
contains the kinetic and spin-current densities, 
to guarantee the Galilean symmetry,
it should include the time-odd densities, such as spin, current, and
spin-tensor densities \cite{Eng75}.
In fact, the nuclear EDFs usually respect an even stronger
symmetry, the local gauge invariance, which is satisfied for
systems with local interactions.
The HPT and the Galilean invariance can be regarded as its special cases.
The local gauge transformation modifies the one-body density matrix as
$\rho(\vec{r},\vec{r}') \rightarrow 
\exp[i\{\chi(\vec{r})-\chi(\vec{r}')\}]\rho(\vec{r},\vec{r}')$.
The local density $\rho(\vec{r})=\rho(\vec{r},\vec{r})$ is apparently
invariant, however, the kinetic and spin-current densities are not,
because the transformation creates a flow with a velocity field,
$\vec{v}(\vec{r},t)=\vec{\nabla}\chi(\vec{r},t)/m$.
These densities appear with characteristic combinations with
the time-odd densities to satisfy the local gauge invariance \cite{DD95}.
Note that the local gauge invariance is guaranteed if the nonlocal effect
is small, but it is not required by the principles.
It has been utilized to restrict the functional form of nuclear EDFs
\cite{CDK08}.

The local $U(1)$ gauge transformation,
$\hat{\psi}(\vec{r}\sigma,t)\rightarrow
e^{i\chi(\vec{r},t)} \hat{\psi}(\vec{r}\sigma,t)$,
with a real function $\chi(\vec{r},t)$
changes the phase of $U$ and $V$ components
with opposite signs (see Eq. (\ref{psi})).
Thus, the transformation reads
\begin{equation}
\bar{\Psi}(t)
=
\begin{pmatrix}
e^{i\chi} & 0\\
0 & e^{-i\chi}
\end{pmatrix}
\Psi(t)
= \exp\{i\chi(t){\cal N}\} \Psi(t) ,
\end{equation}
with
${\cal N}\equiv\begin{pmatrix}
1 & 0 \\
0 & -1
\end{pmatrix}$.
Under this transformation, the generalized density and the Hamiltonian
should be transformed as
\begin{equation}
\label{R_and_H_eff}
\begin{Bmatrix}
\bar{R}(t) \\
\bar{H}_s(t)
\end{Bmatrix}
= e^{i\chi(t){\cal N}}
\begin{Bmatrix}
R(t) \\
H_s(t)
\end{Bmatrix}
e^{-i\chi(t){\cal N}} .
\end{equation}
The transformation (\ref{R_and_H_eff}) 
keeps the density $\rho(\vec{r},t)$ invariant, but multiplies 
$\kappa(\vec{r},t)$ by a local phase $e^{2i\chi(\vec{r},t)}$.
The transformation of the kinetic term can be obtained by shifting
the momentum $\vec{p}$ to $\vec{p}-m\vec{v}(\vec{r},t)$.
The local gauge invariance of the density functionals
guarantees that 
$H_s[R](\vec{r},\vec{p};t)\rightarrow
\bar{H}_s[\bar{R}](\vec{r},\vec{p};t)=
H_s[\bar{R}](\vec{r},\vec{p}-m\vec{v};t)$,
in which the replacement of $\vec{p}\rightarrow\vec{p}-m\vec{v}$
is performed only for the kinetic energy term,
$|\vec{p}|^2/(2m) \rightarrow 
|\vec{p}-m\vec{v}(\vec{r},t)|^2/(2m)$.
The transformed TDBdGKS equations for $\bar{R}$ and $\bar{\Psi}(t)$
are identical to Eqs. (\ref{TDBdGKS_R_eq}) and (\ref{gTDBdGKS_eq}),
but the Hamiltonian is replaced by
$\bar{H}_s[\bar{R}](t)
-{\partial\chi}/{\partial t} \cdot {\cal N}$.
For instance, Eq. (\ref{TDBdGKS_R_eq}) now reads
\begin{equation}
\label{TDBdGKS_R_eq_2}
i\frac{\partial}{\partial t}\bar{R}(t)=
\left[ \bar{H}_s[\bar{R}](t)
-\frac{\partial\chi}{\partial t} {\cal N},
 \bar{R}(t) \right] .
\end{equation}
This is the TDBdGKS equation
in a frame of a gauge function $\chi(\vec{r},t)$.
Under the presence of the local gauge invariance in the EDF,
the functional form of $\bar{H}_s$
is the same as $H_s$ except that the momentum $\vec{p}$ is
replaced by $\vec{p}-m\vec{v}$ in the kinetic term.

\subsection{Equations for decoupled collective motion}
\label{sec:equations_in_the_moving_frame_of_reference}

In this section, we derive an equation for the decoupled collective motion.
In order to elucidate the idea,
let us start with the translational motion.
In this case, the decoupling is exact.
The boosted ground state with the center of mass at
$\vec{R}_\textrm{cm}(t)=\vec{v}t$
has the density,
$\rho(t)=\rho(\vec{R}_\textrm{cm})$,
which depends on time through $\vec{R}_\textrm{cm}(t)=\vec{v}t$.
The total momentum $\vec{P}_\textrm{cm}=Nm\vec{v}$ is a constant of motion.
The TDKS equation (\ref{TDKS_rho_eq}) can be written as
\begin{equation*}
i\vec{v}\cdot\frac{\partial}{\partial \vec{R}_\textrm{cm}}
 \rho(\vec{R}_\textrm{cm})
= \left[ h_s[\rho(\vec{R}_\textrm{cm})], \rho(\vec{R}_\textrm{cm}) \right] .
\end{equation*}
Using the expression,
$\rho(t)= e^{-i\vec{R}_\textrm{cm}(t)\cdot\vec{p}} \bar{\rho}
e^{i\vec{R}_\textrm{cm}(t)\cdot\vec{p}}$
where $\bar{\rho}$ is time-independent,
it leads to
\begin{equation}
\left[ h_s[\rho(t)]-\vec{v}\cdot\vec{p}, \rho(t) \right] = 0 ,
\label{TDKS_mv_2}
\end{equation}
which looks like a stationary equation.
In fact, since $\vec{R}_\textrm{cm}=\vec{v}t$ depends on time,
$\rho(\vec{R}_\textrm{cm})$ is moving in time.
We call Eq. (\ref{TDKS_mv_2}) ``moving-frame'' equation in the following.
It should be noted that the EDFs with the Galilean symmetry is essential
to reproduce the correct total mass $Nm$,
which also influences properties of other collective motions.

\subsubsection{Collective motion in general}

Now, let us generalize the idea and
assume that there are a pair of canonical variables $(q(t),p(t))$
corresponding to a collective motion, 
which determine the time dependence of the generalized density $R(q(t),p(t))$.
This means that the motion described by $(q(t),p(t))$ is decoupled from
the other {\it intrinsic} degrees of freedom.
In the TDBdGKS equation (\ref{TDBdGKS_R_eq}),
the time derivative is now written in terms of the collective variables as
$\dot{q}\partial/\partial q +\dot{p}\partial/\partial p$.
This leads to the moving-frame equation,
\begin{equation}
\left[ H_s[R]-\dot{q}\overcirc{P}(q,p)+\dot{p}\overcirc{Q}(q,p) , R(q,p) \right] = 0,
\label{TDBdGKS_R_moving_frame}
\end{equation}
where $\overcirc{P}(q,p)$ and $\overcirc{Q}(q,p)$
are generators of the collective variables
and defined by
\begin{eqnarray}
\label{generator_P_0}
i\frac{\partial}{\partial q} R(q,p)&=&\left[ \overcirc{P}(q,p),R(q,p)\right], \\
\label{generator_Q_0}
-i\frac{\partial}{\partial p} R(q,p)&=&\left[ \overcirc{Q}(q,p),R(q,p)\right] .
\end{eqnarray}
Note that $\dot{q}$ and $\dot{p}$ are not constant, in general.
In Sec.~\ref{sec:collective_submanifold},
we develop this idea and derive equations of motion for 
a collective motion decoupled from other intrinsic degrees of freedom.

Equation (\ref{TDBdGKS_R_moving_frame}) looks like a stationary equation
with constraints, $\overcirc{Q}(q,p)$ and $\overcirc{P}(q,p)$.
However, it is important to note that
the density $R(q,p)$ in Eq. (\ref{TDBdGKS_R_moving_frame})
still varies in time because the variables $(q,p)$ depend on time.
Because of this time dependence,
the ``cranking terms'', $-\dot{q}\overcirc{P}+\dot{p}\overcirc{Q}$,
in Eq. (\ref{TDBdGKS_R_moving_frame})
are not just the constraint terms in static equations, but
plays a role beyond that.

To explain this point, we go back again to the translational motion.
The equation (\ref{TDKS_mv_2}) looks identical to
the static equation with a constraint operator, $\vec{p}$.
However, the cranking term induces the linear momentum,
$\textrm{Tr}[\rho\vec{p}] = Nm\vec{v}$,
and the density is never static.
During the time evolution $t\rightarrow t+\delta t$,
the center of mass moves as 
$\vec{R}_\textrm{cm}\rightarrow \vec{R}_\textrm{cm}+\vec{v} \delta t$.
Accordingly, the density also evolves,
$\rho \rightarrow \rho+\delta\rho$.
$\vec{R}_\textrm{cm}\rightarrow \vec{R}_\textrm{cm}+\delta\vec{R}$.
This density variation is described by Eq. (\ref{TDKS_mv_2}).
\begin{equation}
\left[
h_s - \vec{v}\cdot\vec{p} , \delta\rho
\right]
+\left[
\frac{\delta h_s}{\delta\rho}\delta\rho , \rho
\right]
 = 0 ,
\label{CM_RPA}
\end{equation}
in the first order in $\delta\rho$.
This is nothing but the random-phase approximation (RPA)
for the translational motion.
If Eq. (\ref{TDKS_mv_2}) is a constrained stationary equation,
obviously, it does not lead to the RPA equation.

If we define the particle (unoccupied) and hole (occupied)
orbitals for $h_s[\rho]-\vec{v}\cdot\vec{p}$,
the particle-particle and hole-hole components,
$\vec{p}_{pp'}$ and $\vec{p}_{hh'}$, contribute to
the determination of $\delta\rho$ in Eq. (\ref{CM_RPA}).
In contrast, for the constrained mean-field equation \cite{RS80},
the particle-particle and hole-hole matrix elements of the
constrained operator are irrelevant.
We think it worth emphasizing that
the cranking terms in Eqs. (\ref{TDKS_mv_2}) and (\ref{TDBdGKS_R_moving_frame})
are different from constraint terms in the static equation
\cite{NWD99,HNMM07,Nak12}.
The issue will be addressed in Sec.~\ref{sec:collective_submanifold}.

\subsubsection{ANG modes and quasi-stationary solutions}
\label{sec:quasi-stationary-solutions}

The ANG modes provide examples of decoupled collective motion to which
Eq. (\ref{TDBdGKS_R_moving_frame}) is applicable.
In these cases, one of the variables becomes cyclic (constant), and
the generators do not depend on the variables $(q,p)$.
They are given by known one-body operators globally defined.

\leftline{1. \underline{\it Translational motion}}

In this case, the generators $(\overcirc{Q}(q,p),\overcirc{P}(q,p))$
correspond to the center-of-mass coordinate and the total momentum,
with $\dot{q}=\vec{v}$ and $\dot{p}=0$.
Thus, we naturally derive Eq. (\ref{TDKS_mv_2}) from
Eq. (\ref{TDBdGKS_R_moving_frame}).
The Galilean invariance guarantees that the translational motion
with a constant velocity does not influence the {\it intrinsic} state.
In fact, the local gauge transformation with
$\chi(\vec{r})=-m \vec{v}\cdot\vec{r}$
removes the cranking term, $-\vec{v}\cdot\vec{p}$.
Then, using the ground-state solution $\rho_0$,
which satisfies the static equation
$[h_s[\rho_0],\rho_0]=0$ ($\vec{v}=0$),
we may construct a solution of Eq. (\ref{TDKS_mv_2}),
$\rho(\vec{R}_\textrm{cm})=e^{im\vec{v}\cdot\vec{r}} \rho_0(\vec{R}_\textrm{cm})
      e^{-im\vec{v}\cdot\vec{r}}$.

\leftline{2. \underline{\it Rotational motion}}

A spatially rotating system
with a constant angular velocity $\vec\omega$
can be described by a solution of Eq. (\ref{TDBdGKS_R_moving_frame})
with $\dot{q}=\dot{\vec\theta}=\vec\omega$, $\dot{p}=\dot{\vec{I}}=0$.
The generator $\overcirc{P}(q,p)$ corresponds to
the angular momentum operator $\vec{j}$.
Though we do not know the conjugate angle operator, it disappears because of
the angular momentum conservation $\dot{\vec{I}}=0$.
Then, it ends up the cranking model \cite{Ing54,Ing56}:
\begin{equation}
\left[ h_s[\rho] - \vec{\omega}\cdot \vec{j},
 \rho \right] = 0 ,
\label{TDKS_rot}
\end{equation}
where the density $\rho(\vec\theta)$ is a function of the angle
$\vec\theta(t)=\vec\omega t$.
Since there is no Galilean symmetry in the rotational motion,
it is impossible to remove the cranking term by a gauge transformation.
In this case, the decoupling is only approximate.
In fact, the rotational motion influences the intrinsic state
in non-trivial ways, such as the centrifugal stretching and
the Coriolis coupling effects.

\leftline{3. \underline{\it Pair rotation}}

In the superconducting phase with $\kappa(t)\neq 0$,
in which the global gauge symmetry is broken,
one may find another rotating solution in the gauge space
with a constant angular velocity $\mu$.
The generator $\overcirc{Q}(q,p)$ corresponds to the particle number
operator\footnote{
Here, we assume that $q$ ($p$) is the time-even (time-odd) variable.
},
$q(t)=N_0$ ($\dot{q}=0$), and $p(t)=\mu t$.
Equation (\ref{TDBdGKS_R_moving_frame}) leads to
\begin{equation}
\label{TDBdGKS_pair_rot}
\left[ H_s[R] -\mu {\cal N}, R \right]  = 0 ,
\end{equation}
where $R(\theta)$ is a function of $\theta(t)=\mu t$.
In terms of the time-dependent formalism, the appearance of the
chemical potential $\mu$ in the stationary BdGKS equation (\ref{BdGKS_eq})
comes from the rotation in the gauge space.

When we study intrinsic excitations perpendicular to the ANG modes,
we should extend the density $R(q,p)$ either
by introducing the second set of variables $(q'(t),p'(t))$,
or by allowing additional time dependence, $R(q,p;t)$.
The former method will be adopted in Sec.~\ref{sec:collective_submanifold}.
The latter method changes the right hand side of
Eq.~(\ref{TDBdGKS_R_moving_frame}) to $i\partial R/\partial t$.
For the case of the pair rotation, this leads to
\begin{equation}
\label{TDBdGKS_pair_rot_intr}
i\frac{\partial}{\partial t}R(\theta;t)
=\left[ H_s[R] -\mu {\cal N}, R(\theta;t) \right] ,
\end{equation}
where $\mu$ is a function of the particle number $N_0$.

\subsection{Recent development in nuclear EDF}
\label{sec:NEDF}

Finding the best density functionals is always a big challenge in the DFT,
not only in nuclear systems but also in electronic systems.
Since we do not know the exact interaction among nucleons,
even for the uniform matter at low-density and high-density limits,
the exact functional is not available.
Thus, strategies in nuclear DFT is somewhat different from those in
electronic systems (Sec.~\ref{sec:electronic}).
Recent developments involve extension of the functional form and
the new optimization to fit reliable calculations and experimental data.
The optimization has been performed mainly for static properties,
including fission isomers and barrier heights.
Here, we present some efforts to improve the EDF, after those shown
in \textcite{BHR03}.

The systematic optimization of the Skyrme EDF was performed to
construct the functionals of UNEDF0-2 \cite{Kor10,Kor12,Kor14},
which produce the root-mean-square deviation from the experimental
binding energies of $1.5-2$ MeV.
These studies also show a clear deviation pattern common to all the
EDFs.
This indicates a necessity of novel functional forms for further
improvements.
The idea based on the density matrix expansion \cite{NV72,Neg82}
is under development to create new functionals \cite{CD10,CDK08,Sto10}.
Other forms of EDF without the derivative terms have been also
developed and produce similar accuracy \cite{BSV08,BRSV13}.
Although some phenomenological corrections significantly improve
the reproduction of the binding energy \cite{GCP13},
those corrections are not applicable to TDDFT calculations.

The Gogny EDF was also improved by fitting nuclear structure and
neutron matter properties, leading to D1N \cite{CGH08} and D1M
\cite{GHGP09}.
A new type of the Gogny EDF has been recently proposed, which 
extends the density-dependent term to the one with
finite range \cite{CPGB15}.
Another type of EDF based on the Yukawa-type potential was also proposed
\cite{Nak13,NI15}.

The modern covariant EDFs adopt either nonlinear meson coupling or
density-dependent coupling constants.
In addition, there are two types of the covariant EDF:
the finite-range meson-field and the point-coupling models.
Each EDF type had recent extensions of the functional form,
such as inclusion of the $\delta$ meson \cite{RVCRS11},
the cross-coupling terms \cite{FHPS10},
the exchange terms \cite{LSGM07},
and new version of the point-coupling models \cite{ZLYM10,NVR08}.

The pairing EDF responsible for the pair potential $\Delta$
is another issue.
The pairing energy in the Gogny EDF is calculated with the same
interaction.
In contrast, most of the Skyrme and covariant EDFs independently treat
the pairing EDF.
Different forms of the pairing EDF have been recently proposed
\cite{YB03,MSH08,YSN09,YMSH12,TMR09}.

Currently, it is difficult to judge which type of nuclear EDF
is the best.
Their accuracy for the mass prediction is rather similar
to each other among the Skyrme, the Gogny, and the covariant EDFs.
Since we know none of them is perfect,
the error analysis on the model is important \cite{DNR14,Erl12}.
Furthermore,
in contrast to the optimization of EDFs with respect to stationary
properties,
the one with respect to dynamical properties has not been performed
in a systematic manner \cite{BDEN02}.
To our knowledge, possibilities beyond the adiabatic approximation
(Sec.~\ref{sec:EF})
have never been examined in nuclear physics.

\section{Linear density response}
\label{sec:linear_response}

The linear density response of interacting systems can be
rigorously formulated, in principle, on the basis of the TDDFT.
The formulation is basically identical to the one known as
the quasiparticle-random-phase approximation (QRPA)
in nuclear physics \cite{RS80,BR86}.

Since the pair rotation inevitably takes place with a finite $\mu$,
the density $R(t)=R(\theta(t))$ is not stationary even for the ground state.
In order to avoid complications in deriving
the QRPA linear response equations,
we should start either with Eq.~(\ref{TDBdGKS_pair_rot_intr}),
or with Eq.~(\ref{TDBdGKS_R_eq_2}) of a gauge function
$\chi(t)=\theta(t)=\mu t$,
in which the time dependence through $\theta(t)$ is hidden.
The following external potential,
multiplied by a parameter $\eta$, is added to $H_s[R]$.
\begin{equation*}
\eta{V}(t)\equiv \eta
\begin{pmatrix}
v_\textrm{ext}(t) & \Delta_\textrm{ext}(t) \\
-\Delta^*_\textrm{ext}(t) & -v^*_\textrm{ext}(t)
\end{pmatrix}
.
\end{equation*}
See Sec.~\ref{sec:useful_notations} for the corresponding
operator form.
It is convenient to introduce a small parameter $\eta$ to elucidate
the linearization.
The time-dependent density and the Hamiltonian are linearized with
respect to $\eta$ as
$R(t)=R_0+\eta\delta R(t) + O(\eta^2)$ and
 $H_s(t)=H_s[R_0]+\eta\delta H(t) + O(\eta^2)$.
The Fourier transform of Eq. (\ref{TDBdGKS_pair_rot_intr}) leads to
\begin{equation}
\label{TDBdGKS_R_lin}
\omega \delta R(\omega) =
\left[
H_s[R_0] -\mu {\cal N},\delta R(\omega)
\right]
+
\left[
{V}(\omega) + \delta H(\omega), R_0
\right],
\end{equation}
in the linear order.
This equation plays a central role in this section.

\subsection{Linear response equations and matrix representation in
the quasiparticle basis}

In order to evaluate Eq. (\ref{TDBdGKS_R_lin}), it is customary to 
adopt the quasiparticle eigenstates at the ground state
in Eq.~(\ref{BdGKS_eq}).
Those with positive [negative] energies, $\Psi^0_i$ [$\tilde{\Psi}_i^0$]
satisfy
$(H_s -\mu{\cal N})\Psi_i^0 = E_i \Psi_i^0$
[$(H_s -\mu{\cal N})\tilde{\Psi}_i^0 = -E_i \tilde{\Psi}_i^0$].
We may write the time-dependent quasiparticle states as
$\Psi_i(t)=e^{-iE_it} (\Psi_i^0+\eta\delta\Psi_i(t))$.
Since the generalized density $R(t)$ is written in terms of the
quasiparticle states $\Psi(t)$ as in Eq.~(\ref{R}),
the fluctuating part $\delta R(t)$ in the linear order
is given by
\begin{equation}
\begin{split}
\delta R(t) 
 &=-\sum_{i}\left\{
\delta\Psi_i(t) \Psi_i^{0\dagger} +\Psi_i^0 \delta\Psi_i^\dagger (t) \right\}  \\
&=-\delta\Psi(t) \Psi^{0\dagger} -\Psi^0 \delta\Psi^\dagger (t) \\
&=\delta\tilde{\Psi}(t) \tilde{\Psi}^{0\dagger}
+ \tilde{\Psi}^0 \delta\tilde{\Psi}^\dagger(t)
,
\end{split}
\label{delta_R_t}
\end{equation}

Using the notation in Sec.~\ref{sec:useful_notations},
we calculate the matrix elements of Eq. (\ref{TDBdGKS_R_lin}) between
these quasiparticle basis.
From the orthonormal relations,
it is easy to see
$\delta R^{(++)} = \delta R^{(--)} = 0$.
Then, only the matrix elements
of $(+-)$ and $(-+)$ types are relevant
for Eq. (\ref{TDBdGKS_R_lin}).
Since these matrix are anti-symmetric,
the $(+-)$ and $(-+)$ matrix elements
of Eq. (\ref{TDBdGKS_R_lin}) read, for $i<j$,
\begin{equation}
\label{TDBdGKS_R_lin_eq1}
\begin{split}
(E_i+E_j-\omega) \delta R_{ij}^{(+-)}(\omega) + \delta H_{ij}^{(+-)}(\omega)
&= -{V}_{ij}^{(+-)}(\omega) , \\
(E_i+E_j+\omega) \delta R_{ij}^{(-+)}(\omega) + \delta H_{ij}^{(-+)}(\omega)
&= -V_{ij}^{(-+)}(\omega) .
\end{split}
\end{equation}
The residual fields $\delta H(\omega)$
are induced by the density fluctuation
$\delta R(\omega)$, as
$\delta H(\omega) = \partial H_s/\partial R|_{R=R_0} \cdot
\delta R(\omega)$.
Expanding their matrix elements as
\begin{equation}
\label{expansion}
\delta H^{(+-)}_{ij}(\omega)=\sum_{k<l} w_{ij,kl} \delta R_{kl}^{(+-)}(\omega)
+ \sum_{k<l} w'_{ij,kl} \delta R_{kl}^{(-+)}(\omega) ,
\end{equation}
we obtain the QRPA linear response equations in the matrix form.
\begin{equation}
\label{QRPA}
\left\{
\begin{pmatrix}
A   & B \\
B^* & A^*
\end{pmatrix}
- \omega
\begin{pmatrix}
1 & 0 \\
0 & -1
\end{pmatrix}
\right\}
\begin{pmatrix}
\delta R^{(+-)} \\
\delta R^{(-+)}
\end{pmatrix}
= -
\begin{pmatrix}
V^{(+-)} \\
V^{(-+)}
\end{pmatrix} ,
\end{equation}
where
$A_{ij,kl}\equiv (E_i+E_j)\delta_{ik}\delta_{jl} + w_{ij,kl}$
and
$B_{ij,kl}\equiv w'_{ij,kl}$.
$R^\dagger = R$ and
$(H_s)_{ij}^{(\pm\mp)}=\delta E/\delta R^{(\mp\pm)}_{ji}$
provide that
$w_{ij,kl}$ are Hermitian and $w'_{ij,kl}$ are symmetric.

When the external potential $V$ is identical to a one-body operator $F$,
the strength function is given by
\begin{equation}
\label{S_F}
S(\omega;F)\equiv
\sum_{n>0} \left| \bra{n} F \ket{0} \right|^2 \delta (\omega-E_n)
= -\frac{1}{\pi} \textrm{Im} R(\omega+i\epsilon;F) ,
\end{equation}
where $\epsilon$ is a positive infinitesimal and
\begin{equation}
\label{R_F}
\begin{split}
R(\omega;F) &=
\sum_{i<j}
\left\{ F^{(+-)*}_{ij} \delta R^{(+-)}_{ij} (\omega) 
+F^{(-+)*}_{ji} \delta R^{(-+)}_{ij}(\omega)
\right\} \\
&=
\frac{1}{2} \textrm{Tr}
\left[ F \delta R(\omega) \right] .
\end{split}
\end{equation}

\subsection{Normal modes and eigenenergies}
\label{sec:normal_modes}

The QRPA normal modes are defined by the eigenvalue problem
setting ${V}=0$ for Eq. (\ref{QRPA}).
We denote the $n$th eigenvalue and eigenstate by $\Omega_n$ and
a column vector ${Z}_n$ of the dimension $2D$, respectively;
$D$ being the number of independent two-quasiparticle pairs $(ij)$ ($i<j$).
It is easy to show that there is a conjugate-partner eigenstate
$\tilde{Z}_n$ with the eigenenergy $-\Omega_n$.
\begin{equation*}
{Z}_n\equiv 
\begin{pmatrix}
X_n \\
Y_n
\end{pmatrix}
,\quad
\tilde{Z}_n\equiv 
{\cal I}
{Z}_n^*
=
\begin{pmatrix}
Y_n^* \\
X_n^*
\end{pmatrix}
,
\end{equation*}
where ${\cal I} =
\begin{pmatrix}
0 & 1 \\
1 & 0
\end{pmatrix}
$.
The QRPA eigenvalue equations are
\begin{equation}
\label{RPA_eigenvalue_eq}
{\cal N} {\cal H} {Z}_n = \Omega_n {Z}_n ,
\quad
{\cal N} {\cal H} \tilde{Z}_n = -\Omega_n \tilde{Z}_n ,
\end{equation}
with the $2D\times 2D$ Hermitian matrices,
\begin{equation}
{\cal H}\equiv
\begin{pmatrix}
A & B\\
B^* & A^* 
\end{pmatrix}
, \quad
{\cal N}\equiv
\begin{pmatrix}
1 & 0 \\
0 & -1
\end{pmatrix}
.
\label{H_and_N}
\end{equation}
The eigenvectors are normalized as
${Z}_n^\dagger {\cal N} {Z}_m
=-\tilde{Z}_n^\dagger {\cal N} \tilde{Z}_m = \delta_{nm}
$.

Let us define 
the following $2D\times 2D$ matrices,
\begin{equation}
{\cal Z}\equiv ({Z},\tilde{Z})
=
\begin{pmatrix}
X & Y^*\\
Y & X^*
\end{pmatrix}
,\quad
\Omega \equiv 
\begin{pmatrix}
\Omega_D & 0\\
0 & \Omega_D
\end{pmatrix}
,
\label{Z_and_Omega}
\end{equation}
where $\Omega_D$ is the $D\times D$
diagonal matrix containing the eigenvalues $\Omega_n$.
Then, Eq. (\ref{RPA_eigenvalue_eq}) can be written as
\begin{equation}
{\cal N H Z} = {\cal Z} \Omega {\cal N} .
\label{NHZ=ZOmegaN}
\end{equation}
Using the Hermicity of ${\cal H}$ and Eq. (\ref{NHZ=ZOmegaN}),
one can prove $[{\cal Z^\dagger N Z},\Omega {\cal N}]=0$,
which indicates that $\Omega{\cal N}$ and ${\cal Z^\dagger N Z}$ are both
diagonal.
Therefore, the normalization condition is written as
${\cal Z^\dagger N Z}={\cal N}$,
which we call ``${\cal N}$-orthonormalization''.
The matrix ${\cal N}$ plays a role of the norm matrix.
Since this also means ${\cal NZN}=({\cal Z}^\dagger)^{-1}$,
it leads to the completeness relation,
${\cal Z N Z^\dagger}={\cal N}$ \cite{RS80}.
The QRPA matrix ${\cal H}$ can be written as
\begin{equation}
\label{H=NZOmegaZN}
{\cal H}={\cal N}
\sum_n \left(
{Z}_n  \Omega_n {Z}_n^\dagger
+ {\tilde{Z}}_n  \Omega_n {\tilde{Z}}_n^\dagger
\right) {\cal N} 
= {\cal NZ}\Omega{\cal Z^\dagger N}
.
\end{equation}
From this, it is easy to find
$
{\cal Z^\dagger HZ} = \Omega
$.

For a given one-body Hermitian operator $\hat{F}$,
 we define a vector
$F_v$ by their $(+-)$-type matrix elements, $F^{(+-)}_{ij}$
with $i<j$,
and its RPA conjugate partner $\tilde{F}_v={\cal I}{F}_v^*$.
\begin{equation*}
F_v=
\begin{pmatrix}
F^{(+-)} \\
0
\end{pmatrix}
,\quad
\tilde{F}_v=
\begin{pmatrix}
0 \\
F^{(+-)*} 
\end{pmatrix}
.
\end{equation*}
The transition amplitude of $F$ between the ground and the $n$th
excited state is given by
\begin{equation}
\begin{split}
\bra{n}\hat{F}\ket{0} &= \sum_{i<j} \left\{
F^{(+-)}_{ij} X_n(ij) + F^{(+-)*}_{ij} Y_n(ij) \right\} \\
&=
{Z}_n^\dagger ({F}_v+\tilde{F}_v)
= ({F}_v+{\tilde{F}_v})^T {Z}_n^* .
\end{split}
\label{nF0=ZF}
\end{equation}

In most of numerical applications,
the QRPA eigenvalue problem is solved by constructing the QRPA
matrices in the quasiparticle- or
canonical-basis representations.
We may transform the non-Hermitian eigenvalue problem of 
Eq. (\ref{RPA_eigenvalue_eq}) to a Hermitian one \cite{RS80}.
For spherical nuclei, the matrix is block-diagonal with respect to
the angular momentum
and the parity of two-quasiparticle states, $[ij]^{J\pi}_M$.
Thus, the numerical cost is moderate in this case and
many calculations were performed
(See review papers by \textcite{BHR03,VALR05}).
In recent years, the QRPA calculations with modern EDFs
have become available for deformed nuclei
\cite{AKR09,YG08,PG08,TE10,Los10}.
The truncation of the two-quasiparticle space $(ij)$ is usually
adopted with respect to either the energy, $E_i+E_j$,
or the occupation of the canonical states $(\rho_i,\rho_j)$.
The calculation of the residual kernels, $w_{ij,kl}$ and $w'_{ij,kl}$,
is very demanding, because they have four quasiparticle indices.

If the residual kernel is written in a separable form with
a Hermitian one-body operator $\hat{F}$,
\begin{equation*}
w_{ij,kl}=\kappa F^{(+-)}_{ij} F^{(-+)}_{lk} ,
\quad
w'_{ij,kl}=\kappa F^{(+-)}_{ij} F^{(+-)}_{lk} ,
\end{equation*}
the computational cost may be significantly reduced
because the QRPA eigenvalue problem can be
cast into a dispersion equation \cite{RS80}.
For a given set of operators $\{ F^{(n)} \}$, the coupling constants
$\kappa^{(mn)}$ are derived from the Skyrme EDFs \cite{NKR02}.
This separable RPA calculation has been performed for deformed nuclei
to give a reasonable description of giant resonances \cite{Nes06}.

When the continuous symmetry is broken in the ground state,
there is another ``ground state'' degenerate in energy
whose density, $R_0+\delta R$,
is infinitesimally deviated from $R_0$.
Since both $R_0$ and $R_0+\delta R$ satisfy the stationary equation
(\ref{TDBdGKS_pair_rot}),
one can immediately derive Eq. (\ref{TDBdGKS_R_lin}) with
$\omega=V=0$.
Therefore,
the ANG modes appear as the zero-mode solution with $\Omega_{\rm ANG}=0$.
In this case, it is useful to rewrite Eq. (\ref{RPA_eigenvalue_eq})
in the momentum-coordinate (PQ) representation \cite{RS80}.
For the ANG mode (translation/rotation/pair-rotation),
the momentum ($P/J/N$) corresponds to a known operator
(Sec.~\ref{sec:quasi-stationary-solutions}).
Then, it ends up the famous equation by \textcite{TV62},
which determines the inertial mass and the coordinate of the ANG mode
\footnote{
We note here that
there have been some other attempts to explain the finite value of
moment of inertia as an analogue of the Higgs mechanism
with the SSB \cite{FU86}.
}.
A modern technique to solve the Thouless-Valatin equation and numerical
examples are presented in \textcite{Hin15}.

\subsection{Finite amplitude method}

Instead of explicitly calculating the residual kernels with four
quasiparticle indices, 
$w_{ij,kl}$ and $w'_{ij,kl}$,
it is possible to compute them in an implicit manner.
A possible approach is the finite amplitude method \cite{NIY07}.
The essential idea comes from the fact that
the linear response equation
(\ref{TDBdGKS_R_lin_eq1}), which is identical to Eq. (\ref{QRPA}),
only contains the ``one-body'' quantities with two quasiparticle indices.
The residual fields $\delta H^{(+-)}(\omega)$ and $\delta {H}^{(-+)}(\omega)$
can be uniquely determined for given $\delta R^{(\pm\mp)}_{ij}$.
The linear expansion in Eq. (\ref{expansion}) is
achieved by a numerical finite difference method,
and $\delta H$ in the left hand side is obtained
without calculating $w_{ij,kl}$ and $w'_{ij,kl}$.

\subsubsection{Basic idea}

The Fourier component $\delta R(\omega)$ can be written
in terms of their matrix elements as
\begin{equation}
\label{delta_R_omega}
\delta R(\omega)=\sum_{i,j} \left\{
\Psi_i^0 \delta R^{(+-)}_{ij}(\omega) \tilde{\Psi}_j^{0\dagger}
+ \tilde\Psi_i^0 \delta R^{(-+)}_{ij}(\omega) \Psi_j^{0\dagger}
\right\} .
\end{equation}
Here, the summation with respect to $i$ and $j$
is taken over all the positive-energy quasiparticles.
Comparing Eqs. (\ref{delta_R_t}) and (\ref{delta_R_omega}),
we find $\delta\Psi_i(\omega)=-\sum_j \tilde\Psi_j^0 \delta R^{(-+)}_{ji}$
and $\delta\Psi^\dagger_i(\omega)=-\sum_j \delta R^{(+-)}_{ij}
 \tilde\Psi_j^{0\dagger}$.
Using quasiparticle states slightly modified from $\Psi_i^0$,
\begin{equation*}
\begin{split}
\Psi_i(\omega)&=\Psi_i^0+\eta\delta\Psi_i(\omega)
=\Psi_i^0-\eta \sum_{j>0} \tilde\Psi_j^0 \delta R^{(-+)}_{ji}(\omega), \\
\Psi_i'^\dagger(\omega)&=\Psi_i^{0\dagger}+\eta\delta\Psi_i^\dagger(\omega)
 =\Psi_i^{0\dagger} -\eta \sum_{j>0} \delta R^{(+-)}_{ij}(\omega) \tilde{\Psi}_j^{0\dagger} ,
\end{split}
\end{equation*}
the density $R_\eta(\omega) \equiv R_0+\eta \delta R(\omega)$ can be
written as
\begin{equation*}
R_\eta(\omega) 
= 1-\sum_i\Psi_i(\omega) \Psi_i'^\dagger(\omega) +O(\eta^2) .
\end{equation*}
Note that, since the Fourier component $\delta R(\omega)$ is no longer
Hermitian, $\Psi_i(\omega)$ and $\Psi_i'^\dagger(\omega)$ are not
Hermitian conjugate to each other.
The induced fields are now calculable in the following way.
\begin{equation}
\label{FAM_formula}
\begin{split}
\delta H^{(+-)}_{ij}(\omega) &=
\Psi_i^{0\dagger}  \frac{1}{\eta}
\left\{ H_s[R_\eta(\omega)]- H_s[R_0] \right\}
\tilde{\Psi}_j^0 ,\\
\delta H^{(-+)}_{ij}(\omega) &=
\tilde{\Psi}_i^{0\dagger}  \frac{1}{\eta}
\left\{ H_s[R_\eta(\omega)]- H_s[R_0] \right\}
{\Psi}_j^0 .
\end{split}
\end{equation}
Rigorously speaking, the limit of $\eta\rightarrow 0$ should be taken.
However, in practice, we may use a small but finite value of $\eta$.
Using Eq. (\ref{FAM_formula}) with a small value of $\eta$,
calculation of the induced residual fields $\delta H^{(\pm\mp)}$
can be achieved by
calculation of matrix elements of the BdGKS Hamiltonian
$H_s[R]$.
This is much easier task than calculation of the residual kernels,
$w_{ij,kl}$ and $w'_{ij,kl}$.
It should be noted that
$H_s[R_\eta]$ should be constructed self-consistently with
the quasiparticles $\Psi_i$ and $\Psi_i'^\dagger$,
namely, $(U,V)$ with a small mixture of $\delta R^{(-+)}$ and
$(U^*,V^*)$ with a small mixture of $\delta R^{(+-)}$.
In order to obtain the solution $\delta R^{(\pm\mp)}$,
we solve Eq. (\ref{TDBdGKS_R_lin_eq1}) iteratively,
starting from initial
values for $(\delta R^{(+-)},\delta R^{(-+)})$.

\subsubsection{Strength functions}

For calculation of strength functions, one can solve the linear
response equation with a given frequency $\omega$
by choosing the external potential $V$ identical to the operator $F$.
Then, according to Eqs. (\ref{S_F}) and (\ref{R_F}),
the strength function $S(F,\omega)$ with respect to $F$ is obtained.
To obtain an energy profile of $S(F;\omega)$, we need to repeat the
calculation with different values of $\omega$.

There is another approach based on the iterative construction of
the subspace in which the diagonalization is performed \cite{OJJJ88,TINC09}.
The Krylov subspace generated by a pivot vector with respect to
the one-body operator $F$ 
preserves the energy-weighted sum rule (EWSR) values.
Therefore, it is suitable for calculating a gross energy profile
of the strength function by a small number of iterations.
Some more details of these iterative methods will be discussed in
Sec.~\ref{sec:iterative_methods}.
Applications of the finite amplitude method to calculation of the
strength functions have been performed for the Skyrme EDFs
\cite{INY09,INY11,Sto11,INY13,IHSN14,MSZE14,PKZX14,NAEIY11-P,Nak14-P1}
and the covariant EDFs 
\cite{LNNM13,NKTVR13,LNNM14-P}.
The finite amplitude method is also applied to calculation of the sum rules,
which suggests approximate validity of the Thouless theorem
for nuclear EDFs \cite{HKNO15}.

\subsubsection{Normal-mode eigenstates}

It is often our interest to obtain the QRPA eigenmodes.
These eigenmodes are, in principle, obtained if the matrix ${\cal H}$
in Eq.~(\ref{NHZ=ZOmegaN})
is explicitly constructed.
The finite amplitude method can also be used for this purpose,
to facilitate the calculation of the residual kernels \cite{AN13}.
Suppose we set
$\delta R^{(+-)}_{kl}=1$ for a specific pair $(kl)$ and
the rest all zero.
Then, the calculation of $\delta H_{ij}^{(+-)}$
using the formula (\ref{FAM_formula}) provides $w_{ij,kl}$.
On the other hand, 
setting $\delta R^{(-+)}_{kl}=1$ and
the rest zero, 
the calculation of $\delta H_{ij}^{(+-)}$ produces $w'_{ij,kl}$.
This can be easily understood from Eq. (\ref{expansion}).
In this way, the QRPA matrix can be calculated without a complicated coding
process.
The usefulness of the method is demonstrated for
the Skyrme \cite{AN13} and the covariant EDFs \cite{LNNM13}.

When the matrix dimension becomes too large to directly handle,
there are other approaches.
For instance,
solution of the linear response equations (\ref{QRPA}) with complex
frequencies combined with the contour integral serves for this purpose
\cite{HKN13}.
This is based on the idea that the contour integral around the $n$th
eigenenergy provides
\begin{equation*}
\begin{split}
(2\pi i)^{-1} \int_{C_n} \delta R^{(+-)}_{ij}(\omega) d\omega
= X_n(ij)\bra{n}F\ket{0} ,\\
(2\pi i)^{-1} \int_{C_n} \delta R^{(-+)}_{ij}(\omega) d\omega
= Y_n(ij)\bra{n}F\ket{0} \\
\end{split}
\end{equation*}
for an external potential $V=F$.
The contour must be chosen to enclose a single pole.
This has been tested also for the charge-changing modes \cite{MSZE14}.

The truncation of the space by an iterative procedure
is another possible option.
See Sec.~\ref{sec:iterative_methods} for some more details.

\subsection{Iterative methods for solutions}
\label{sec:iterative_methods}

In the finite amplitude method,
the numerical solution of the linear response equation is
obtained by using an iterative algorithm.
This significantly saves computational resources, especially
the necessary memory size, 
because all we need to calculate are one-body quantities,
not two-body ones.

\subsubsection{Solution for fixed energy}

A possible
iterative procedure for the solution of 
Eq. (\ref{TDBdGKS_R_lin_eq1}) is given as follows:
For a given external potential $V^{(\pm\mp)}_{ij}$,
we assume a certain initial value for $\delta R^{(\pm\mp)}_{ij}$
for which the residual induced fields $\delta H^{(\pm\mp)}_{ij}$
are calculated according to Eq. (\ref{FAM_formula}).
$H_s[R_\eta]$ can be calculated with the quasiparticle
states $\Psi_i^0$ and $\Psi_i'^{0\dagger}$ replaced by $\Psi_i(\omega)$
and $\Psi_i^\dagger(\omega)$, respectively.
Then, the left hand side of Eq. (\ref{TDBdGKS_R_lin_eq1})
which is identical to that of Eq. (\ref{QRPA}),
are computed.
If these equations are not satisfied,
we update the densities,
$\delta R^{(\pm\mp)}_{ij}$,
according to an adopted iterative algorithm and repeat the calculation
until the convergence.
When the frequency $\omega$ is complex,
one should adopt an iterative algorithm which can be applied to
a linear algebraic equation with a non-Hermitian matrix.

\subsubsection{Diagonalization in Krylov subspace}

There are recent developments based on the iterative diagonalization
based on the Krylov space techniques.
This is especially useful for calculations of the strength function, 
because it conserves the energy-weighted sum-rule (EWSR) value
of odd moments.
Basically, they resort to the transformation of the matrix with
dimension $2D$
into the one in the Krylov subspace
with dimension $2d \ll 2D$.

Using Eq.~(\ref{H=NZOmegaZN}), we have
\begin{equation}
\label{NH^L}
({\cal NH})^L
= {\cal Z}(\Omega{\cal N})^{L-1} \Omega {\cal Z^\dagger N}
= {\cal Z}\Omega^L {\cal N}^{L-1} {\cal Z^\dagger N}
.
\end{equation}
Using the expression of transition amplitudes of Eq. (\ref{nF0=ZF}),
the EWSR value of order $L$ is given by
\begin{equation*}
m_L\equiv \sum_n \Omega_n^L |\bra{n} F \ket{0}|^2
=\frac{1}{2}
({F}_v+\tilde{F}_v)^\dagger
{\cal Z}\Omega^L{\cal Z}^\dagger
({F}_v+\tilde{F}_v)
.
\end{equation*}
For odd-$L$, using Eq. (\ref{NH^L}), this can be written as
\begin{equation}
m_L
=\frac{1}{2}
({F}_v+\tilde{F}_v)^\dagger
({\cal NH})^L{\cal N}
({F}_v+\tilde{F}_v)
.
\label{EWSR_odd_L}
\end{equation}
Therefore, starting from a pivot vector ${F}_v$ and its conjugate
$\tilde{F}_v$, the Krylov subspace of dimension $2d>L$,
\begin{equation}
\label{Krylov_subspace}
\{
F_v, {\tilde{F}_v},
({\cal NH}){F}_v,({\cal NH}) {\tilde{F}_v},
 \cdots, 
({\cal NH})^{d-1} {F}_v,({\cal NH})^{d-1} {\tilde{F}_v}
\}
\end{equation}
can span the intermediate space in Eq. (\ref{EWSR_odd_L}).
In Appendix~\ref{sec:appendix_krylov}, we show that
the reduction from the $2D$ into the $2d$ RPA subspace (\ref{Krylov_subspace})
conserves
the sum rules $m_L$ with odd $L$ and $L<2d$ \cite{JBH99}.

To construct the subspace (\ref{Krylov_subspace}), 
one can adopt the Lanczos iteration algorithm.
The Lanczos iteration produces an orthonormal basis set for the Krylov
subspace, which makes $a$ and $b$ matrices tridiagonal.
This works nicely for the case of
a schematic separable interaction \cite{JBH99}.
However, since numerical errors are accumulated
during the iterations,
other algorithms, such as the non-Hermitian Arnoldi iteration,
have been adopted for realistic Skyrme energy functionals \cite{Toi10}.
Even for low-lying eigenstates, 
the method successfully works \cite{CTP12}.
The conjugate gradient algorithm may be another possible solver,
which was used for low-lying RPA solutions
in the coordinate-space representation \cite{IH03,Ina05,Ina06}.

\subsection{Green's function method}
\label{sec:Green's_function_method}

It becomes increasingly important to study unbound and weakly bound
nuclei in physics of rare isotopes near the drip lines.
There have been a number of developments for treatment of the resonance and
continuum, including the continuum shell model \cite{OPR03},
the Gamow shell model \cite{MNPB02,BLSV02},
the Gamow HFB method \cite{MMS08},
the complex scaling method \cite{AMKI06}, and
the $R$-matrix theory \cite{DB10}.
In the linear response calculation based on the TDDFT,
the one-body continuum (and a part of two-body continuum) can be
taken into account by the use of Green's function.
In this section, we recapitulate the general formalism for
superconducting cases (``continuum QRPA'').

\subsubsection{Response function}

The QRPA linear response equation (\ref{QRPA})
can be rewritten as
\begin{equation}
\label{QRPA_2}
\left(\Pi_0^{-1}(\omega) - {\cal W} \right)
\begin{pmatrix}
\delta R^{(+-)}(\omega) \\
\delta R^{(-+)}(\omega)
\end{pmatrix}
 = 
\begin{pmatrix}
V^{(+-)}(\omega) \\
V^{(-+)}(\omega) 
\end{pmatrix}
,
\end{equation}
with
\begin{equation*}
\Pi_0^{-1}(\omega)\equiv
\begin{pmatrix}
\omega - A_0 & 0 \\
0 & -\omega -A_0
\end{pmatrix}
, \quad
{\cal W}\equiv
\begin{pmatrix}
w  & w' \\
w'^* & w^*
\end{pmatrix}
,
\end{equation*}
where $(A_0)_{ij,kl}=(E_i+E_j)\delta_{ik}\delta_{jl}$
in the quasiparticle representation.
Equation (\ref{QRPA_2}) is inverted by the QRPA response function
$\Pi(\omega)$ as
$\delta R(\omega)=\Pi(\omega){V}(\omega)$, where
\begin{equation}
\label{Pi}
\Pi(\omega) = (\Pi_0^{-1}(\omega) - {\cal W} )^{-1}
=(1-\Pi_0(\omega) {\cal W} )^{-1} \Pi_0(\omega) .
\end{equation}
Here, $\Pi_0(\omega)$ can be schematically written as
\begin{equation}
\Pi_0(\omega)= \sum_{i} \left\{
{\cal G}_0(\omega-E_i) \tilde\Phi_i\tilde\Phi_i^\dagger
+\tilde\Phi_i\tilde\Phi_i^\dagger {\cal G}_0(-\omega-E_i)
\right\}
,
\label{Pi_0}
\end{equation}
using the Green's function ${\cal G}_0(E)$.
Its derivation is given in Appendix~\ref{sec:appendix_response}.
The precise forms of Eq.~(\ref{Pi_0}) are given by
Eqs.~(\ref{Pi_0_4_rep}) and (\ref{Pi_0_2_rep}).

The strength function with respect to the operator $F$ is
obtained according to Eqs. (\ref{S_F}) and (\ref{R_F}).
For $\omega\geq 0$,
\begin{equation*}
S(\omega;F)
=-\frac{1}{2\pi}\textrm{Im}\left[
F^\dagger \Pi(\omega+i\eta) F
\right] .
\end{equation*}
From Eq.~(\ref{delta_R_0_spectrum}), one can see that,
without the residual interaction ${\cal W}=0$,
this leads to the unperturbed strength function, 
$(1/2)\sum_{ij}|V_{ij}^{(+-)}|^2\delta(\omega-E_i-E_j)$.

Since the response function has four indices, in general,
their calculation and inverse operation in Eq. (\ref{Pi})
are very difficult tasks.
It becomes practical when we need only their diagonal elements.
The functional of local densities, such as Skyrme functionals
with local potentials,
provides an example in which the coordinate-space representation
$\{ \vec{r} \}$
allows us the diagonal representation.
The presence of the spin-orbit and finite-range exchange terms makes
its application more difficult.

\subsubsection{Boundary condition}
\label{sec:boundary_condition}

One of the motivation of the Green's function formalism is
the exact treatment of the continuum.
This can be done by imposing the proper boundary condition in
the Green's functions in Eq. (\ref{Pi_0}).
The density response in the time domain can be given by
\begin{equation*}
\delta R(t)= \int \Pi(t-t') {V}(t') dt' .
\end{equation*}
Here, $\Pi(t-t')$ should be zero for $t<t'$;
$\Pi(t)=\theta(t)\Pi(t)$.
This causality condition is achieved by adding a positive infinitesimal
to $\omega$ in its the Fourier component $\Pi(\omega)$.
Thus, the replacement of $\omega\rightarrow \omega+i\eta$ leads to
the retarded (outgoing) boundary condition for ${\cal G}_0(\omega-E_i)$
and the advanced (incoming) boundary condition for ${\cal G}_0(-\omega-E_i)$
in the expression of $\Pi_0$.
For $\omega>E_i$, the outgoing asymptotic behavior is
important for the former Green's function, which describes
escaping of a particle or a Cooper pair.
This provides an exact treatment of the continuum in the
linear density response.

For superconducting systems with the ground-state BdGKS solution with
$\kappa\neq 0$,
the Green's function with the outgoing (incoming) boundary condition
can be constructed for a spherical system
using the partial-wave expansion \cite{BSTF87}.
The quasiparticle states $\Psi_i^0$
whose energy is smaller than the absolute value of the chemical potential,
$E_i<|\mu|$, are bound and discrete,
while those with $E_i>|\mu|$ are unbound with continuum spectra.
The summation over the quasiparticle states in Eq. (\ref{Pi_0})
must be performed with respect to all the
negative-energy states $\tilde\Psi_i^0$.
This is not trivial because the index $i$ are not
discrete but continuous.
To overcome this difficulty, the contour integral in the complex energy plane
is useful \cite{Mat01}.
The spectral representation of the Green's function (\ref{G_0})
leads to
\begin{equation}
\sum_i f(-E_i) \tilde\Phi_i \tilde\Phi_i^\dagger
= (2\pi i)^{-1} \int_C f(E) {\cal G}_0(E) ,
\label{sum_to_contour_integral}
\end{equation}
for arbitrary function $f(E)$.
Here, the contour $C$ is chosen to enclose the negative part of the
real axis.
Replacing the summation in Eq. (\ref{Pi_0}) by the contour
integral of Eq. (\ref{sum_to_contour_integral}),
the response function is able to describe escaping of
one-particle and two-particle decays from excited states.
Therefore, the QRPA linear response theory with the Green's function
can describe correlations among two escaping particles.

the negative-energy quasiparticles $\tilde\Psi_i^0$
are nothing but hole states and the summation over $i$ runs over
only the hole states.
This method is known as the continuum RPA, and much easier than the
continuum QRPA.
The numerical applications were first achieved for spherical systems
\cite{SB75,ZS80}.
The continuum RPA calculations with the Gogny EDFs have been
recently achieved for spherical systems,
by transforming the RPA eigenvalue equation
(\ref{RPA_eigenvalue_eq}) into those for the channel functions \cite{DCAL11}.

For deformed systems, decomposing the BdGKS Hamiltonian into
its spherical and deformed parts,
$H_s=H_\textrm{sph} + V_\textrm{def}$,
we can use the identity
\begin{equation*}
{\cal G}_0^{(\pm)}(E) = {\cal G}_\textrm{sph}^{(\pm)}(E)
+ {\cal G}_\textrm{sph}^{(\pm)}(E) V_\textrm{def} 
{\cal G}_0^{(\pm)}(E) ,
\end{equation*}
where 
${\cal G}_\textrm{sph}^{(\pm)}(E)$ is the Green's function for
the spherical Hamiltonian $H_\textrm{sph}$.
This method with the three-dimensional coordinate-space representation
has been applied to normal systems, such as photoabsorption
in molecules \cite{NY01,NY03,YNIB06} and light nuclei \cite{NY05},
however, not to superconducting systems.
For deformed superconducting nuclei, although the full continuum
linear response calculation has not been achieved yet, the construction of the
Green's function has been carried out by using the coupled-channel
scheme \cite{OM09}.
The similar method was developed earlier for normal systems,
and applied to
linear density response in axial symmetric molecules \cite{LS83,LS84,Lev84}.

\subsection{Real-time method}
\label{sec:linear_real_time}

Another approach to the linear response
is to solve the TDBdGKS equation (\ref{TDBdGKS_eq}) directly in real time,
with a weak perturbative external field.
In the calculation, we do not linearize the equation.
Thus, the same numerical code could serve for studies of
the non-linear dynamics (Sec.~\ref{sec:real_time}).
This is particularly convenient for calculation of the strength
function $S(F;E)$ for a wide range of energy,
associated with a one-body operator $F$
which does not excite the ANG modes.
On the contrary, the method is not suitable for obtaining
information on a few excited normal modes.
This is due to the uncertainty principle;
The achieved energy resolution $\Delta E$ is inversely proportional to
the duration of time evolution $T$.

A bulk property of the linear response is determined by time evolution
of a short period of time.
For instance, the EWSR value associated with a one-body operator $F$
is obtained instantly as
\begin{equation*}
m_1 = \sum_n \Omega_n |\bra{n}F\ket{0}|^2 
=\frac{1}{2\eta} \left.\frac{d}{dt} \textrm{Tr}\left[F R(t)\right]
\right|_{t=0} ,
\end{equation*}
where the initial state is boosted by the operator $F$ as
$\Psi_i(t=0) 
=\begin{pmatrix}
e^{i\eta F} U_i \\
e^{-i\eta F} V_i
\end{pmatrix}
$,
where the parameter $\eta$ is a small number.
This is generally true for all odd-$L$ moments,
$m_L \propto d^L \textrm{Tr}[FR(t)]/dt^L |_{t=0}$.

\subsubsection{Strength functions}

The real-time calculation of the strength function is performed in the
following way.
The initial state is the ground state,
and an external potential ${V}(t)=f(t) F$, which is proportional
to the operator $F$, is activated at time $t=0$.
In the linear regime, the function $f(t)$ 
should be small to validate the linear response.
The strength function (\ref{S_F}) can be obtained as
\begin{equation}
S(\omega;F) 
=
\frac{-1}{2f(\omega)}
\textrm{Im}\int_{-\infty}^\infty \textrm{Tr}[F R(t)] g(t) e^{i\omega t} dt ,
\label{S(E)_fourier}
\end{equation}
where $f(\omega)$ is a Fourier transform of $f(t)$.
If we choose $f(t)=f_0 \delta(t)$, we have $f(\omega)=f_0$ which excites
all the normal modes with equal strength.
In the linear regime, $\int \textrm{Tr}[FR(t)] e^{i\omega t} dt$
is proportional to $f(\omega)$.
Thus, Eq. (\ref{S(E)_fourier}) gives a unique result.

In order to get a smooth energy profile $S(\omega;F)$,
the time dependence in the integrand in Eq. (\ref{S(E)_fourier})
must vanish at $t=T$.
In practice, 
it is customary to include the damping factor $g(t)$ in the integrand
in Eq. (\ref{S(E)_fourier}), e.g.,
the exponential damping associated with
a smearing width $\gamma$;
$g(t)=\theta(t)\theta(T-t)e^{-\gamma t/2}$.
The idea of the real-time method was proposed in
\textcite{BF79} to calculate the
energies of the giant resonances.
The strength functions are calculated with modern Skyrme EDFs
\cite{NY05,FSS12,UO05,Mar05},
and including pairing effects \cite{SBMR11,Eba10,ENI14,SL14,HN07,Has12,TU02}.

\subsubsection{Absorbing boundary condition}
\label{sec:ABC}

In general,
an external potential $V(t)$ excites the system into a superposition
of many different elementary modes of excitation.
Therefore, the particle decays simultaneously occur at different
energies. 
In contrast to the linear response equation with fixed frequency $\omega$,
we do not know the asymptotic form in the real-time method.
Nevertheless, in the linear regime,
there is a useful method to realize an approximate outgoing boundary
condition for normal systems.

A key is that the ground-state KS orbitals $\varphi_i(\vec{r})$
and the transition density in the linear response
$\delta\rho(\vec{r},t)=\sum_i \varphi_i(\vec{r}) \delta \varphi_i^*(\vec{r},t)
+ \textrm{c.c.}$ are both localized in space.
During the time evolution, we may simply {\it absorb} the
outgoing waves from $\varphi_i(\vec{r},t)$ in an outer region ($r>r_0$)
where $\varphi_i(\vec{r})|_{r>r_0}=0$.
This can be approximately done
by choosing a proper absorbing imaginary potential
in the outer region.
Note that,
in the linear regime, the particle number is still conserved,
because $\int_\textrm{out} \delta\rho(\vec{r},t) d\vec{r} = 0$.
This absorbing boundary condition has been adopted in nuclear TDDFT
calculations \cite{NY02-P2,NY04-P1,NY05,RSAMS06}
and treated in a rigorous manner \cite{PS13}.
It is also used in other fields of physical problems
\cite{MPNE04,YKNI11}.
For the superconducting case, 
even at the ground state of finite localized systems,
most of $U_i(\vec{r},\sigma)$ are not localized in space.
Thus, 
the application of the absorbing boundary condition is not trivial
in this case.

\subsection{Extension: Particle-vibration coupling}
\label{sec:PVC}

The QRPA calculation is successful to reproduce a variety of properties of
nuclear excitations, especially of high-lying giant resonances.
However, it has known limitations too.
For instance, the widths of giant resonances
in heavy nuclei are not well accounted for, although
the peak energy and summed strength are well reproduced.
The continuum QRPA is capable of calculating the escaping width of neutrons,
however, it does not describe the spreading
associated with coupling to complex configurations, such as
many-particle-many-hole states.
A possible improvement is explicit inclusion of higher-order terms and
two-body correlations, which will be presented in Sec.~\ref{sec:TDDM}.
Another approach, which is discussed here, is the particle-vibration
coupling (PVC) scheme.
The PVC is also supposed to be responsible for the fact that
the experimental single-particle level density near the Fermi level is
higher than that in modern EDFs whose effective masses are smaller than unity.

\begin{figure}
\includegraphics[width=0.48\textwidth]{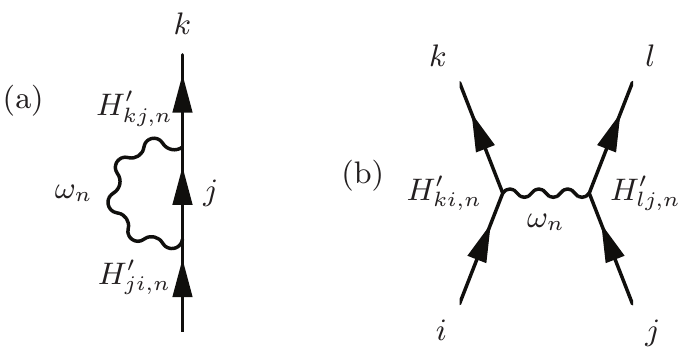}
\caption{
Second-order diagrams for particle-vibration coupling,
contributing to
(a) self-energy part $\Sigma_{ki}(E)$,
and (b) effective two-particle interactions.
}
\label{fig:PVC}
\end{figure}

The idea of the PVC is very old,
and connected to the essential concept of the Bohr-Mottelson's
unified model.
That is to say,
the single-particle motion and the vibrational (collective) motion 
in nuclei are coupled and influence each other.
In earlier times, a phenomenological potential with a schematic separable
interaction, $(\kappa/2) \hat{F}\hat{F}$, was used in many applications, 
which is essentially inspired by
the field coupling, $H'=\kappa \alpha \hat{F}$,
of \textcite{BM75}.
The PVC produces dressed (renormalized) single-particle states.
This affects many kinds of single-particle properties, including
self-energies, single-particle moments, transfer matrix elements,
and fragmentation of single-particle strengths
(See Fig.~\ref{fig:PVC} (a)).
It is also expected to contribute to effective two-particle interactions,
as Fig.~\ref{fig:PVC} (b),
which may be partially responsible for the attractive pairing interaction.

The causal single-particle Green's function obeys the Dyson's equation
\begin{equation*}
G(E)=G_0(E)+G_0(E)\Sigma(E) G(E) ,
\end{equation*}
where $G_0(E)$ is the unperturbed Green's function similar to
Eq. (\ref{G_0}) with the causal boundary condition, and
$\Sigma(E)$ is the proper self-energy part.
The self-energy is alternatively denoted as $M(E)$ and called
``mass operator'' \cite{MBBD85}.
In the PVC, $\Sigma(E)$ takes account of coupling to collective vibrations.
Normally, low-lying collective vibrational states are selectively
included in $\Sigma(E)$.
The lowest-order contribution to $\Sigma(E)$ is in the second order
coupling in $H'$,
as seen in Fig.~\ref{fig:PVC} (a).
The diagonal approximation is often adopted for the Dyson's equation,
namely, only the diagonal matrix elements of $\Sigma(E)$
in the quasiparticle basis are taken into account.

Recently, the PVC calculation has been carried
out with modern EDFs \cite{CSB10,BCB12,CCSB14,LR06,LA11,NCV14}.
It is extended to the quasiparticle-vibration coupling \cite{LRT08,Lit12,Yos09}.
They have shown successful description of various kinds of nuclear
phenomena, though there exist some ambiguities due to selection of
vibrational modes to be taken into account.
For weakly bound systems, vibrational states as well as the
single-particle states may be in the continuum.
As we discussed in Sec.~\ref{sec:boundary_condition},
this can be handled by the proper boundary condition for the Green's function.
The Dyson's equation in the coordinate-space representation
provides a scheme to treat the continuum boundary condition,
using a causal response function $\Pi$ also with the continuum \cite{MCV12}.
This was done for spherical normal systems, so far.

It is not so straightforward to formulate the PVC consistent with
the principle of the DFT.
A subtraction prescription is proposed \cite{Tse07,Tse13} and
applied to the PVC \cite{LRT10} and the second RPA \cite{GGE15}.
For the Skyrme EDF (zero-range effective interactions),
some attempts have been recently made to renormalize the divergent
second-order diagrams and to produce new EDFs for PVC calculations
\cite{MGRC12,BCR14}.
To our knowledge,
full respect of the Pauli principle and construction of the
{\it DFT-based}
particle-vibration coupling theory remain as challenging subjects.

\subsection{Illustrative examples}
\label{sec:linear_response_appli}

Recent trends in the linear response studies for nuclei are
calculations with all the residual fields (interactions),
continuum, pairing, and deformed ground states.
Let us show some examples.

\subsubsection{Giant resonances and ground-state deformation}

\begin{figure}[t]
\includegraphics[width=0.48\textwidth]{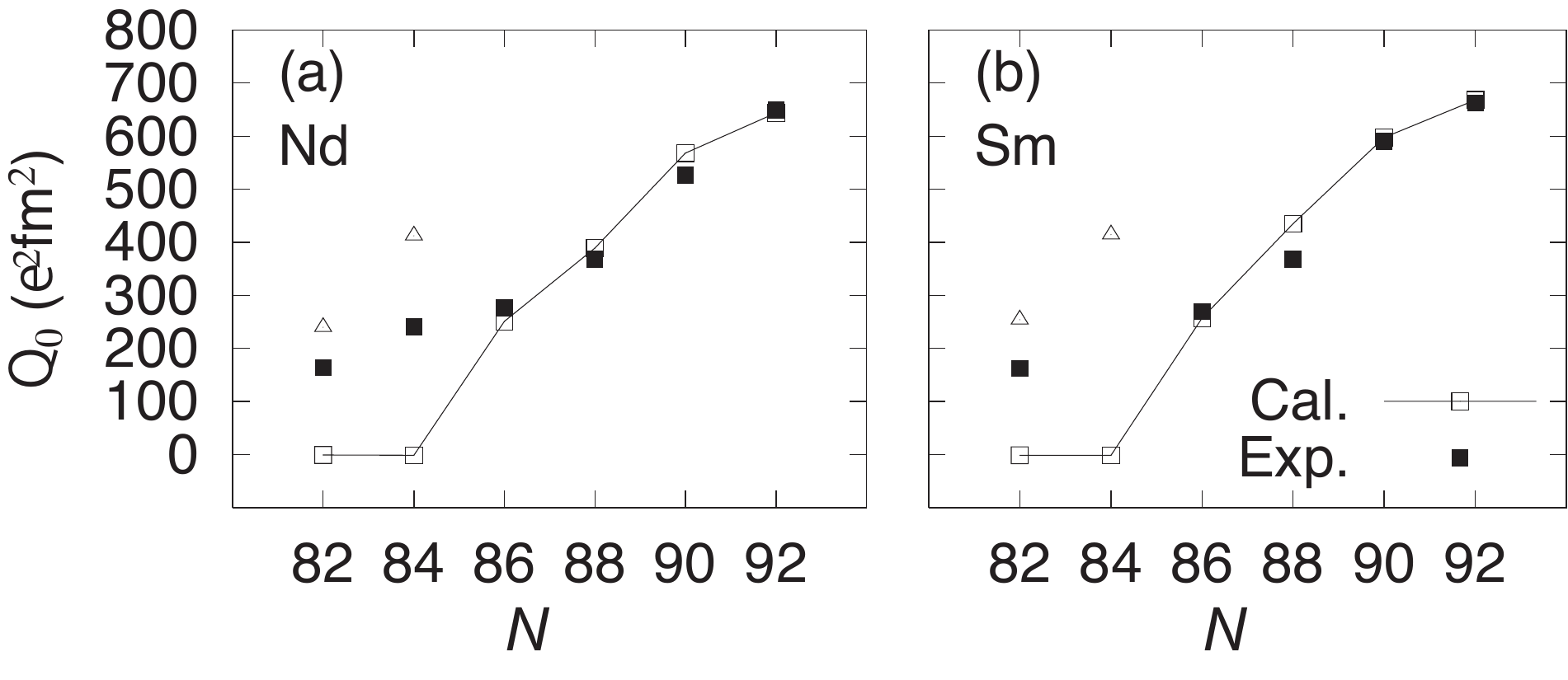}
\includegraphics[scale=0.66]{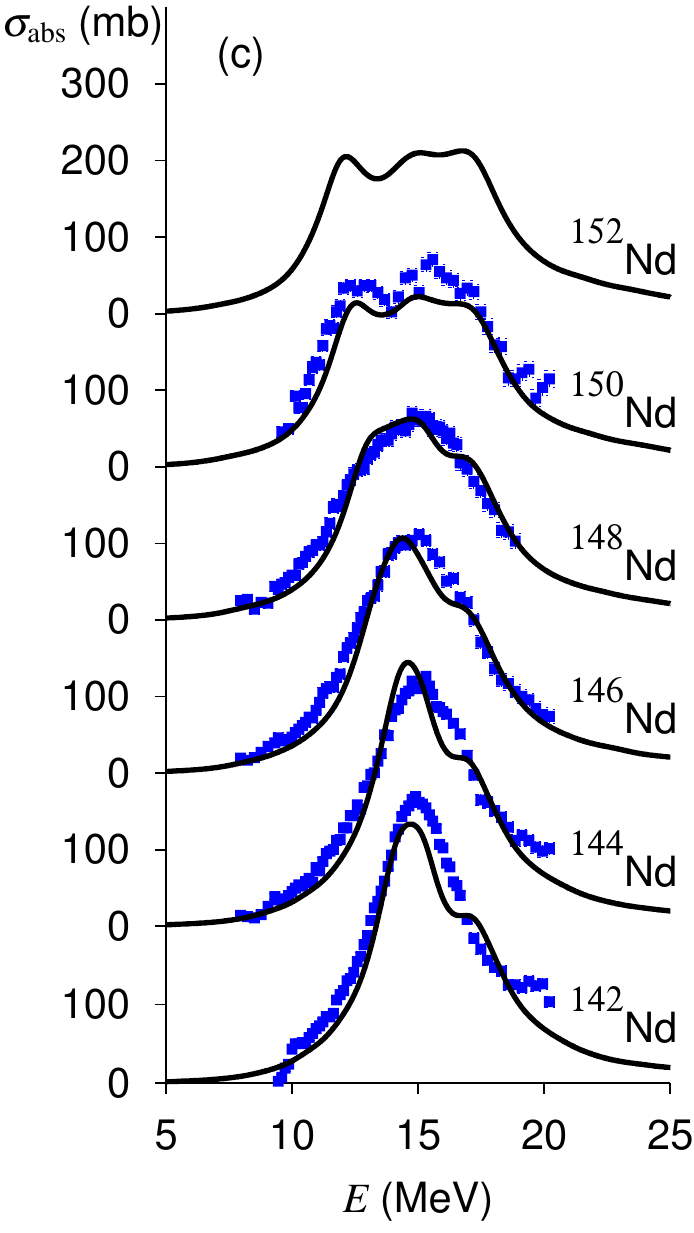}
\includegraphics[scale=0.66]{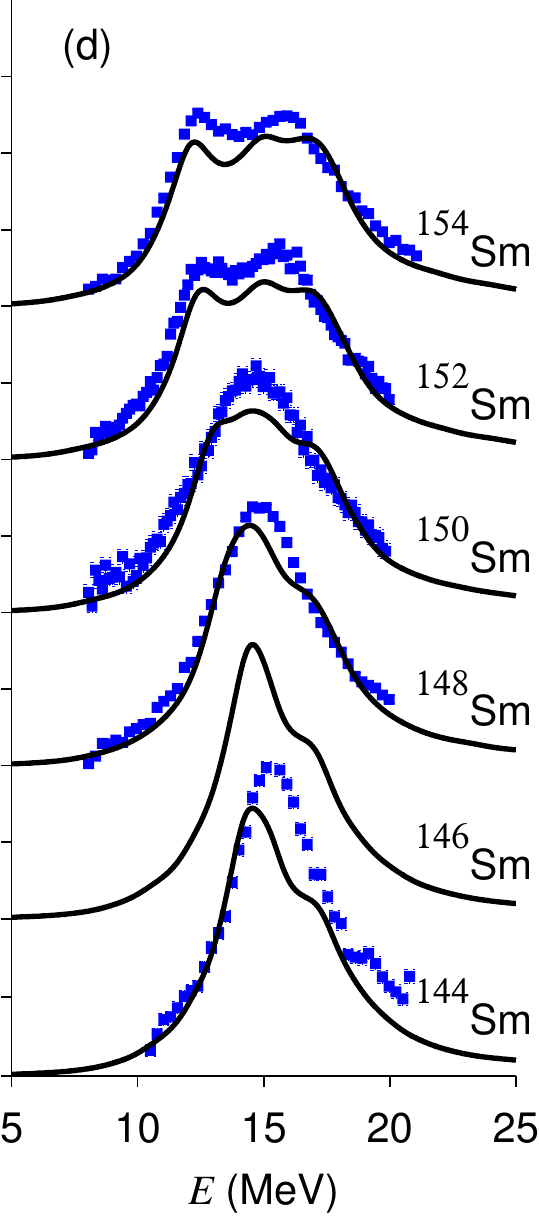}
\caption{
(Color online)
Deformation and photoabsorption in Nd and Sm isotopes, calculated
with the Skyrme energy functional of SkM$^*$.
Calculated and experimental intrinsic quadrupole moments
are denoted by open and closed symbols, respectively, for (a) Nd
and (b) Sm isotopes.
For spherical nuclei with $N=82$ and 84, we also plot the
values (triangles)
extracted from the QRPA calculation for $B(E2;2^+\rightarrow 0^+)$.
Photoabsorption cross sections for (c) Nd and (d) Sm isotopes.
for Nd and Sm isotopes.
The solid lines show the calculation and blue symbols are
experimental data \cite{CBBLV71,Car74}.
Adapted from \textcite{YN11}.
}
\label{fig:GDR}
\end{figure}

One of the successful applications of the nuclear EDF 
to linear response is the study of giant resonances.
The giant resonances are high-frequency collective modes of
excitation in nuclei, which exhausts a major part of the
energy-weighted sum-rule of the transition strengths.
They are usually classified according to the spin $S$,
isospin $T$, and multipolarity $L$.
Their properties are supposed to reflect some basic quantities of
the nuclear matter, such as the incompressibility, the symmetry energy, and
the effective mass \cite{HW01,RS80}.
Among them, the isovector giant dipole resonance ($S=0$, $T=1$, $L=1$),
which is excited by the photoabsorption,
is best known for a long time.
The giant dipole resonance is simply characterized by
the out-of-phase oscillation
of neutrons and protons.
The symmetry energy plays a major role in
determination of its peak position.
Figure~\ref{fig:GDR} shows the photoabsorption cross section for
Nd and Sm isotopes.
These isotopes are classical examples in the rare-earth region
exhibiting the spontaneous shape transition in the ground state from
spherical to prolate-deformed shapes, with increasing the neutron
number from $N=82$ ($^{142}$Nd and $^{144}$Sm) to $N=92$
($^{152}$Nd and $^{144}$Sm) \cite{BM75}.
The experimental intrinsic quadrupole moment $Q_0$
is estimated from $B(E2; 2_1^+\rightarrow 0_\textrm{gs}^+)$ values,
assuming the strong-coupling rotor \cite{BM75,RS80}.
The self-consistent calculation with SkM$^*$ and
the pairing energy functional \cite{YSN09}
nicely reproduces these values for $N\geq 86$.
The development of nuclear deformation leads to a broadening
and peak splitting in the photoabsorption cross section.
It is the well-known deformation splitting associated with
two oscillation modes parallel to the symmetry axis
($K^\pi=0^-$) and perpendicular to that ($K^\pi=1^-$).

The calculation involves solving the eigenvalue problem of 
Eq.~(\ref{NHZ=ZOmegaN}) within the space truncated with the
two-quasiparticle energies $E_i+E_j \leq 60$ MeV.
The photoabsorption cross section is obtained from
the $E1$ transition strengths, according to Eq. (\ref{nF0=ZF}),
smeared with the Lorentzian width of 2 MeV.
This smearing width is the only free parameter in the calculation,
which accounts for the spreading effect
beyond the present QRPA treatment
(See Sec.~\ref{sec:PVC}).
It should be noted that, for light systems ($A\lesssim 40$),
the agreement is not as good as in heavy nuclei \cite{EKR10}.
This may suggest an insufficient surface symmetry energy
in current EDFs.

\begin{figure}[t]
\includegraphics[width=0.35\textwidth]{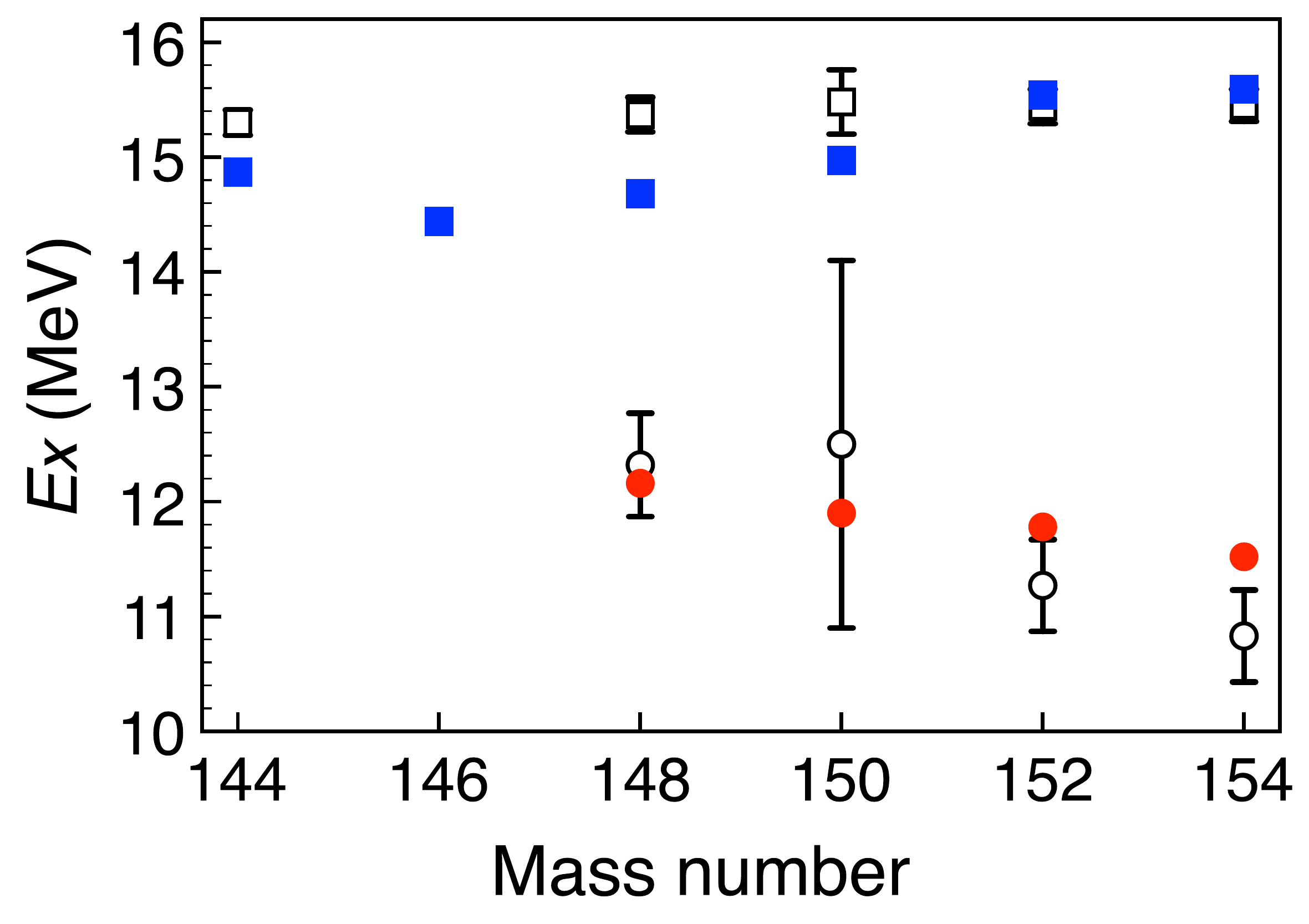}
\caption{
(Color online)
The excitation energies of the isoscalar giant monopole resonances in
the Sm isotopes;
calculated values (closed symbols) and
experimental data (open symbols) \cite{Ito03}.
From \textcite{YN13}.
}
\label{fig:Sm_GMR}
\end{figure}

The isoscalar and isovector giant monopole resonances ($L=0$) also show
the deformation splitting for $N\geq 86$,
which is consistent with the experimental data.
The excitation energies of the split peaks 
are shown in Fig.~\ref{fig:Sm_GMR} for Sm.
This splitting is due to the coupling to the $K^\pi=0^+$ component of
the giant quadrupole resonance ($L=2$).
The monopole and quadrupole are decoupled for spherical nuclei.
However, they are coupled in deformed nuclei,
the lower peak in Fig.~\ref{fig:Sm_GMR}
appears at the $K^\pi=0^+$ peak of the corresponding
giant quadrupole resonance.

The deformation of the momentum distribution (Fermi sphere)
plays an essential role in
the restoring force for the isoscalar giant resonances
\cite{RS80}.
A typical well-studied example is the giant quadrupole resonance whose energy
is approximately fit by 64$A^{-1/3}$ MeV.
The nuclear EDFs in the KS scheme nicely account for this
effect of the quantum Fermi liquid,
producing the correct mass number dependence.
For deformed systems, in addition to this, the deformation splitting
among $K^\pi=0^+$, $1^+$, and $2^+$ peaks is well reproduced.
The simple pairing-plus-quadrupole interaction produces
the $K$ splitting, $E_{K=2}-E_{K=0}$, of about 7 MeV for $^{154}$Sm.
This is too large and inconsistent with experiments \cite{Kis75}.
It is due to the violation of the nuclear self-consistency
between the shapes of the potential and the density distribution.
The calculation of the SkM$^*$ functional predicts the $K$ splitting
of 2.8 MeV \cite{YN13}.

Systematic calculations with Skyrme EDFs for spherical nuclei
have been performed using the canonical-basis QRPA \cite{TE06,PRNV03}.
The QRPA computer codes for deformed nuclei based on the
matrix diagonalization have been developed
for the Skyrme EDF \cite{TE10,YG08,Los10,YN11},
the Gogny EDF \cite{PG08,Per11},
and the covariant EDF \cite{AKR09}.
The calculations for deformed systems require large computational
resources for construction and storage of matrix ${\cal H}$ in
Eq. (\ref{H_and_N}).
Systematic calculations for a wide range of nuclei have been performed
by avoiding explicit calculations of ${\cal H}$,
with the finite amplitude method \cite{INY11,INY09-P2,NAEIY11-P},
and with the real-time method in Sec.~\ref{sec:Cb-TDHFB} \cite{ENI14,SL13-2}.

\subsubsection{Low-lying quadrupole states}

Low-lying states associated with the quadrupole vibrations have been
one of the major interests in nuclear structure problems.
Systematic analysis of the QRPA calculations for the first excited
$J^\pi=2^+$ states in spherical nuclei,
and for the gamma vibrations ($K^\pi=2^+$) in deformed rare-earth nuclei
have been performed by \textcite{TEB08,TE11} using the Skyrme EDFs.
They 
qualitatively agree with
the trend of experimental data for spherical nuclei.
Overall agreement of the QRPA results with experiments are better than
that of other approaches based on the generator coordinate method
\cite{SBBH07}.
However, the agreement is not quite as good for deformed nuclei.
The five-dimensional collective Hamiltonian for
the large amplitude quadrupole motion
may give a better description \cite{Ber07,Del10}.
The problems in the description of low-frequency quadrupole modes of
excitation will be
discussed in Sec.~\ref{sec:collective_submanifold}.

\subsubsection{Charge-exchange modes}

The isovector excitations have charge-changing ($\tau_\pm$) modes.
For spherical nuclei,
the calculations have been performed mostly with the Skyrme EDFs,
\cite{Eng99,BDEN02,FC05,PVKC07},
but also with the covariant EDFs \cite{PNVR04,LGM08,Niu13}.
The deformed QRPA calculations for the charge-exchange
modes have been performed with the separable approximation \cite{SME01,Sar12}.
Very recently, the full QRPA calculations
have become available too \cite{ME13,Yos13,MPG14}.
The neutrino-nucleus reaction was also studied including
inelastic neutral-current scattering \cite{DP12}.
The Gamow-Teller strength distribution ($S=1$, $T=1$)
significantly affects the $\beta$-decay half-lives and
the waiting point of the rapid neutron capture process (r-process).
To determine the r-process path far away from the stability line,
the reliable theoretical estimates are highly desired.

\subsubsection{Nuclear response in the continuum}

\begin{figure}[t]
\includegraphics[width=0.35\textwidth]{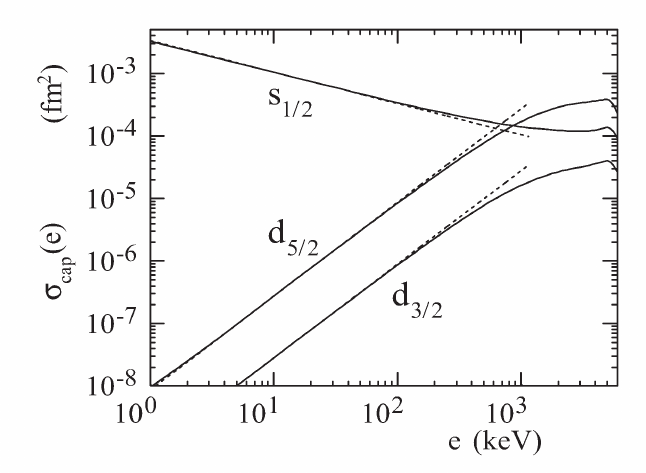}
\caption{
Neutron capture cross sections $n+^{141}$Sn as functions of
neutron energy $e$, for
incident neutrons in $s_{1/2}$, $d_{3/2}$, and $d_{5/2}$ states.
The Skyrme SLy4 EDF and the density-dependent pairing EDF is used.
The dotted lines indicate the power law scaling
$\propto e^{l-1/2}$.
From \textcite{Mat15}.
}
\label{fig:142Sn_n_capture}
\end{figure}

For nuclei near the neutron drip line, 
weakly bound neutrons may produce large transition strength just
above the threshold.
The examples were observed in low-energy electric dipole ($E1$) strength
in light halo nuclei, such as $^{11}$Be \cite{Nak94}
and $^{11}$Li \cite{Iek93,Shi95,Zin97,Nak06}.
The enhancement is not associated with the collectivity, but
due to the quantum mechanical ``threshold effect''.
Whether the {\it collective} low-energy dipole resonances 
exist in heavier neutron-rich nuclei is
still an open question \cite{HJ87,Ike92}.
In order to properly address these issues in which the continuum plays
an important role, the Green's function method
in Sec.~\ref{sec:Green's_function_method} is a powerful tool.
For doubly-closed spherical nuclei, we may neglect the pairing, and
the continuum RPA calculations have been extensively performed
to study a variety of strength functions (See \textcite{Sag01,PVKC07}
 and references therein).
However, for open-shell and heavier systems, we need to simultaneously
treat the deformation and the paring correlations.
This has not been achieved yet, however, partially done
with modern EDFs;
the Green's function method for deformed systems \cite{NY05}
and that for superconducting systems
 \cite{MMS09,SM09,DR11,Mat15}.

The photo-absorption of neutron-rich nuclei  leads to
neutron decays if the excitation energy exceeds the neutron separation
energy, which is very low in  nuclei near the neutron-drip line. 
It has been known that one can decompose the strength function
(the photo-absorption cross section) 
in the continuum RPA into partial strength functions for 
individual channels of particle escape \cite{ZS80,NY01}.
\textcite{Mat15} has recently extended the idea
to the continuum QRPA. The decomposition allows, using the reciprocity theorem
for the inverse processes, 
 to compute the cross section of the
direct neutron capture cross sections 
for different entrance channels separately.
Figure~\ref{fig:142Sn_n_capture} shows those for $n+^{141}$Sn,
calculated from the E1 strength functions in the continuum QRPA.
In this example, the cross section follows the power-law scaling rule.
This would not be the case
if there was a low-energy resonance.

\section{Real-time calculations beyond the linear regime}
\label{sec:real_time}

In nuclear physics, the real-time real-space calculations of
the TDDFT have been explored since 1970's, starting with
simplified energy functionals \cite{BKN76}.
It became the primary approach for studying low-energy heavy-ion collisions.
Since the Pauli blocking hinders the two-body collisions,
the method was thought to work well at low energy,
typically lower than the Fermi energy of about 40 MeV.
There is an excellent review paper on these developments in early years,
before 1982 \cite{Neg82}.
In recent years, we have observed important progresses in the
real-time calculations with respect to several aspects.

\leftline{1. \underline{\it Realistic EDF}}

In earlier works,
it was common to adopt simplified EDFs such that 
the spin-orbit term is neglected.
Recent calculations remove these restrictions and incorporate
the full EDF self-consistently.
The adopted EDFs for time-dependent calculations
have become as realistic as those for static calculations.
Most time-dependent calculations
beyond the linear regime
have been performed with Skyrme
energy functionals \cite{KOB97,UO06-1,Sim12,MRSU14}.
These changes produce even qualitative differences in nuclear
dynamics.
For instance, the famous fusion window anomaly was
significantly hindered by the inclusion of
the spin-orbit term in EDFs
\cite{USR86,RUDSL88}.
Extensive studies have been performed recently for studies of
nuclear dynamics, such as quasi-fission
\cite{OUS14,UOS15,SSL15},
charge equilibration \cite{IOMI10,IOMI10-2},
and high-spin rotation \cite{IMMI14}.

\leftline{2. \underline{\it TDBdGKS (TDHFB) scheme}}

Until very recently, the dynamical pairing correlations were
always neglected in the real-time calculations.
In the TDBdGKS scheme,
the number of quasiparticle orbitals is
identical to the dimension of the
single-particle model space we adopt.
Therefore, the real-time solution of the TDBdGKS (TDHFB)
equations requires extremely heavy computational tasks.

Applications of the full TDBdGKS scheme for realistic nuclear EDF
were performed with the spherical symmetry restriction \cite{ASC08}.
Later, it has been achieved with no assumption on the
spatial symmetry with the  Skyrme \cite{SBMR11,SBBMR15,Bul13} and
Gogny EDFs \cite{Has12}.
However, the applications are very limited at present, because of
its high computational demands and some problems inherent in the TDBdGKS
including the preparation of the initial state and treatment of
the non-vanishing wave functions at the boundary.
An approximate feasible approach is shown in Sec.~\ref{sec:Cb-TDHFB}.

\leftline{3. \underline{\it Nucleus-nucleus potential and friction parameters}}

To get insight into nuclear dynamics with energy dissipation,
several ideas have been proposed in late 1970's and 1980's
to extract ``macroscopic''
quantities, such as the nucleus-nucleus potential and the
friction parameter associated with the one-body dissipation
\cite{Koo77,BS81,CRSMG85}.
These ideas, which have been combined with realistic EDFs and
recent computational advances, lead to further developments
producing a number of new results in recent years.
Two different approaches will be presented in Sec.~\ref{sec:NN_pot_friction}.

\leftline{4. \underline{\it Transfer reaction and fluctuations}}

The particle-number projection method in a restricted coordinate space 
has been proposed to study the mass (charge) distribution in transfer reactions
\cite{Sim10}.
It is identical to the method based on the decomposition of the
Slater determinant proposed in \textcite{Koo77}, however,
the former has a significant computational advantage for heavier systems.
The recent calculations with realistic EDFs show qualitative agreements
with experiments.
See Sec.~\ref{sec:transfer}.

The TDDFT simulations for heavy-ion collision in early days
showed that, although the average values of one-body observables were well
reproduced, their fluctuations were underestimated.
Accordingly, for the transfer reaction, 
the calculated production rates are well reproduced in major
channels, however, not good in rare channels.
In order to overcome the difficulties, the fluctuation around the
TDDFT path is taken into account
(Sec.~\ref{sec:fluctuations}).

\subsection{Approximate schemes for TDBdGKS equations}
\label{sec:Cb-TDHFB}

Although the real-time calculation based on the full TDBdGKS equations
in the three-dimensional space becomes available for a few cases
\cite{SBMR11,Has12,SBBMR15}, it is still a very demanding task.
Thus, its approximate schemes
are useful at present.

The easiest and old one is
introduction of the fixed fractional occupation
numbers for KS orbitals.
For the stationary BCS (ground) state,
each orbital $\phi_i$ has a time-reversal-conjugate partner, $\phi_{\bar{i}}$,
and the occupation probability $\rho_i=|v_i|^2$.
Then, for the time evolution, we simply neglect the pair potential,
$\Delta_{ij}(t)=0$.
The TDKS equations for $N_c$ orbitals ($N_c>N$)
are solved in real time.
Thus, the pairing effect is taken into account only in the fractional
occupation which is completely determined at the preparation of the
initial state.
In this scheme,
the pair potentials play no role in the time evolution.

To include the dynamical pairing in a minimum way, we may keep the
diagonal form of the Hamiltonian, but with the pair potential
$\Delta_{i\bar{j}}(t)=-\Delta_i(t) \delta_{ij}$.
The quasiparticles are given by the canonical pair of orbitals
$\phi_i$ and $\phi_{\bar{i}}$ multiplied by
complex factors ($|u_i|^2 + |v_i|^2 = 1$).
\begin{equation*}
\Phi_i=\begin{pmatrix}
u_i\phi_i \\
-v_i^*\phi_{\bar{i}}^*
\end{pmatrix}
,\quad
\Phi_{\bar{i}}=\begin{pmatrix}
u_i\phi_{\bar{i}} \\
v_i^*\phi_i^*
\end{pmatrix}
.
\end{equation*}
Then, the TDBdGKS equations (\ref{TDBdGKS_eq})
are factorized into $2\times 2$ form.
Using the relation $\Delta_{i\bar{i}} \phi^*_{\bar{i}} = -\Delta_i \phi_i$,
the TDBdGKS equations are split into
\begin{eqnarray}
i\frac{\partial}{\partial t} \phi_i(t)
 &=& \left\{ h_s[\rho(t)] - \mu - \eta_i(t) \right\} \phi_i(t)
,\quad i\leftrightarrow \bar{i}, \nonumber \\
i\frac{d}{dt} \rho_i(t)&=&\kappa_i(t)\Delta_i^*(t) - \kappa_i^*(t) \Delta_i(t)
, 
\label{CbTDHFB}
\\
i\frac{d}{dt} \kappa_i(t)&=&\left\{\eta_i(t)+\eta_{\bar{i}}(t)\right\}
 \kappa_i(t) +\Delta_i(t) \left\{ 2\rho_i(t) -1 \right\}  , \nonumber
\end{eqnarray}
where $\rho_i(t)\equiv |v_i(t)|^2$,
$\kappa_i(t)\equiv u_i(t)v_i(t)$,
and $\eta_i(t)$ are parameters to control the phase of
the canonical orbitals $\phi_i(t)$.
The $\eta_i(t)$ are arbitrary, if the diagonal form of the pair potential
is consistent with the gauge invariant EDFs.
When it is violated in practice,
a choice of the minimal phase change was proposed \cite{Eba10}.

When the pair potential is calculated from
the anti-symmetrized two-body interaction $\bar{v}$,
$\Delta_i(t)=-\sum_{j>0} \kappa_j(t) \bar{v}_{i\bar{i},j\bar{j}}$.
The densities are constructed as
\begin{equation*}
\begin{split}
\rho(\alpha\beta,t)&=\sum_i \rho_i(t)
 \phi_i(\alpha;t)\phi_i^*(\beta;t) , \\
\kappa(\alpha\beta,t)&=\sum_{i>0} \kappa_i(t) \left\{
 \phi_i(\alpha;t)\phi_{\bar{i}}(\beta;t) -
 \phi_{\bar{i}}(\alpha;t)\phi_i(\beta;t) \right\} .
\end{split}
\end{equation*}

Equations similar to Eq. (\ref{CbTDHFB}) were derived using the time-dependent
variational principle some time ago \cite{BF76}
and revisited in terms of the TDBdGKS equations \cite{Eba10}.
The conservation of the average particle number is guaranteed for
arbitrary choice of $\mu$, however, the energy conservation depends on
the choice of the parameter $\eta_i(t)$ \cite{Eba10},
and the current conservation is violated in this approximation
\cite{SLBW12}.
The equations may describe dynamical pairing effects, coupled to
motion of the canonical orbitals.
The method has been applied to real-time calculations for
linear response \cite{Eba10,ENI14,SL13-2},
neutron transfer reactions \cite{SL13-1},
and fusion/fission reactions \cite{EN14-P,EN15-P,SSL15}.

\subsection{Heavy-ion collision: Nucleus-nucleus potential
 and one-body dissipation}
\label{sec:NN_pot_friction}

In real-time calculation of heavy-ion collision,
so far,
the TDKS equations with $\kappa=\Delta=0$ are solved in most applications.
The initial state is prepared as two
nuclei in their ground states, placed well separated in space.
First, we locate the two nuclei, $(N_L,Z_L)$ in the left and $(N_R,Z_R)$ in 
the right, with respect to the $z$ coordinate.
In this initial state,
each KS orbital $\ket{\phi_i}$ belongs to either
``left'' or ``right'', and those in the left nucleus are boosted toward
right by $\ket{\phi_i(t=0)} = e^{ik_L z} \ket{\phi_i}$, while
those in the right by $\ket{\phi_i(t=0)} = e^{-ik_R z} \ket{\phi_i}$.
Then, the time evolution of KS orbitals is computed
to obtain the density
$\rho(\alpha\beta;t)=\sum_i \rho_i(t) \phi_i(\alpha;t) \phi_i^*(\beta;t)$.

Recently,
there are a number of works
to extract the nucleus-nucleus potential and the friction
from non-empirical TDDFT calculations.
To achieve this, we should divide the total system into two parts,
one associated with a small number of collective degrees of freedom,
and the rest of the Hilbert space called ``intrinsic'' space.
To our understanding, so far,
this division is guided by ``a priori'' assumptions,
not by the TDDFT dynamics itself.

\subsubsection{Density-constraint calculation}
\label{sec:density_constrained}

Among many kinds of densities,
for the colliding nuclei under consideration,
the normal density distribution $\rho(\vec{r};t)$ and
the current density, 
$\vec{j}(\vec{r};t) = 1/(2i)
(\vec{\nabla}-\vec{\nabla'} ) \rho(\vec{r},\vec{r'};t)|_{\vec{r}=\vec{r'}}$,
are regarded as
quantities associated with collective motion.
Then, the collective energy associated with the collisional motion
is assumed to be a functional of $\rho(\vec{r})$ and $\vec{j}(\vec{r})$,
which is defined as the minimization with constraints on the density and
the current.
\begin{equation}
E_\textrm{coll}\left[\rho(\vec{r},t),\vec{j}(\vec{r},t)\right] =
 \min_{\rho\rightarrow (\rho(\vec{r}),\vec{j}(\vec{r}))}
F[\rho] - E_L - E_R,
\label{E_coll}
\end{equation}
where $E_L$ and $E_R$ are the ground-state energies of two nuclei.
For the initial state with two nuclei far apart ($t=0$),
this approximately corresponds to the sum of
the kinetic energy of center of mass of each nucleus,
$P_L^2/(2A_L m)+P_R^2/(2A_R m)$,
and the Coulomb energy between the two,
$Z_L Z_R e^2/|\vec{R}_L-\vec{R}_R|$.
Since the total energy is conserved during the time evolution,
we have $E_\textrm{total}\approx E_L + E_R + E_\textrm{coll}(t=0)$.

The TDDFT simulation of the heavy-ion collision produces
the time-dependent density $\rho(\vec{r},t)$ and current $\vec{j}(\vec{r},t)$.
From these,
the intrinsic excitation energy during the collision is given by
\begin{equation}
E^*(t) = E_\textrm{total}
-E_\textrm{coll}\left[\rho(\vec{r};t),\vec{j}(\vec{r};t)\right] -E_L-E_R .
\label{E*}
\end{equation}
Furthermore, the collective energy is divided into two;
$E_\textrm{coll}[\rho(\vec{r}),\vec{j}(\vec{r})]=
E_\textrm{kin}[\rho(\vec{r}),\vec{j}(\vec{r})] +
V_\textrm{pot}[\rho(\vec{r})]$
and the nucleus-nucleus potential is defined by the latter, obtained by
minimization with a constraint on $\rho(\vec{r})$.
\begin{equation}
V_\textrm{pot}\left[\rho(\vec{r})\right] = \min_{\rho\rightarrow \rho(\vec{r})}
F[\rho] - E_L -E_R .
\label{V_pot}
\end{equation}
This minimization automatically produces $\vec{j}(\vec{r})=0$
for even-even nuclei.
In practice, since the density and current constraint calculation
of Eq. (\ref{E_coll}) is computationally demanding,
the density constraint calculation of Eq. (\ref{V_pot})
is performed.
Then, the collective kinetic energy is assumed to be
\begin{equation}
E_\textrm{kin}\left[ \rho(\vec{r}),\vec{j}(\vec{r}) \right]
= \frac{1}{2m} \int \frac{|\vec{j}(\vec{r})|^2}{\rho(\vec{r})} d\vec{r}  .
\label{E_kin}
\end{equation}
So far, all the quantities are calculated as functions of time $t$.
A possible mapping from $t$ to a collective coordinate $R(t)$ is given
in Sec.~\ref{sec:1D_mapping}.

The idea and computational algorithm of this method are proposed in
\textcite{CRSMG85}.
Extensive studies have been performed in recent years by
Oberacker, Umar, and coworkers
\cite{UO06-2,UO06-3,UO07,UO08,UOMR09,OUMR10,UMIO10,UOH12,UOMR12,OU13,SKUO13}.
The TDDFT naturally provides dynamical change of the
nuclear structure during collisions.
Therefore, the potential $V_\textrm{pot}$ in Eq. (\ref{V_pot})
contains such polarization effects.
However, the separation between the collective energy (\ref{E_coll}) and
the dissipation energy (\ref{E*}) is less reliable
when two nuclei are significantly overlapped.
Even without any dissipation,
the current density $\vec{j}(\vec{r})$ is
reduced in the overlapping region
because two nuclei are moving to opposite directions.
This leads to the reduction of $E_\textrm{coll}$ ($E_\textrm{kin}$)
and to overestimation of $E^*$.

\subsubsection{Mapping to one-dimensional
Hamilton equations of motion}
\label{sec:1D_mapping}

Another even simpler method is based on the explicit introduction of the
one-dimensional (1D) collective coordinate and momentum.
We recapitulate here the method presented in \textcite{WL08} to extract the
nucleus-nucleus potential $V_\textrm{pot}(R)$ and friction parameter
$\gamma(R)$.
Similar methods are proposed earlier \cite{Koo77,BS81}.
We introduce the relative distance between two nuclei,
$R(t)$, calculated as the distance between two centers of mass in
the left and the right.
Assuming the head-on collision on the $z$-axis,
$R(t)=(1/A_R) \int_R z \rho(\vec{r},t) d\vec{r} 
-(1/A_L) \int_L z \rho(\vec{r},t) d\vec{r}$.
The momentum $P(t)$ is calculated as
$P(t)=(A_L \int_R j_z(\vec{r},t) d\vec{r} 
-A_R \int_L j_z(\vec{r},t) d\vec{r})/(A_L+A_R)$.
Here, the integration $\int_{R(L)}d\vec{r}$  are defined by
$\int_{R(L)} d\vec{r} f(\vec{r}) = \int d\vec{r} f(\vec{r})
 \theta(\pm (z-z_0))$.
The $z=z_0$ plane can be chosen, for instance, as the plane of the lowest
density (neck position).
The TDDFT calculation produces $R(t)$ and $P(t)$ as functions of time,
which are assumed to obey the 1D classical Hamilton equation of motion:
\begin{equation}
\frac{dR}{dt}=\frac{P}{\mu(R)} , \quad
\frac{dP}{dt}=-\frac{dV_\textrm{pot}}{dR} - \gamma(R) \frac{dR}{dt} ,
\label{1_dim_classical_eq}
\end{equation}
where the first equation provides the definition of the reduced mass $\mu(R)$.
There are two unknown quantities remaining,
the force $dV_\textrm{pot}/{dR}$ and
the friction parameter $\gamma(R)$.
Assuming weak energy dependence of these quantities, we can estimate
these by performing the TDDFT simulation with two slightly different
initial energies.
Note that, because of the head-on assumption,
the parameter $\gamma(R)$ may represent only the radial friction,
not the tangential one.

Since the density-constraint calculation at different $R(t)$
is not necessary in this approach,
it is computationally easier than the previous one.
Similarly to the density constrained calculation,
the calculated relative momentum decreases after two nuclei touch,
even if no dissipation takes place.
In addition,
the assumption, that $R$ and $P$ are canonical conjugate variables,
becomes questionable as well.

\subsection{Heavy-ion collision: Transfer reaction}
\label{sec:transfer}

\subsubsection{Number projection}

The mass number distribution after the collision was estimated
for a schematic EDF \cite{Koo77}.
It is based on the decomposition of the single Slater determinant 
in a restricted space, and has been used for electron transfer processes
in atomic collisions \cite{LD83,NYTA00}.
Recently, an alternative expression has been given using the
particle number projection \cite{Sim10}.
They are identical in principle, however,
the latter has a computational advantage over the previous expression.

Let us divide the space $V$ into two regions; one is $V_f$ and the rest
$V_{\bar f}=V-V_f$.
The particle number in the space $V_f$, $N_f$, is defined by
$\hat{N}_f=\int_f \hat{\psi}^\dagger(\vec{r}\sigma) \hat{\psi}(\vec{r}\sigma)
d\vec{r}$.
The particle number projection in the right space, $\hat{P}_f(N)$, is given
by
\begin{equation*}
\hat{P}_f(N)=\frac{1}{2\pi} \int_0^{2\pi} d\theta e^{i\theta (N-\hat{N}_f)}
\end{equation*}
Let us define the matrix $B_{ij}(\theta)$ as
$
B_{ij}(\theta) \equiv
 \inproduct{\phi_i}{\phi_j}_{\bar f} +
 e^{-i\theta} \inproduct{\phi_i}{\phi_j}_f
$,
with the overlap in the spaces $V_f$ and $V_{\bar f}$ given by
\begin{equation*}
\inproduct{\phi_i}{\phi_j}_{f({\bar f})}\equiv \sum_\sigma \int_{f({\bar f})}
\phi_i^*(\vec{r}\sigma) \phi_j(\vec{r}\sigma)
d\vec{r} .
\end{equation*}
The probability that the $N$ particles are present in $V_f$
is given by
\begin{equation}
P_N = 
\frac{1}{2\pi} \int_0^{2\pi} d\theta e^{i\theta N}
\det B(\theta)
\label{P_N}
\end{equation}

In the real-time simulation,
after the two nuclei collide and separate
again, we specify the region $V_f$ where one of the nucleus is located.
Then, the mass number distribution is calculated according to Eq. (\ref{P_N}).
The production cross section of the nucleus with $N$ particles
is estimated by repeating the same calculation
with different impact parameter $b$.
\begin{equation}
\sigma(N)=2\pi \int db b P_N(b) .
\label{sigma_N_transfer}
\end{equation}
This is most useful for the calculations of transfer reaction cross section
$\sigma_{\pm N, \pm Z}$.
When the pair potential is present, the number projection is required
for the initial state too.

The expectation value of the operator in each reaction
product can also be evaluated with the present technique \cite{SY14,SY15,Son15}.
For instance, the one-body operator local in the coordinate,
$\hat{O}= \sum_\sigma \int O(\vec{r}) \hat{\psi}^\dagger (\vec{r}\sigma)
 \hat{\psi}(\vec{r}\sigma)d\vec{r}$,
is given by
\begin{equation*}
\begin{split}
O_N=&\frac{1}{2\pi P_N} \int_0^{2\pi} d\theta e^{i\theta N} \det B(\theta) \\
&\sum_i \left( \bra{\phi_i} O \ket{\phi_i}_L +
e^{-i\theta} \bra{\phi_i} O \ket{\phi_i}_R \right)
\end{split}
\end{equation*}

\subsubsection{Fluctuations}
\label{sec:fluctuations}

The TDDFT provides feasible approaches to nuclear collective dynamics
in a large-amplitude nature,
and has been successful to describe mean values of one-body observables.
However, it has been known for some time that
 it underestimates
fluctuations \cite{Koo77,Neg82}.
As far as we calculate the one-body observables according to the
KS orbitals,
a severe limitation comes from mainly two sources:
One is missing effect of two-body collisions.
The inclusion of the nucleon-nucleon collision is treated by
a stochastic approaches \cite{Aic91} or by explicit inclusion of
two-body correlations \cite{WC85}.
Although the two-body collision becomes less important at lower energy,
there is another well-known limitation, which we address here.
The TDDFT is
described by a single time-dependent mean-field (KS) potential.
The collision of nuclei, in general, leads to superposition of different
final states, $\ket{\Phi^{(1)}}$, $\ket{\Phi^{(2)}}$,  $\cdots$,
for which different mean fields should exist,
$v_s^{(1)}(\vec{r})$, $v_s^{(2)}(\vec{r})$, $\cdots$.
Since these dynamics in multi-channels are
described by a single average mean field $v_s^{\textrm{(av)}}$, 
the TDDFT naturally hinders the fluctuation.
This may be crucial at low energies, in which
one-body dynamics are supposed to play a dominant role
\cite{IYY86}.

A practical way to improve the situation is given by
replacing the quantum fluctuation by
classical statistical ensemble in the initial state.
Each state is evolved in time with its own mean field.
This is often called ``stochastic mean field theory'' \cite{Ayi08,LA14}.

The quantum fluctuation at the initial state is estimated by
the fluctuation of one-body operator $\hat{A}$ in a Slater determinant
\begin{equation*}
\sigma_A^2 \equiv \langle \hat{A}^2 \rangle - \langle \hat{A} \rangle^2 \\
= \sum_{ij} \left|\bra{\phi_i}\hat{A}\ket{\phi_j}\right|^2 \rho_i (1-\rho_j) .
\end{equation*}
For normal systems at zero temperature, the occupation is integer number,
$\rho_i=0$ or 1.
In order to describe this quantum fluctuation
by the classical statistical average
as $\langle \hat{A} \rangle = \textrm{tr}[\overline{\rho^{(n)}A}]$ and
$\sigma_A^2 = \overline{(\textrm{tr}[\delta\rho^{(n)}A])^2}$,
we use random Gaussian numbers for one-body density,
$\rho^{(n)}=\overline{\rho^{(n)}}+\delta\rho^{(n)}$,
which satisfies the ensemble average values
\begin{eqnarray}
\label{bar_rho}
\overline{\rho^{(n)}_{ij}} &=& \rho_i \delta_{ij} ,\\
\label{bar delta_rho_2}
\overline{\delta\rho^{(n)}_{ij} \delta\rho^{(n)}_{kl}} &=&
 \frac{1}{2} \delta_{il}\delta_{jk}
 \left\{ \rho_i (1-\rho_j)+\rho_j(1-\rho_i)\right\} .
\end{eqnarray}
Starting from each initial configuration $\rho^{(n)}$,
$\rho^{(n)}(t)$ evolves in time
following the TDKS equation, Eq. (\ref{TDKS_rho_eq}),
with the density given by
\begin{equation*}
\rho^{(n)}(t)=
\sum_{ij} \ket{\phi_i^{(n)}(t)}\rho^{(n)}_{ij}\bra{\phi_j^{(n)}(t)} .
\end{equation*}
Since the off-diagonal elements of $\rho^{(n)}_{ij}$ are non-zero,
we need to solve the time evolution of not only the hole states,
but also the particle states.

For calculations of 
small fluctuations around the TDKS trajectory
in the observable $\hat{A}$ at $t=t_f$,
instead of performing the forward time evolution of $\delta\rho(t)$
with the initial fluctuation of Eqs. (\ref{bar_rho}) and
(\ref{bar delta_rho_2}), we may utilize a backward time evolution.
The time evolution of $\delta\rho(t)$ is described by a unitary operator $U(t)$
as $\rho(t)=U(t)\rho(t=0)U^\dagger(t)$ in general.
Note that the linear approximation with respect to $\delta\rho^{(n)}$
leads to the operator $U(t)$ independent of the event label $(n)$,
Thus, the fluctuating part of the observable $\hat{A}$ can be written as
\begin{equation*}
\delta A^{(n)}(t_f)
= \textrm{tr}\left[ \hat{A} \delta\rho^{(n)}(t_f) \right]
= \textrm{tr}\left[ \hat{B} \delta\rho^{(n)}(0) \right] ,
\end{equation*}
where $t_f$ represents the final time when the observation is made and
$t=0$ is the initial time.
Here, the one-body Hermitian operator $\hat{B}$ is given by
$B_{ij}=\left\{U^\dagger(t_f) \hat{A} U(t_f) \right\}_{ij}$.
The fluctuation of $\hat{A}$ at $t=t_f$ is now given by the ensemble average
at $t=0$.
\begin{eqnarray}
\sigma_A^2 &=& \overline{\left\{\delta A^{(n)}(t_f)\right\}^2}
= \sum_{ijkl} B_{ij} B_{kl} 
\overline{\delta\rho^{(n)}_{ji}\delta\rho^{(n)}_{lk}} \\
&=& \sum_{ij} |B_{ij}|^2 \rho_i (1-\rho_j)
\label{SMF_A}
\end{eqnarray}

In fact, Eq. (\ref{SMF_A}) is the same as the one previously derived
with the variational approach by Balian-V\'en\'eroni \cite{BV85}.
It is easy to see that Eq. (\ref{SMF_A}) can be alternatively written 
as $-\textrm{tr}[[B,\rho(0)]^2]/2$ with $\rho_{ij}(0)=\rho_i\delta_{ij}$.
Thus, modifying the TDDFT density $\rho(t)$ at $t=t_f$ as
$\eta_\epsilon(t_f)\equiv e^{i\epsilon \hat{A}} \rho(t_f)
e^{-i\epsilon \hat{A}}$, the backward evolution of $\eta_\epsilon(t)$
up to $t=0$ gives the following expression
\begin{equation}
\sigma_A^2=\lim_{\epsilon\rightarrow 0}
\frac{1}{2\epsilon^2} \textrm{tr}\left[ \left\{
U^\dagger(t_f) \eta_\epsilon(t_f)  U(t_f) - \rho(0) \right\}^2 \right] .
\label{BV_A}
\end{equation}
This is useful for practical calculations \cite{Sim11,Sim12}.
The KS wave functions $\ket{\phi_i(t_f)}$ are modified
to $e^{i\epsilon \hat{A}} \phi_i(t_f)$ with small $\epsilon$,
then, calculate the backward time evolution to $t=0$.
This will provide $U^\dagger(t_f) \eta_\epsilon(t_f) U(t_f)$.
Several different values of $\epsilon$ may be enough
to identify its quadratic dependence.
More details and derivation can be found in \textcite{Sim12}.

\subsection{Illustrative examples}
\label{sec:real_time_appli}

In this section, we present some examples of recent calculations
in studies of nuclear collision dynamics.
The full TDBdGKS calculation of collision dynamics has not been
achieved, but is under progress \cite{SBBMR15}.
Most of recent calculations beyond the linear regime
have been performed
based on the TDKS equations with the Skyrme EDFs.

\begin{figure}[t]
\includegraphics[width=0.3\textwidth]{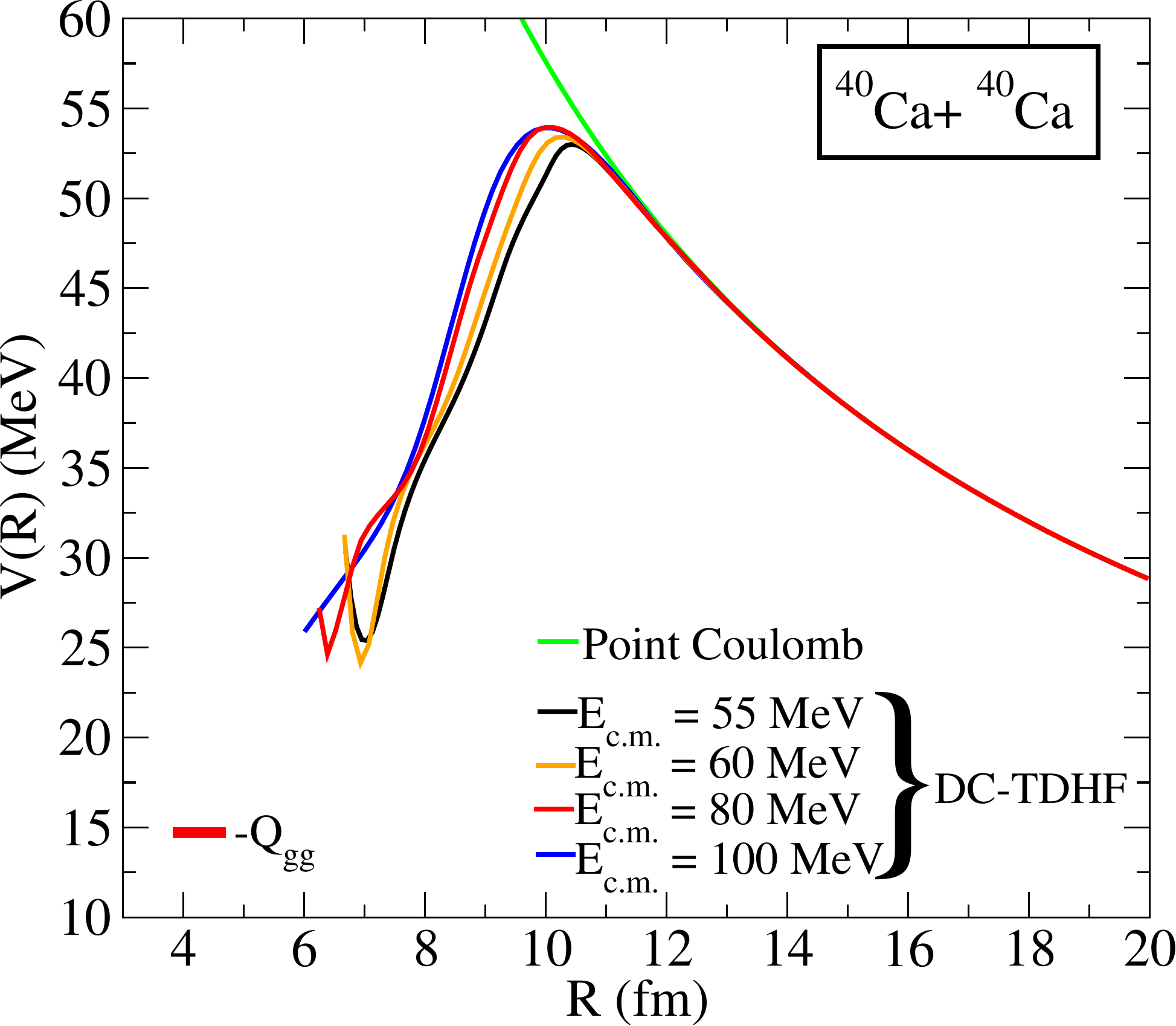}
\caption{
(Color online)
Internucleus 
potential for $^{40}$Ca$+^{40}$Ca for
different energies.
The horizontal axis is the distance between two $^{40}$Ca nuclei.
The 
green
line indicates the Coulomb potential assuming the point charge
$20e$ at the center of each nucleus.
``DC-TDHF'' is an abbreviation of the density-constrained TDHF.
From \textcite{UOMR09}.
}
\label{fig:40Ca_pot}
\end{figure}

\subsubsection{
Internucleus 
potential and precompound excitation}

Extensive studies using the real-time simulation have been performed
for microscopic derivation of the nucleus-nucleus potential and
dissipation energy at initial stages of the nuclear fusion.
This can be done with the density-constraint calculation
shown in Sec.~\ref{sec:NN_pot_friction}.
The real-time simulation for the fusion reaction
produces the time evolution of the density $\rho(\vec{r},t)$,
the current $\vec{j}(\vec{r},t)$, etc.
At the beginning, 
the total energy is given by
$E_\textrm{total}=E_\textrm{coll}(t=0)+E_L+E_R$.
After the two nuclei touch, $E_\textrm{total}$ is also
shared by the intrinsic excitation energy $E^*$.

\begin{figure}[t]
\includegraphics[width=0.3\textwidth]{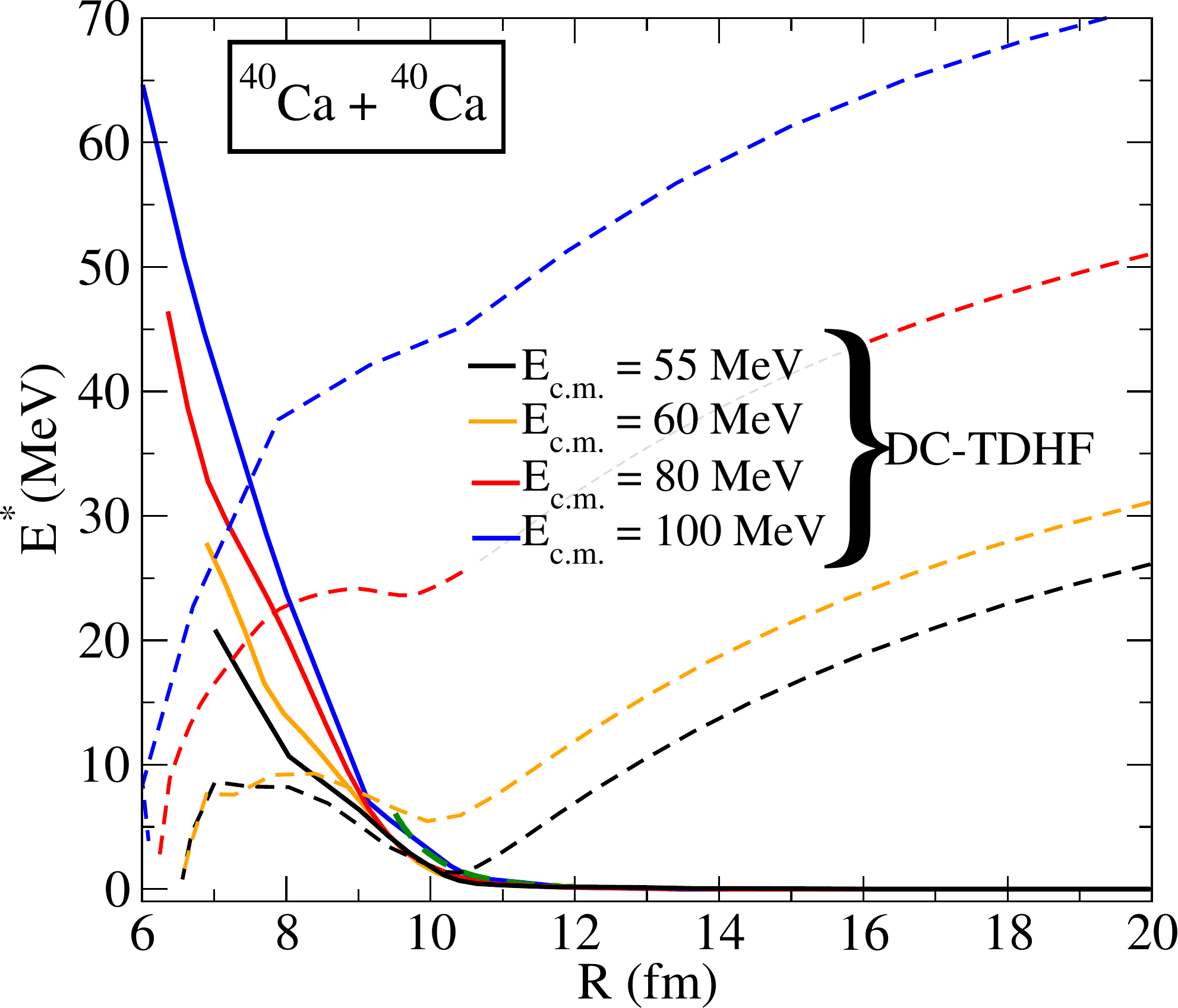}
\caption{
(Color online)
Excitation energy $E^*$ (solid lines) in Eq. (\ref{E*})
for $^{40}$Ca$+^{40}$Ca collision at
different energies.
The dashed lines indicate the collective kinetic energy $E_\textrm{kin}$
in Eq. (\ref{E_kin}).
From \textcite{UOMR09}.
}
\label{fig:40Ca_E*}
\end{figure}

According to Eq. (\ref{V_pot}), (\ref{E*}) and (\ref{E_kin}),
\textcite{UOMR09} estimated
the nucleus-nucleus potential $V(R)$ and the intrinsic excitation $E^*(R)$
for $^{40}$Ca$+^{40}$Ca.
These are illustrated in
Figs.~\ref{fig:40Ca_pot} and \ref{fig:40Ca_E*}.
The amount of the dissipative energy $E^*$
is roughly proportional to the bombarding energy
$E_\textrm{cm}=E_\textrm{coll}(t=0)$,
while the potential $V_\textrm{pot}$ is approximately independent
of the choice of $E_\textrm{cm}$.
The excitation energy for the fused system $^{80}$Zr is expected to be
$E_\textrm{total}-E_\textrm{g.s.}(^{80}\textrm{Zr})$ at the end.
The calculated $E^*$ at the capture point near $R=6$ is still lower than
this value by about 20 MeV.
It is confirmed that this 20 MeV is due to the difference in the density
distribution between at the ground state of $^{80}$Zr and at the
capture point.

The
internucleus 
potential obtained from the mapping to the
one-dimensional (1D) classical equation of motion (\ref{1_dim_classical_eq})
seems to be similar to the one of the density-constrained
calculation for some light systems \cite{WL08}.
However, in heavier systems where the dissipation becomes more
relevant, they may produce different potentials.
In fact, for the heavy systems with $Z_L Z_R \gtrsim 1600$, it is known that
the fusion probability is significantly hindered.
An example is given by the fact that the fusion cross section of
$^{96}$Zr$+^{124}$Sn ($Z_L Z_R=2000$) is much smaller than
that of $^{40}$Ar$+^{180}$Hf ($Z_L Z_R=1296$), both leading to
the same fused system, $^{220}$Th \cite{Sah85}.
This was supposed to be due to the strong energy dissipation inside
the Coulomb barrier \cite{Swi82}.
The quasi-fission before the formation of a compound nucleus
may play a primary role in the fusion hindrance.
Although the TDDFT cannot fully take into account the collisional
damping, it reproduces some features of the quasi-fission process
\cite{Sim12,OUS14}.

\begin{figure}[t]
\includegraphics[clip,width=0.3\textwidth]{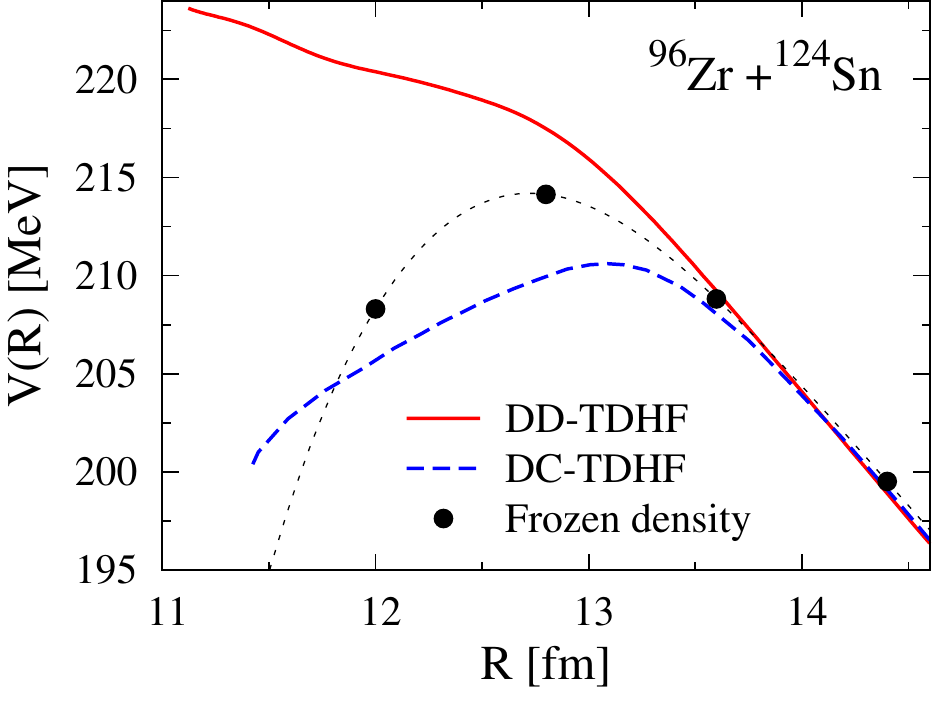}
\caption{
(Color online)
Internucleus 
potential for $^{96}$Zr$+^{124}$Sn calculated at
$E_\textrm{cm}=230$ MeV.
The red solid curve is based on Eq. (\ref{1_dim_classical_eq}),
while the blue dashed line is the one of the density-constrained
calculation of Eq. (\ref{V_pot}).
The frozen density potential is plotted by the filled circles.
``DD-TDHF'' is an abbreviation of the dissipative-dynamics TDHF.
From \textcite{Was15}.
}
\label{fig:96Zr_124Sn_pot}
\end{figure}

Figure~\ref{fig:96Zr_124Sn_pot}
shows the calculated potential for $^{96}$Zr$+^{124}$Sn.
The potential of the density-constrained calculation shows a maximum
around $R=13.1$ fm and decreases at $R<13$ fm.
This is very different from the one obtained by mapping to the 1D
classical equations, which keeps rising even at $R<13$ fm.
This must be attributed to the difference in the decomposition
of the total energy
into $V(R)$, $E_\textrm{kin}(R)$, and $E^*(R)$.
Since we can expect the $E_\textrm{kin}$ in these two methods are 
rather similar, the intrinsic excitation $E^*(R)$ should compensate
the difference in $V(R)$.
The relation between the two methods in \ref{sec:density_constrained} and
\ref{sec:1D_mapping} is not clear at present.
The further studies are desired to clarify the microscopic origin
of the fusion hindrance \cite{Sim12,GN12-P}.
It is also related to a conceptual question:
{\it
What are the collective variables, the potential, and the inertial mass
for proper description of
many kinds of nuclear reaction?%
}
This is the main subject of Sec.~\ref{sec:collective_submanifold}.

\subsubsection{Multi-nucleon transfer reaction}

Another example of low-energy nuclear reaction is the multi-nucleon
transfer reaction for heavy-ion collisions.
At energies near the Coulomb barrier, this reaction involves many kinds of
quantum non-equilibrium many-body dynamics,
such as shell effects, neck formation, and tunneling.
The {\sc grazing} model \cite{Win94}
is frequently used to describe the multi-nucleon transfer reaction.
This model is based on statistical treatment of the single-particle
transfer processes
and a semi-classical formulation of coupled-channel method.
The TDKS (TDHF) simulation may provide an alternative microscopic
approach to the low-energy transfer reaction and
help our fundamental understanding of the quantum dynamics.

\begin{figure}[t]
\includegraphics[width=0.5\textwidth]{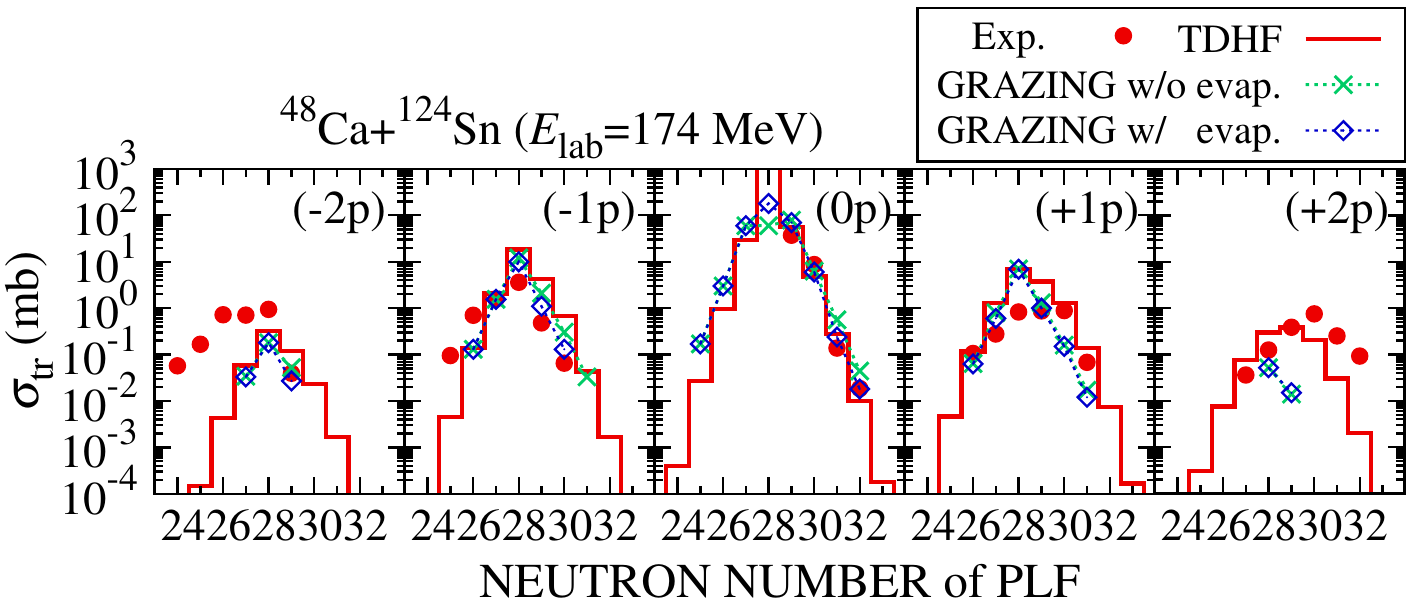}
\caption{
(Color online)
Transfer cross sections for $^{48}$Ca$+^{124}$Sn reaction at
$E_\textrm{lab}=174$ MeV.
The red solid lines show the TDDFT results with the particle-number
projection.
The experimental data and the {\sc grazing} results are taken from
\textcite{Cor97}.
See text for details.
From \textcite{SY13}.
}
\label{fig:48ca_124Sn_transfer}
\end{figure}

After the real-time simulation at the impact parameter $b$,
the transfer probability $P_{N,Z}(b)$
for each channel of $(N,Z)$ can be calculated
according to Eq. (\ref{P_N}).
Repeating the calculation with different values of $b$,
the cross section $\sigma_{N,Z}$
is calculated as Eq. (\ref{sigma_N_transfer}).
An example for the $^{48}$Ca$+^{124}$Sn reaction
is presented  in Fig.~\ref{fig:48ca_124Sn_transfer},
showing the production cross sections
of Ar ($-2p$), K ($-1p$), Ca ($0p$), Sc ($+1p$), Ti  ($+2p$) isotopes.
The horizontal axis corresponds to the neutron number of fragments.
In the major channels of $0p$ and $\pm 1p$, the experimental data are
well reproduced.
The calculated mass distribution is rather symmetric with respect to
the neutron number around $N=28$.
The experimental data seem to suggest that this symmetry is broken
for the $\pm 2p$ channels.
In general, the discrepancy becomes more prominent for rarer channels
with large number of exchanged nucleons \cite{SY13}.
Nevertheless, the quality of the agreement is the same as the
{\sc grazing} calculation.
It should be noted that the simulation was carried out using
the Skyrme SLy5 EDF and there were no free parameters.

\begin{figure}[t]
\includegraphics[width=0.35\textwidth]{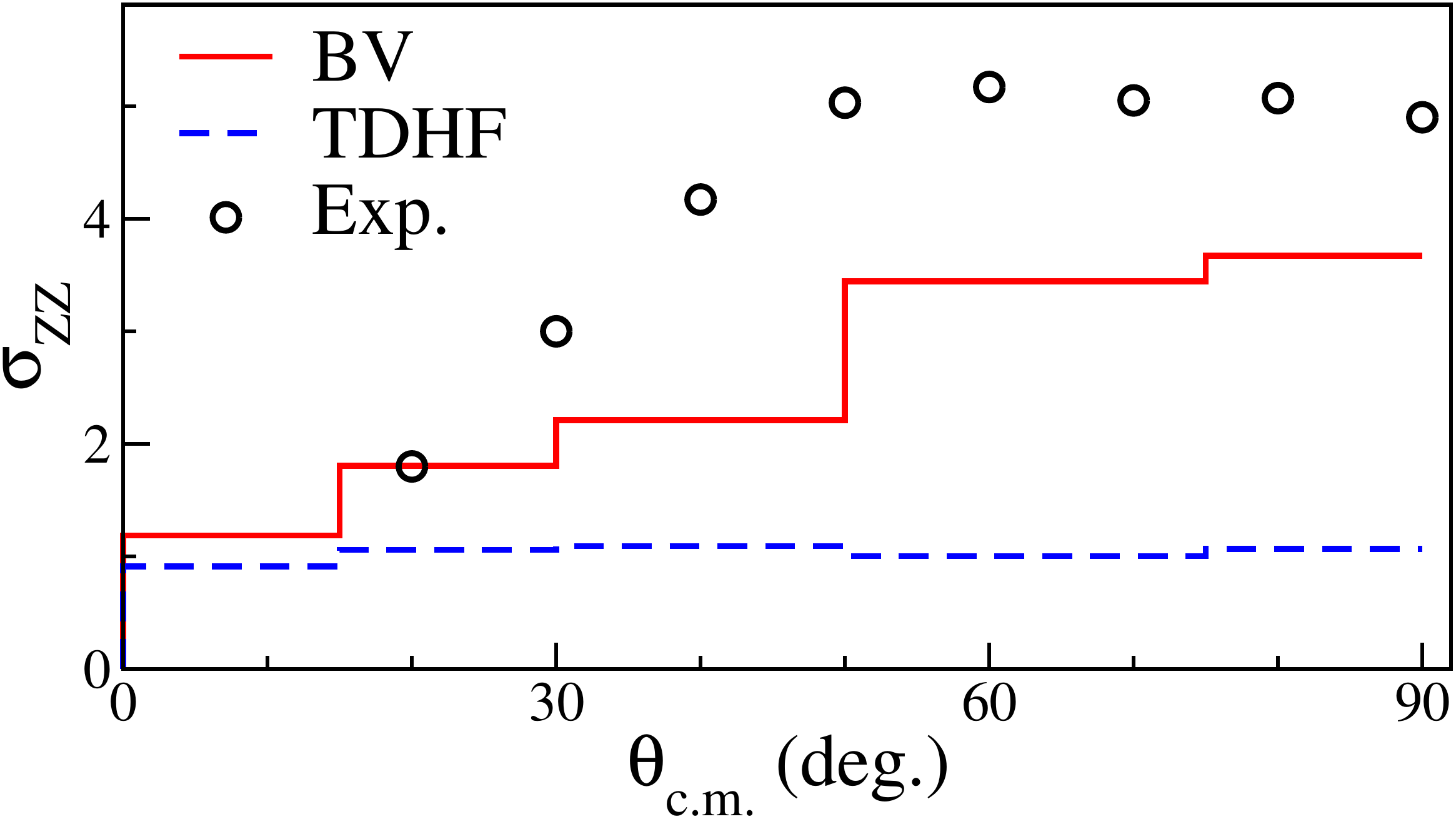}
\caption{
(Color online)
The fluctuation in the proton number distribution,
$\sigma_{ZZ}\equiv (\< Z^2 \> - \<Z\>^2)^{1/2}$,
for final
fragments in the $^{40}$Ca$+^{40}$Ca reaction.
The horizontal axis corresponds to the scattering angle that is
determined by the impact parameter in the TDDFT calculation.
The circles show experimental data \cite{Roy77}.
From \textcite{Sim11}.
}
\label{fig:40Ca_sigma_Z}
\end{figure}

As we have seen in Fig.~\ref{fig:48ca_124Sn_transfer},
in the $\pm 2p$ channels,
the neutrons tend to
move together with the protons,
which is not reproduced in the calculation.
This is due to the fact that the TDDFT calculation does not have
correlations between neutron and proton distributions,
namely, $P_{N,Z}=P_N P_Z$.
This missing correlation and fluctuation has been studied
by \textcite{Sim11} for $^{40}$Ca$+^{40}$Ca at $E_\textrm{cm}=128$ MeV,
using the Balian-V\'en\'eroni formula analogous to Eq. (\ref{BV_A}).
For small impact parameter $b$,
he has found strong correlation between proton and neutron distributions.
In addition, the fluctuation of the proton distribution is compared with the
available experiment in Fig.~\ref{fig:40Ca_sigma_Z}.
The conventional TDDFT simulation significantly underestimates the fluctuation.
It is enhanced by the formula (\ref{BV_A}) getting closer to the
experimental data, though it is not enough for the perfect reproduction.

\section{Collective submanifold and requantization of TDDFT}
\label{sec:collective_submanifold}

In this section, we introduce an assumption that 
the time-evolution of the densities are determined 
by a few collective coordinates and momenta, $R(q,p)$,
as we have done in Sec.~\ref{sec:equations_in_the_moving_frame_of_reference}.
This leads to a microscopic derivation of the collective Hamiltonian 
describing large amplitude collective phenomena. 
We then quantize the collective variables and obtain 
the collective Schr\"odinger equation.   
Numerical examples are given for low-frequency quadrupole collective
excitations  
which are dominating in low-lying states in almost all nuclei.   
We focus on recent advances and basic ideas of the approaches based 
on the TDDFT but relations to other 
time-independent approaches are also briefly discussed.

\subsection{Problems in large-amplitude collective motion}
\label{sec:problems_LACM}

First, let us discuss conceptual problems
in TDDFT studies beyond the linear regime in nuclear physics.
We have presented in Sec.~\ref{sec:linear_response} that
excitation energies and transition amplitudes can be obtained
in the linear response.
For instance, the  Fourier analysis on the time
evolution of the density, such as Eq. (\ref{S(E)_fourier}),
allows us to extract those quantities.
In this case,
when we scale the external field by a parameter $f$
as ${V}\rightarrow f{V}$,
the density fluctuation $\delta R(\omega)$ is invariant 
except for the same linear scaling,
$\delta R(\omega) \rightarrow f \delta R(\omega)$.
This allows us to uniquely define the transition densities.

In principle, the TDDFT can describe exact dynamics
of many-body systems (see Sec.~\ref{sec:TDDFT}).
However,
in nuclear EDFs, at least, we do not know in practice how to extract
information on excited states and 
genuine quantum phenomena 
which involve large-amplitude many-body dynamics.
Perhaps, most typical example is given by spontaneous fission phenomena.
Even if the nucleus is energetically favored by dividing it into two
fragments, 
the non-linearity of the TDDFT forbids the tunneling.

Beyond the linear regime,
as the oscillating amplitudes become larger, the nonlinear effects
play more important roles.
In fact, there are some attempts to quantify the nonlinear coupling
strengths between different modes of excitation using real-time TDDFT
calculations \cite{SC03,SCF01,SCF07}.
In addition to the linear response, the quadratic dependence is
identified to extract the coupling between dipole and quadrupole modes
\cite{SC09}.
Nevertheless, the practical real-time method to quantify energy spectra of
anharmonic vibrations has not been established.

Our strategy to these difficulties is to adopt the ``requantization''
procedure.
Perhaps, this is not perfectly consistent with the original principle of TDDFT
which should be ``exact'' and does not require additional quantum
fluctuation in the theory.
However, as we have noted in Secs.~\ref{sec:ANG} and \ref{sec:practices}, 
since the present nuclear EDF is reliable within a certain time scale
(typically the SSB time scale), the quantum fluctuations associated with
longer time scales should be addressed additionally.
The TDDFT dynamics of Eq. (\ref{TDBdGKS_R_eq}) can be 
parameterized with classical canonical variables $(\xi^\alpha,\pi_\alpha)$
which obey the classical Hamilton equations \cite{BR86}.
The space spanned by these variables are called ``TDHFB phase space'',
whose dimension is twice the number of two-quasiparticle pairs.
Therefore, in nuclear physics, the issue has been often discussed in terms
of the requantization of the TDDFT dynamics.
Further arguments on the requantization
are given by the stationary phase approximation
to the functional integral formulation of the many-body quantum theory
\cite{Neg82}.

To describe long time-scale slow motion in nuclei, 
we introduce small number of collective variables.
In low-energy spectra in nuclei, we observe a number of states
which possess properties very difficult to quantify with the real-time
TDDFT simulations;
for instance, states with fluctuating shapes,
those with mixture of different shapes,
anharmonic nature of many phonon states,
quasi-rotational spectra which show features between phonon-like
and rotational-like excitations.
Nuclear fission also provides another typical example of
nuclear large amplitude collective motion.
These low-energy dynamics in nuclei requires us to develop practical
theories applicable to nuclear large-amplitude collective motion (LACM).

\subsection{Fundamental concepts for low-energy nuclear dynamics
and historical remarks}

In Sec.~\ref{sec:linear_response},
we present the QRPA method,
as a small-amplitude approximation of the TDDFT,
for microscopically describing 
various kinds of collective excitations
around the equilibrium points, given by
$[H_\textrm{eff}[R_0]-\mu{\cal N},R_0]=0$.
In this Sec.~\ref{sec:collective_submanifold},
we review the recent advances of the approaches aiming at microscopic 
description of LACM by extending the basic
ideas of the QRPA to 
non-equilibrium states far from the local minima of the EDF.
Construction of microscopic theory of LACM
has been a difficult long-standing subject in nuclear structure theory.
The issues in 1980's were discussed in a proceedings \cite{AS83},
including the one by 
\textcite{Vil83} 
which summarized problems and
questions for theories of nuclear collective motion.
Since then, we have achieved
a significant progress in theoretical formulation and applications to
real nuclear phenomena in recent years.

\subsubsection{Basic ideas}

The basic idea for constructing,
on the basis of the time-dependent density functional method, 
a microscopic theory of large-amplitude collective phenomena
(at zero temperature)    
is to introduce an assumption that time evolution of the density is determined 
by a few collective coordinates $q(t)=\{q_1(t), q_2(t), \cdots, q_f(t)\}$ 
and collective momenta $p(t)=\{p_1(t), p_2(t), \cdots, p_f(t)\}$.
We assume that the time-dependent density can be written as 
$
R(t)=R(q(t),p(t))
$.
At this stage, $p$ and $q$ are introduced as parameters
in place of the time $t$. 
We shall see, however, that it is possible to formulate a theory such that 
they are canonical variables obeying the Hamilton equations of motion,   
i.e, they are classical dynamical variables.
Accordingly, we call them ``collective variables''. 
The great merit of this approach is, obviously, that they are
readily quantized, according to the standard canonical quantization.
In this way, we can derive, microscopically and self-consistently, 
the quantum collective Hamiltonian describing LACM.
Because of developments in the nuclear-theory history,
we call this canonical quantization procedure 
``collective quantization of time-dependent self-consistent mean field''. 
In the TDDFT terminology,
this can be regarded as the inclusion of missing correlations in
long time scales.
In Sec.~\ref{sec:derivation_of_H_coll},
we develop this idea in a more concrete form.

\medskip
\leftline{\underline{\it Notes on terminology and notation}}

Because of these practical and historical reasons,
it is customary to use the terminology and the notation of mean-field theories,
such as TDHF and TDHFB instead of TDDFT (TDKS, TDBdGKS).
We follow this tradition in this section.
The theory presented here takes account of correlations and
fluctuations beyond the mean field, which correspond to those
missing in current nuclear EDFs.

In order to help comprehensibility, we introduce
the TDHF(B) state $\ket{\phi(t)}$
in Sec.~\ref{sec:derivation_of_H_coll}, which is defined by the
quasiparticle vacuum $a_i(t) \ket{\phi(t)}=0$ at every time $t$
(time-dependent version of Eq. (\ref{vacc})).
Accordingly, we also use the Hamiltonian $\hat{H}$ 
which is related to the EDF as $E[\rho]=\bra{\phi(t)} \hat{H} \ket{\phi(t)}$.

\subsubsection{ANG modes associated with broken symmetries and
quantum fluctuations in finite systems}

We discussed in Sec.~\ref{sec:quasi-stationary-solutions}
how to treat the collective motions restoring the symmetries 
spontaneously broken in the mean fields for three typical examples
(center-of-mass motion, pair rotation in gauge space, and three-dimensional
rotation in coordinate space). 
Let us recall, in particular: 
\newline\noindent 1.
The ANG modes restoring the gauge invariance broken in the BCS theory of
superconductivity 
have been experimentally observed in nuclei as the pairing rotational modes
\cite{BB05}.
\newline\noindent 2.
The rotational spectra widely seen in nuclei can be regarded as ANG modes 
restoring the spherical symmetry spontaneously broken in the mean field
\cite{Ald56, BM75, Fra01}.
\newline\noindent 3.
We know generators of the collective variables for the ANG modes,
at least for one of $\overcirc{Q}$ and $\overcirc{P}$
in Sec.~\ref{sec:equations_in_the_moving_frame_of_reference}.
Those are given by global one-body operators, such as
$\hat{\vec{R}}_\textrm{cm}$ and $\hat{\vec{P}}_\textrm{cm}$ for the translation,
$\hat{\vec{J}}$ for the rotation,
and $\hat{N}$ for the pair rotation.
However, the generators conjugate to
$\hat{\vec{J}}$ and $\hat{N}$ are not trivial.

On the other hand, we should keep in mind that the mean fields of
finite quantum systems 
always accompany quantum fluctuations.
One of the most important characteristics of 
low-energy excitation spectra of nuclei is that the amplitudes of
the quantum shape fluctuation 
often become very large.
Among large-amplitude shape fluctuation phenomena, we should be, of course,
referred to the well-known spontaneous fission,
which can be regarded as macroscopic quantum tunnelings  
through the potential barrier generated by the self-consistent mean field. 
To construct a microscopic theory capable of describing
such large-amplitude shape 
fluctuations/evolutions has been a challenge in nuclear structure theory. 
Historically, such attempts started in 1950's to formulate a microscopic
theory of collective model of Bohr and Mottelson.
The major approach at that time is to introduce collective coordinates 
explicitly as functions of coordinates of individual nucleons and
separate collective shape degrees of 
freedom from the rest (see, {\it e.g.}, \textcite{Tom55-1}). 
This turned out to fail in description of low-energy modes of
shape fluctuations.
One of the important lessons we learned from these early attempts is that, 
in contrast to the ANG modes,  
it is not trivial at all to define microscopic structure of collective
coordinates appropriate for low-energy shape vibrations.  
The low-energy collective vibrations are associated  with fluctuations
of order parameters 
characterizing the mean field
\cite{Str79}.
In this sense, it may be categorized as a kind of
Higgs amplitude modes \cite{PV15},
but we need to go beyond the small-amplitude approximation for fluctuations  
about the equilibrium points in order to describe them.        

After the initial success of the BCS+QRPA approach for small amplitude
oscillations
in 1960's,
attempts to construct a microscopic theory of LACM started in mid 1970's. 
At that time, real-time TDHF (TDDFT) calculations
for heavy-ion collisions also started, using semi-realistic EDFs.
These attempts introduced
collective coordinates as parameters specifying 
the time-evolution of the self-consistent mean field,
instead of explicitly defining them as 
functions of coordinates of individual nucleons.    
This was a historical turning point
in basic concept of collective coordinate theory: 
In these new approaches, it is unnecessary to define {\it global}
collective operators, 
as functions of coordinates of individual nucleons. 
As we shall see in Sec.~\ref{sec:derivation_of_H_coll},
it is sufficient to determine infinitesimal generators for 
time-evolution of the self-consistent mean field, {\it locally} 
at every point of the the collective variables $(q,p)$.
Note that we use, in this section, the term {\it local}
to indicate the neighbor of a point $(q,p)$ in the collective space, 
instead of the conventional coordinate $\vec{r}$
in the three-dimensional coordinate space.
In general, the microscopic structures of the infinitesimal generators
for shape evolution may change as functions of $(q,p)$. 
From this point of view, 
it is not only unnecessary but also inappropriate
to introduce the global operators in order 
to describe low-energy shape fluctuations. 
This is in sharp contrast with the high-frequency giant quadrupole resonances
for which the small-amplitude approximation works well and the mass-quadrupole 
operator can be regarded as an approximate collective coordinate operator.

\subsubsection{Characteristics of quadrupole excitation spectra
in low-lying states}

Low-frequency quadrupole vibrations of the nucleus may be regarded as 
collective surface excitations of a finite superfluid system.
Accordingly, pairing correlations and varying shell structure of the
self-consistent mean field play essential roles in their emergence
\cite{BM75,AFN90,BHR03,RW10,MHS13}.
For nuclei in the transitional region from spherical to deformed,
amplitudes of quantum shape fluctuation remarkably increase.
This corresponds to soft modes of the quantum phase transition
towards symmetry-violating equilibrium deformations of the mean filed.
The gain in binding energies
due to the symmetry breaking is comparable in magnitude
to the vibrational zero-point energies.
The transitional region is prevalent in nuclear chart,
and those nuclei exhibit a rich variety of excitation spectra
in systematics.

In finite quantum systems like nuclei, the rotational ANG modes may couple 
rather strongly with quantum shape fluctuation modes. 
For instance, even when the self-consistent mean field acquires  
a deep local minimum at a finite value of $\beta$, 
the nucleus may exhibit a large-amplitude shape fluctuation 
in the $\gamma$ degree of freedom, 
if the deformation potential is flat in this direction. 
Here, as usual, $\beta$ and $\gamma$ represent the magnitudes of 
axially symmetric and asymmetric quadrupole deformations, respectively.  
Such a situation is widely observed in experiments and called $\gamma$-soft
nuclei. 
The rotational degrees of freedoms about three principal axes are 
all activated once the axial symmetry is dynamically broken due to the 
quantum shape fluctuation.
Consequently, rotational spectra in such $\gamma$-soft nuclei 
do not exhibit a simple $I(I+1)$ pattern. 
Such an interplay of the shape fluctuation and rotational modes may be regarded as 
a characteristic feature of finite quantum systems and provides an invaluable opportunity 
to investigate the process of the quantum phase transition through analysis of quantum spectra. 

Thus, we need to treat the two kinds of collective variables, {\it i.e.}, those associated with 
the symmetry-restoring ANG modes and those for quantum shape fluctuations, 
in a unified manner 
to describe low-energy excitation spectra of nuclei.

\subsection{Microscopic derivation of collective Hamiltonian}
\label{sec:derivation_of_H_coll}

\subsubsection{Extraction of collective submanifold}

As we have mentioned in Sec.~\ref{sec:problems_LACM},
the TDHFB dynamics is represented as the classical Hamilton equations
for canonical variables in the TDHFB phase space,
$(\xi^\alpha,\pi_\alpha)$
\cite{Neg82,YK87,Kur01}.
The dimension of this phase space is huge,
$\alpha=1,\cdots,D$ where $D$ is the
number of all the two-quasiparticle pairs.
The TDHFB state vector $\ket{\phi(t)}=\ket{\phi(\xi(t),\pi(t))}$
is regarded as a generalized
coherent state moving on a trajectory in the TDHFB phase space.  
For low-energy fluctuations in collective motion, however,
we assume that the time evolution is governed by a few collective variables.

During the attempts to construct microscopic theory of LACM
since the latter half of the 1970s, significant progress has been achieved in the 
fundamental concepts of collective motion.
Especially important is the recognition that
microscopic derivation of the collective Hamiltonian is equivalent
to extraction of a collective submanifold embedded in the TDHFB phase space,
which is approximately decoupled
from other ``non-collective'' degrees of freedom.
From this point of view we can say that 
collective variables are nothing but local canonical variables
which can be flexibly chosen on this submanifold.
Here, we recapitulate recent developments achieved on the basis of such
concepts.

Attempts to formulate a LACM theory 
without assuming adiabaticity of large-amplitude collective motion 
were initiated by 
and led to the formulation of
the self-consistent collective coordinate (SCC) method
\cite{MMSK80}.
In these approaches,
collective coordinates and collective momenta are treated on the same footing.  
In the SCC method, basic equations determining the collective submanifold
 are derived 
by requiring maximal decoupling of the collective motion of interest from 
other non-collective degrees of freedom. 
The collective submanifold 
is invariant with respect to the choice of 
the coordinate system, whereas the collective coordinates depend on it.  
The idea of coordinate-independent theory of collective motion
was developed also by \textcite{Row82}, and \textcite{YK87}. 
This idea gave a significant impact on the fundamental question,
``what are the collective variables?''.
The SCC method was first formulated for the canonical form of the TDHF
without pairing.
Later, it is extended to that of TDHFB
for describing nuclei with superfluidity \cite{Mat86}.
 
In the SCC method, the TDHFB state $\ket{\phi(t)}$
is written as $\ket{\phi(q,p)}$ under the assumption that the time evolution
is governed by a few collective coordinates $q=(q^1,q^2,\cdots,q^f)$  and
collective momenta $p=(p_1,p_2,\cdots,p_f)$.
Note that the parameterizing the TDHFB state with the $2f$-degrees of freedom
$(q,p)$ is
nothing but defining a submanifold inside the TDHFB phase space
$(\xi^\alpha,\pi_\alpha)$,
which we call ``collective submanifold''.
The time-dependent densities are readily obtained 
from the TDHFB state $\ket{\phi(q,p)}$ by
\begin{eqnarray*}
\rho(\vec{r}; q,p)  = \sum_{\sigma}\bra{\phi(q,p)}
 \hat\psi^{\dagger}(\vec{r}\sigma)\hat\psi(\vec{r}\sigma)\ket{\phi(q,p)}, \\
\kappa(\vec{r}; q,p)  = \bra{\phi(q,p)}
 \hat\psi(\vec{r}\downarrow)\hat\psi(\vec{r}\uparrow)\ket{\phi(q,p)}.
\end{eqnarray*}

The following basic equations
determine the TDHFB state $\ket{\phi(q,p)}$ parameterized by $(q,p)$
and its time evolution,
which gives the definition of the submanifold.
\newline\noindent 1.
\underline{\it Invariance principle of the TDHFB equation} \\
We require that the TDHFB equation of motion is invariant 
in the collective submanifold.  
This requirement can be written in a variational form as  
\begin{equation}
\delta\bra{\phi(q,p)} \left\{ i\frac{\partial}{\partial t} - \hat{H} \right\} \ket{\phi(q,p)}=0.  
\label{invariance principle}
\end{equation}
Here, the variation $\delta$ is given by 
$\delta \ket{\phi(q,p)} =a_i^\dagger a_j^\dagger \ket{\phi(q,p)}$ 
in terms of the quasiparticle operators $\{a_i^\dagger, a_j\}$, 
which satisfy the vacuum condition $a_i\ket{\phi(q,p)}=0$.  
Under the basic assumption,  the time derivative is replaced by
\begin{equation*}
\frac{\partial}{\partial t} = \sum_{i=1}^f
\left(\dot{q^i}\frac{\partial}{\partial q^i} + 
\dot{p_i}\frac{\partial}{\partial p_i} \right)
=
\dot{q^i}\frac{\partial}{\partial q^i} + 
\dot{p_i}\frac{\partial}{\partial p_i}
. 
\end{equation*}
Hereafter, to simplify the notation,
we adopt the Einstein summation convention
to remove $\sum_{i=1}^f$.
Accordingly, Eq.~(\ref{invariance principle}) is rewritten as
\begin{equation}
\delta\bra{\phi(q,p)} \left\{ 
\dot{q}^i \overcirc{P}_i(q,p) 
- \dot{p}_i\overcirc{Q}^i(q,p)  
 - \hat{H} \right\} \ket{\phi(q,p)}=0
\label{invariance principle2}
\end{equation}
in terms of the local infinitesimal generators defined by 
\begin{eqnarray}
\overcirc{P}_i(q,p) \ket{\phi(q,p)} &=&   i \frac{\del}{\del q^i} \ket{\phi(q,p)}, 
\label{generator_P}
\\
\overcirc{Q}^i(q,p) \ket{\phi(q,p)} &=& -i \frac{\del}{\del p_i} \ket{\phi(q,p)}. 
\label{generator_Q}
\end{eqnarray}
These are one-body operators which can be written as linear combinations 
of bilinear products $\{a_i^\dagger a_j^\dagger, a_i^\dagger a_j, a_j a_i \}$ 
of the quasiparticle operators defined with respect to $\ket{\phi(q,p)}$.  
Equations (\ref{invariance principle2}), (\ref{generator_P}) and (\ref{generator_Q}) 
correspond to Eqs.~(\ref{TDBdGKS_R_moving_frame}),
(\ref{generator_P_0}), and (\ref{generator_Q_0})
in Sec.~\ref{sec:equations_in_the_moving_frame_of_reference},
respectively.
\newline\noindent 2.
\underline{\it Canonicity conditions} \\
We require $q$ and $p$ to be canonical variables. 
According to the theorem of Frobenius and Darboux
 \cite{Arn89},
pairs of canonical variables $(q, p)$ exist 
for the TDHFB states $\ket{\phi(q,p)}$ satisfying the following 
{\it canonicity conditions}, 
\begin{eqnarray}
\bra{\phi(q,p)} \overcirc{P}_i(q,p) \ket{\phi(q,p)} &=& p_i + \frac{\del S}{\del q^i}, 
\label{canonicity conditions1}
\\
\bra{\phi(q,p)} \overcirc{Q}^i(q,p) \ket{\phi(q,p)} &=& -\frac{\del S}{\del p_i}, 
\label{canonicity conditions2}
\end{eqnarray} 
where $S$ is an arbitrary differentiable function of $q$ and $p$
\cite{MMSK80,YK87}.
By specifying $S$ we can fix the type of allowed canonical transformations 
among the collective variables $(q, p)$.
We shall discuss typical examples in subsequent subsections and 
call the canonicity conditions with a specified function $S$ 
``{\it canonical-variable conditions.}"
Taking derivatives of Eqs.~(\ref{canonicity conditions1}) 
and (\ref{canonicity conditions2})
with respect to $p_i$ and $q^i$, respectively, 
we can readily confirm that the local infinitesimal generators satisfy 
the `weakly' canonical commutation relations, 
\begin{equation*}
\bra{\phi(q,p)} \left[ \overcirc{Q}^i(q,p),  \overcirc{P}_j(q,p) \right] \ket{\phi(q,p)} 
= i \delta_{ij}. 
\end{equation*}
 
Taking variations of Eq. (\ref{invariance principle2})
in the direction of the collective variables, $q$ and $p$,  
generated by  $\overcirc{P}_i$ and $\overcirc{Q}^i$,  
we obtain the Hamilton equations of motion, 
\begin{equation}
\frac{dq^i}{dt}=\frac{\partial \Hc}{\partial p_i}
,\quad
\frac{dp_i}{dt}=-\frac{\partial \Hc}{\partial q^i}. 
\label{Hamilton_eq}
\end{equation}
Here, the total energy $\Hc(q,p)\equiv \bra{\phi(q,p)} \hat{H} \ket{\phi(q,p)}$
plays the role of the classical collective Hamiltonian. 
\newline\noindent 3.
\underline{\it Equation of collective submanifold} \\
The invariance principle (\ref{invariance principle2}) and
Eq.~(\ref{Hamilton_eq})
lead to the equation of collective submanifold:
\begin{equation}
\delta\bra{\phi(q,p)} \left\{
\hat{H} -  \frac{\partial \Hc}{\partial p_i}  \overcirc{P}_i(q,p)
- \frac{\partial \Hc}{\partial q^i}\overcirc{Q}^i(q,p)  \right\}
 \ket{\phi(q,p)}=0 .
\label{eq of submanifold1}
\end{equation}
Taking variations $\delta_{\perp}$ in the directions orthogonal to $q$ and $p$, 
we can immediately find 
$\delta_{\perp}\bra{\phi(q,p)} \hat{H} \ket{\phi(q,p)}=0$.  
This implies that the energy expectation value
is stationary with respect to all variations 
except for those along directions tangent to the collective submanifold. 
In other words, the large-amplitude collective motion is decoupled from other modes of excitation.

\subsubsection{Solution with $(\eta, \eta^*)$ expansion}

In the original paper of the SCC method \cite{MMSK80}, 
the TDHFB state $\ket{\phi(q,p)}$ is written as 
\begin{equation*}
\ket{\phi(q,p)} = U(q,p) \ket{\phi_0} = e^{i{\hat G}(q, p)} \ket{\phi_0}. 
\end{equation*}
Here, $U(q,p)$ represents a time-dependent unitary transformation 
from the HFB ground state $\ket{\phi_0}$ taken as an initial state;  
$U(q,p)  =1$ at $q=p=0$.  
It is written in terms of an Hermitian one-body operator $\hat{G}(q,p)$. 

With use of complex variables $\eta=(\eta_1,\eta_2,\cdots,\eta_f)$
defined by    
\begin{equation*}
\eta_i = \frac{1}{\sqrt 2}(q^i+ip_i)
,\quad
\eta_i^* = \frac{1}{\sqrt 2}(q^i-ip_i) ,
\end{equation*}
we can rewrite the TDHFB state as
\begin{equation*}
\ket{\phi(\eta,\eta^*)} = U(\eta,\eta^*) \ket{\phi_0} = e^{i{\hat G}(\eta, \eta^*)} \ket{\phi_0}, 
\end{equation*}
Correspondingly, we define 
local infinitesimal generators, $\overcirc{O}_i^\dag(\eta,\eta^*)$ and $\overcirc{O}_i(\eta,\eta^*)$,  by
\begin{eqnarray*}
\overcirc{O}_i^\dag (\eta,\eta^*) \ket{\phi(\eta,\eta^*)} &=& \frac{\del}{\del \eta_i} \ket{\phi(\eta,\eta^*)} , 
\\
\overcirc{O}_i (\eta,\eta^*)  \ket{\phi(\eta,\eta^*)} &=& -\frac{\del}{\del \eta_i^*}\ket{\phi(\eta,\eta^*)}  . 
\end{eqnarray*}
Replacing $(q,p)$ by $(\eta,\eta^*)$,
the equation of collective submanifold (\ref{eq of submanifold1})
is rewritten as 
\begin{eqnarray}
\delta\bra{\phi_0}U^\dag(\eta,\eta^*)
\left\{
\hat{H} -
\frac{\partial \Hc}{\partial \eta_i^*} \overcirc{O}_i^\dag(\eta,\eta^*)
-\frac{\partial \Hc}{\partial \eta_i}\overcirc{O}_i(\eta,\eta^*)
  \right\}&& \nonumber\\
\times U(\eta,\eta^*)\ket{\phi_0}=0 . \quad\quad\quad &&
\label{eq of submanifold2}
\end{eqnarray}
Here, the variation is to be performed only for the HFB ground state $\ket{\phi_0}$. 

We assume the following canonical-variable conditions, 
\begin{eqnarray}
\bra{\phi(\eta,\eta^*)} \overcirc{O}_i^\dag(\eta,\eta^*)\ket{\phi(\eta,\eta^*)} &=& \frac{1}{2}\eta_i^*, 
\label{canonical-variable conditions1}
\\
\bra{\phi(\eta,\eta^*)} \overcirc{O}_i(\eta,\eta^*) \ket{\phi(\eta,\eta^*)} &=& \frac{1}{2}\eta_i, 
\label{canonical-variable conditions2} 
\end{eqnarray}
which are obtained by a specific choice of  $S=-\frac{1}{2}\sum_i q^i p_i$ in 
the canonicity conditions, (\ref{canonicity conditions1}) and (\ref{canonicity conditions2}). 
From Eqs.~(\ref{canonical-variable conditions1}) and (\ref{canonical-variable conditions2}), 
we can easily obtain the ``weak'' boson commutation relations, 
\begin{equation*}
\bra{\phi(\eta,\eta^*)} \left[ \overcirc{O}_i(\eta,\eta^*),   
\overcirc{O}_j^\dag(\eta,\eta^*)\right] \ket{\phi(\eta,\eta^*)} = \delta_{ij}. 
\end{equation*}
Because only linear canonical transformations among $\eta$ and $\eta^*$ keep  
Eqs.~(\ref{canonical-variable conditions1}) and (\ref{canonical-variable conditions2})  
invariant,  these canonical-variable conditions are suitable to 
a solution of the variational equation (\ref{eq of submanifold2})  
with a power series expansion of ${\hat G}$ with respect to $(\eta,\eta^*)$, 
\begin{eqnarray*}
{\hat G}(\eta,\eta^*) &=&  {\hat G}_i^{(10)}\eta_i^* + {\hat G}_i^{(01)}\eta_i
 \nonumber \\
&&+{\hat G}_{ij}^{(20)}\eta_i^*\eta_j^* +  {\hat G}_{ij}^{(11)}\eta_i^*\eta_j
+  {\hat G}_{ij}^{(02)}\eta_i\eta_j + \cdots \quad\quad
\end{eqnarray*}

Requiring that the variational principle~(\ref{eq of submanifold2})
holds for every power of $(\eta,\eta^*)$, 
we can successively determine the one-body operator ${\hat G}^{(m, n)}$ 
with $m+n=1, 2, 3, \cdots$.  
This method of solution is called the ``$(\eta,\eta^*)$-expansion method."
Because $(\eta,\eta^*)$ are complex canonical variables, 
they are replaced by boson operators after the canonical quantization. 
The lowest linear order corresponds to the QRPA.  
Accordingly, the collective variables
$(\eta_i,\eta_i^*)$ correspond to a specific QRPA mode 
in the small-amplitude limit.
However, in the higher orders,
the microscopic structure of ${\hat G}$ changes
as a function of $(\eta,\eta^*)$ 
due to the mode-mode coupling effects
among different QRPA modes.
In this sense, the $(\eta,\eta^*)$-expansion method may be regarded as
a dynamical extension of the boson expansion method 
\cite{MM85-1}. 
Thus, the SCC method with the $(\eta,\eta^*)$ expansion
is a powerful method of treating anharmonic effects
to the QRPA vibrations 
originating from mode-mode couplings.
This is shown in its application to the two-phonon states 
of anharmonic $\gamma$ vibration \cite{MM85-2,MSM85}. 
The SCC method was also used for derivation of
the 5D collective Hamiltonian and analysis of 
the quantum phase transition from spherical to deformed shapes \cite{Yam93}
and for constructing diabatic representation
in the rotating shell model \cite{SM01}. 
The validity of the canonical quantization procedure, including a treatment of  
the ordering ambiguity problem, was examined in  
\cite{MM85-1}.

\subsubsection{Solution with adiabatic expansion}
\label{sec:ASCC}

The $(\eta,\eta^*)$ expansion about a single HFB equilibrium point is
not suitable for treating 
situations where a few local minima energetically compete
in the HFB potential energy surface. 
It is also difficult to apply the expansion method
to a collective motion which goes far away from the equilibrium,
such as nuclear fission.
These low-energy LACM's in nuclei are often characterized by ``slow'' motion.
For describing adiabatic LACM
extending over very far from the HFB equilibrium,
a new method of solution has been proposed \cite{MNM00}.  
In this method, the basic equations of the SCC method are
solved by an expansion with respect to 
the collective momenta, keeping full orders in the collective coordinates.
It is called ``adiabatic SCC (ASCC) method.''
Similar methods have been developed also 
by \textcite{KDW91} and \textcite{AW04}. 

\begin{figure}[t]
\includegraphics[width=0.2\textwidth]{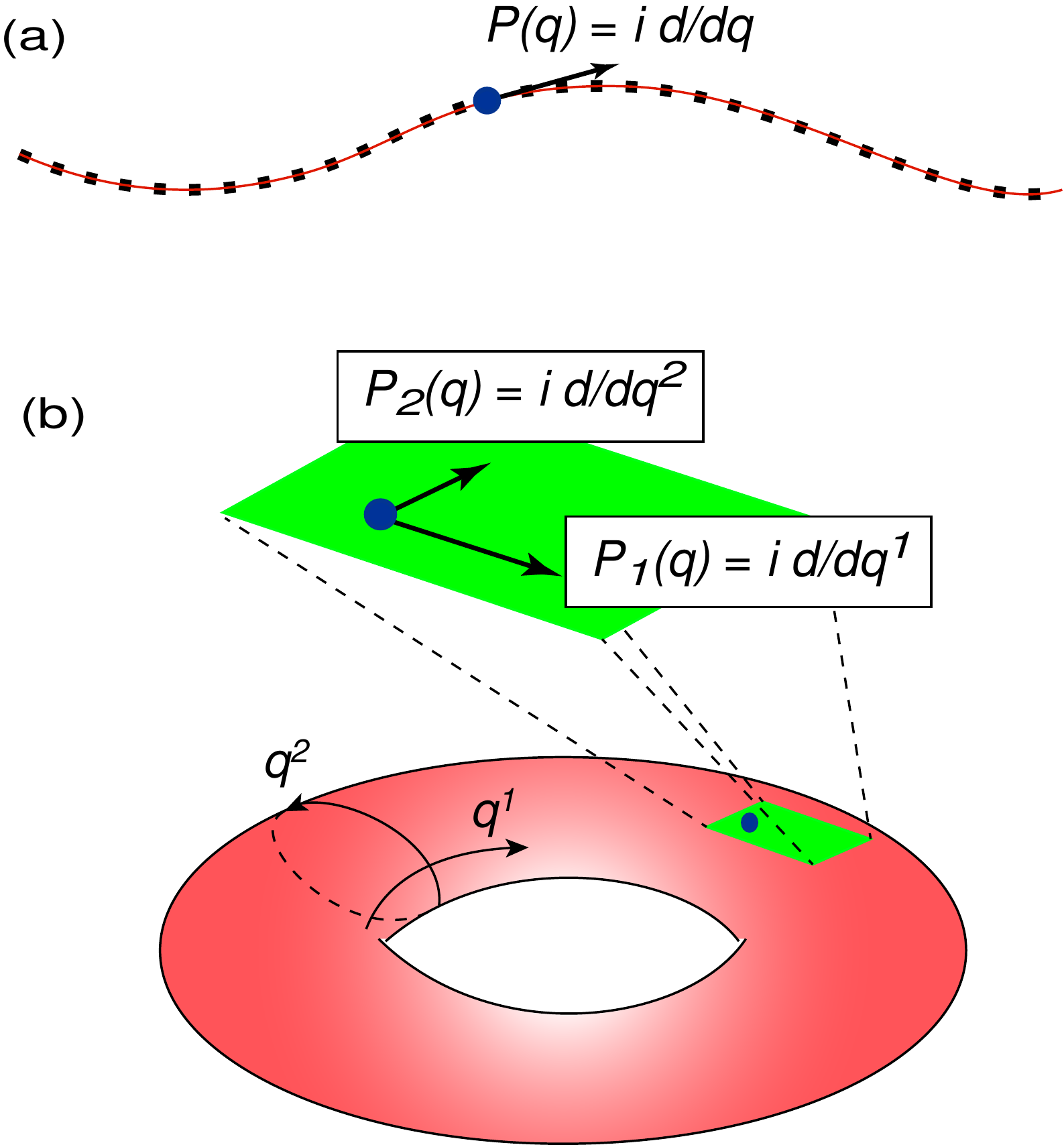}
\caption{
(Color online)
Schematic illustrations of collective submanifold in the TDHFB space.
(a) 1D collective path specified by a series of the states $\ket{\phi(q)}$
with a local generator $P(q)=id/dq$.
(b) A section of a 2D collective hypersurface and
local generators of the collective coordinate, $(\Phat_1(q),\Phat_2(q))$.
}
\label{fig:submanifold}
\end{figure}

A microscopic theory for adiabatic LACM is constructed by the ASCC method
in the following way. 
We assume that the TDHFB state  $\ket{\phi(q,p)}$
can be written in a form
\begin{equation}
 \ket{\phi(q,p)}  =  \exp\left\{ i p_i \Qhat^i(q) \right\}
 \ket{\phi(q)} ,
\label{eq:ASCCstate}
\end{equation}
where $\Qhat^i(q)$ are one-body operators corresponding to
infinitesimal generators of $p_i$ locally defined at the state $\ket{\phi(q)}$ 
which represents a TDHFB state $\ket{\phi(q,p)}$ at $p\rightarrow 0$. 
This state $\ket{\phi(q)}$ is called a ``moving-frame HFB state." 
See Fig.~\ref{fig:submanifold} for illustrations.
We use the canonical-variable conditions different from 
(\ref{canonical-variable conditions1}) and (\ref{canonical-variable conditions2}), 
\begin{eqnarray}
\bra{\phi(q,p)} \overcirc{P}_i(q,p) \ket{\phi(q,p)} &=& p_i, 
\label{canonical-variable conditions3}
\\
\bra{\phi(q,p)} \overcirc{Q}^i(q,p) \ket{\phi(q,p)} &=& 0, 
\label{canonical-variable conditions4}
\end{eqnarray}
which are obtained by putting $S=$const. 
in the canonicity conditions (\ref{canonicity conditions1}) and (\ref{canonicity conditions2}). 
Because (\ref{canonical-variable conditions3}) and (\ref{canonical-variable conditions4}) are 
invariant only against point transformations,
$q\rightarrow q'(q)$
(more generally, similarity transformations) 
which do not mix $p$ and $q$, 
these canonical-variable conditions are suitable for the adiabatic expansion  
with respect to the collective momenta $p$.

We insert the above form of the TDHFB state (\ref{eq:ASCCstate})
into  the equation of collective submanifold (\ref{eq of submanifold2})
and the canonical variable conditions, 
(\ref{canonical-variable conditions3}) and (\ref{canonical-variable conditions4}), 
and make a power-series expansion in $p$. 
We can determine the microscopic structures of $\Qhat^i(q)$ and $\ket{\phi(q)}$  
by requiring that these equations hold for every power of $p$.   
We take into account up to the second order.  
The canonical variable conditions, 
(\ref{canonical-variable conditions3}) and (\ref{canonical-variable conditions4}), 
then yield the `weakly' canonical commutation relations, 
\begin{equation*}
\bra{\phi(q)} \left[ \Qhat^i(q), \Phat_j(q) \right] \ket{\phi(q)} = i\delta_{ij}. 
\end{equation*}
We also obtain $\bra{\phi(q)} \Qhat^i(q) \ket{\phi(q)} = 0$ and  
$\bra{\phi(q)} \Phat_i(q) \ket{\phi(q)} =0$, which are trivially satisfied.   
Here, the displacement operators $\Phat_i(q)$ are defined by 
\begin{equation*}
 \Phat_i(q) \ket{\phi(q)} = i \frac{\del}{\del q^i} \ket{\phi(q)} .
\end{equation*}
Note that  $\Qhat^i(q)$ and $\Phat_i(q)$ operate on $\ket{\phi(q)}$, 
while $\overcirc{Q}^i(q,p)$ and $\overcirc{P}_i(q,p)$ on $\ket{\phi(q,p)}$. 
The time derivatives $\dot{q}^i$ and $\dot{p}_i$  are determined by 
the Hamilton equations of motion (\ref{Hamilton_eq}) 
with the classical collective Hamiltonian
$\Hc(q,p)$ expanded with respect to $p$ up to the second order,
\begin{eqnarray*}
&&\Hc(q,p)= V(q) + \frac{1}{2} B^{ij}(q) p_i p_j , \\ 
&&V(q) = \Hc(q,p=0)
,\quad
B^{ij}(q)=\left.\frac{\partial^2 \Hc}{\partial p_i \partial p_j}\right|_{p=0} .
\nonumber
\end{eqnarray*}
The collective inertia tensors $B_{ij}(q)$ are defined as the inverse matrix
of $B^{ij}(q)$, ~$B^{ij} B_{jk}=\delta^i_k$.
Under these preparation,
the following equations, which constitute the core of the ASCC method,
can be derived \cite{MNM00}.
Here, to further simplify the expression, we show the case for normal systems
with TDHF.

\noindent
\leftline{\underline{1.~{\it Moving-frame HF(B) equation}}}
\begin{equation}
 \delta\bra{\phi(q)}\Hhat_M(q)\ket{\phi(q)} = 0, 
\label{eq:mfHFB}
\end{equation}
where $\Hhat_M(q)$ represents the Hamiltonian in the frame attached to the moving mean field, 
\begin{equation*}
  \Hhat_M(q) = 
 \Hhat 
 -  \frac{\del V}{\del q^i}\Qhat^i(q) .
\end{equation*}

\leftline{\underline{2.~{\it Moving-frame (Q)RPA equations}}}
\begin{eqnarray}
&&\delta\bra{\phi(q)} \left[\Hhat_M(q), \Qhat^i(q)\right]
- \frac{1}{i} B^{ij}(q) \Phat_j(q)   \quad\quad
\nonumber \\
&& \quad\quad\quad\quad\quad
+\frac{1}{2}\left[\frac{\del V}{\del q^j}\Qhat^j(q),
 \Qhat^i(q)\right]
\ket{\phi(q)} = 0, \quad
\label{eq:ASCC1}
\\
&& \delta\bra{\phi(q)} [\Hhat_M(q),
 \frac{1}{i}\Phat_i(q)] - C_{ij}(q) \Qhat^j(q)
\nonumber \\
&&
- \frac{1}{2}\left[\left[\Hhat_M(q), \frac{\del V}{\del
 q^k}\Qhat^k(q)\right], B_{ij}(q) \Qhat^j(q)\right] 
\ket{\phi(q)} = 0, \quad\quad
\label{eq:ASCC2}
\end{eqnarray}
where  
\begin{eqnarray*}
 C_{ij}(q) &=& \frac{\del^2 V}{\del q^i \del q^j} - \Gamma_{ij}^k\frac{\del V}{\del q^k}, \\
 \Gamma_{ij}^k(q) &=& \frac{1}{2} B^{kl}\left( \frac{\del B_{li}}{\del q^j}
 + \frac{\del B_{lj}}{\del q^i} - \frac{\del B_{ij}}{\del q^l} \right). 
\end{eqnarray*} 
The double-commutator term in Eq.~(\ref{eq:ASCC2}) arises from the $q$-derivative 
of the infinitesimal generators $\Qhat^i(q)$ and represents the curvatures of the collective submanifold.  
Diagonalizing the matrix, $B^{ik} C_{kj}$, at each point of $q$,
we may identify the local normal modes and eigen-frequencies 
$\omega_i(q)$ of the moving-frame QRPA equations.

Extension from TDHF to TDHFB for superfluid nuclei
can be achieved by introducing
the particle number $n \equiv N-N_0$
and their conjugate angle $\theta$
as additional collective variables.
See Sec.~\ref{sec:gauge_invariance} 
and \textcite{MNM00} for more details.

Solving Eqs.~(\ref{eq:mfHFB}), (\ref{eq:ASCC1}), and (\ref{eq:ASCC2})
self-consistently, 
we can determine the state $\ket{\phi(q)}$ and the microscopic expressions of 
the infinitesimal generators, $\Qhat^i(q)$ and $\Phat_i(q)$,
in bilinear forms of the quasiparticle creation and annihilation operators 
defined locally with respect to $\ket{\phi(q)}$.
See Fig.~\ref{fig:submanifold}.
Note that these equations
reduce to the HF(B) and (Q)RPA equations at the equilibrium point where 
$\del V/\del q^i=0$. 
Therefore, they are natural extensions of the HFB-QRPA equations 
to non-equilibrium states. 
Here, we remark on some key points of the ASCC method.

\leftline{1. {\it Difference from the constrained HFB equations}}
The moving-frame HFB equation (\ref{eq:mfHFB}) resembles 
the constrained HFB equation. 
An essential difference is that the infinitesimal generators $\Qhat^i(q)$ 
are here self-consistently determined together with $\Phat_i(q)$
as solutions of the 
moving-frame QRPA equations, (\ref{eq:ASCC1}) and (\ref{eq:ASCC2}),  
at every point of the collective coordinate $q$. 
Thus, contrary to constrained operators in the constrained HFB theory, 
their microscopic structure changes as functions of $q$. 
The optimal ``constraining'' operators are
locally determined at each $q$. 
The collective submanifold
embedded in the TDHFB phase space is 
extracted in this way.

\leftline{2. {\it Meaning of the term ``adiabatic''}}
The word of ``adiabatic approximation''
 is frequently used with different meanings.
In the present context, we use this term for 
the approximate solution of the variational equation
(\ref{invariance principle2})
by taking into account up to the second order 
in an expansion with respect to the collective momenta $p$.  
It is important to note that the effects of finite frequency of the LACM
 are taken into account 
through the moving-frame QRPA equation.
No assumption is made, such as that
the kinetic energy of LACM is much smaller 
than the lowest two-quasiparticle excitation energy at every point of $q$.

\leftline{3. {\it Collective inertial mass}}
Although the collective submanifold is invariant against coordinate
transformations, $q \rightarrow q'(q)$,
the collective inertial tensors $B_{ij}(q)$ depends on the adopted
coordinate system. 
The scale of the coordinates can be arbitrary chosen as far as the
canonical-variable conditions are satisfied.
Note, however, that it is convenient to
adopt a conventional coordinate system,
such as the quadrupole $(\beta,\gamma)$ variables,
to obtain physical insights
and to find the effects of
time-odd components in the mean field
(see Sec.~\ref{sec:CHFB+AP}).

\leftline{4. {\it Canonical quantization}}
The collective inertia tensors $B_{ij}(q)$ take a diagonal form  
when the classical collective Hamiltonian is represented in terms of 
the local normal modes of the moving-frame QRPA equations.  
We can then make a scale transformation of the collective coordinates $q$ 
such that  they become unity. 
The kinetic energy term in the resulting collective Hamiltonian 
depends only on $p$. 
Thus, there is no ordering ambiguity between $q$ and $p$ 
in the canonical quantization procedure.

\subsubsection{Inclusion of the pair rotation and gauge invariance}
\label{sec:gauge_invariance}

In the QRPA at the HFB equilibrium,
the ANG modes like the number fluctuation (pairing rotational)
modes are decoupled from other normal modes.
Thereby, the QRPA restores the gauge invariance (number conservation) 
broken in the HFB mean field \cite{BB05}.
It is desirable to keep this nice property
beyond the small-amplitude approximation.   
Otherwise, spurious number-fluctuation modes would heavily mix
in the LACM of interest. 
This can be achieved in the SCC method \cite{Mat86}.

Introducing the number-fluctuation $n=N-N_0$
and their conjugate angle $\theta$
as additional collective variables,
we generalize the TDHFB state (\ref{eq:ASCCstate})
to 
\begin{eqnarray*}
\ket{\phi(q,p,\theta,n)}  &=&  e^{-i\theta\Nhat} \ket{\phi(q,p,n)} ,\\
\ket{\phi(q,p,n)} &=& e^{i \left[ p_i \Qhat^i(q) 
+ 
 n \That(q) \right]} \ket{\phi(q)}, 
\end{eqnarray*}
where $\That(q)$ denotes the infinitesimal generator for the pair-rotation degree of freedom. 
The state vector $\ket{\phi(q,p,n)}$ may be regarded
as an intrinsic state for the pair rotation. 
In practice, $(\Nhat, \That(q))$ should be doubled to treat
both neutrons and protons.
The extension of the equation for the collective submanifold  
(\ref{invariance principle2})
is straightforward;  
\begin{eqnarray*}
\delta \bra{\phi(q,p,\theta,n)}
\left\{
i \dot{q}^i \frac{\partial}{\partial q^i}
+ i\dot{p}_i \frac{\partial}{\partial p_i} \right.
\quad\quad\quad\quad\quad &&\nonumber \\
\left.
+ i\dot{\theta} \frac{\partial}{\partial\theta}
- \Hhat
\right\}
\ket{\phi(q,p,\theta,n)}  = 0. &&
\end{eqnarray*}
Note that $\dot{n}=0$, 
because the Hamilton equations for the canonical conjugate pair $(n, \theta)$ are   
\begin{equation*}
\dot{\theta}=\frac{\partial \Hc}{\partial n}, 
\quad 
\dot{n}=-\frac{\partial \Hc}{\partial \theta}, 
\end{equation*}
and the classical collective Hamiltonian 
$\Hc(q,p,\theta,n)\equiv \bra{\phi(q,p,\theta,n)} \hat{H} \ket{\phi(q,p,\theta,n)}$ 
does not depend on $\theta$. 

Expanding in $n$ as well as $p$ up to the second order, 
we can determine $\That(q)$ simultaneously with $\Qhat^i(q)$ and $\Phat_i(q)$ 
such that the moving-frame equations become invariant 
against the rotation of the gauge angle $\theta$.

\textcite{HNMM07} investigated 
the gauge-invariance properties of the ASCC equations 
and extended the infinitesimal generators $\Qhat^i(q)$ to include
quasiparticle creation-annihilation parts 
in addition to two-quasiparticle creation and annihilation parts.
This is the reason why Eqs. (\ref{eq:ASCC1}) and (\ref{eq:ASCC2}) 
are written in a more general form than those originally given by 
\textcite{MNM00}.
The gauge invariance implies that we need to fix the gauge in numerical applications.
A convenient  procedure of the gauge fixing is discussed in 
\textcite{HNMM07}.
A more general consideration on the gauge symmetry is given 
from a viewpoint of constrained dynamical systems 
\cite{Sato15}. 

\subsection{Relations to other approaches}
\label{sec:other_approaches}

In Sec.~\ref{sec:derivation_of_H_coll},
we reviewed the basics of a microscopic theory of LACM 
focusing on new developments in the ASCC method, achieved after 2000. 
In this section, we discuss the relations of the above formulation
to other approaches to LACM.  
Typical approaches developed up to 1980 are described in detail
in the textbook of \textcite{RS80}, 
and achievements during 1980-2000 are well summarized in the review
by \textcite{DKW00}.

\subsubsection{Constrained HFB + adiabatic perturbation}
\label{sec:CHFB+AP}
 
This method is convenient and widely used in the microscopic description
of LACM.  
The theory is based on the {\it adiabatic} assumption that the 
collective motion is much slower than the single-particle motion
(see remarks in Sec.~\ref{sec:ASCC}).
We first postulate a few one-body operators, $\hat{F_i}$,
corresponding to collective coordinates $\alpha^i$. 
The collective potential energy is given by
the constrained HFB (or constrained HF + BCS) equations
\begin{eqnarray*}
&& \delta \bra{\phi_0(\alpha)} \Hhat -
 \mu^i(\alpha) \hat{F_i}\ket{\phi_0(\alpha)} = 0,
  \\
&&\alpha^i = \bra{\phi_0(\alpha)}\hat{F_i}\ket{\phi_0(\alpha)} ,
\end{eqnarray*}
where $\mu^i(\alpha)$ are the Lagrange multipliers. 
Then, assuming that the frequency of the collective motion is much
smaller than the two-quasiparticle energies, 
we calculate the collective kinetic energy
$T_{\rm{coll}}$ using the adiabatic perturbation theory;
$
T_{\rm{coll}}=(1/2) D_{ij}(\alpha)\dot{\alpha}^{i*}\dot{\alpha}^j
$,
where 
\begin{equation*}
D_{ij}(\alpha)=2\sum_n
\frac{\bra{\phi_0(\alpha)}{\frac{\partial}{\partial \alpha^{i*}}}\ket{\phi_n(\alpha)}
\bra{\phi_n(\alpha)}{\frac{\partial}{\partial \alpha^{j}}} \ket{\phi_0(\alpha)}}
{E_n(\alpha)-E_0(\alpha)}
\end{equation*}
are called Inglis-Belyaev cranking masses \cite{RS80}.
Here $\ket{\phi_0(\alpha)}$ and $\ket{\phi_n(\alpha)}$ represent the ground and 
two-quasiparticle excited states
for a given set of values $\alpha=\{\alpha^i\}$. 
In most of applications, it is simplified furthermore, by an assumption that
the derivatives of the constrained HFB Hamiltonian
with respect to $\alpha^i$ is
proportional to $\hat{F}_i$,
which leads to
\begin{eqnarray*}
D_{ij}(\alpha)&=&\frac{1}{2} \left[ {\cal M}_1^{-1}(\alpha)
 {\cal M}_3(\alpha) {\cal M}_1^{-1}(\alpha) \right]_{ij} \\
{\cal M}_n(\alpha)_{ij}&=&
\sum_n \frac{\bra{\phi_0(\alpha)}\hat{F_i}^\dag\ket{\phi_n(\alpha)}
\bra{\phi_n(\alpha)}\hat{F_j} \ket{\phi_0(\alpha)}}
{(E_n(\alpha)-E_0(\alpha))^n} ,
\end{eqnarray*}
These cranking masses were used in conjunction with phenomenological 
mean-field models in the study of fission dynamics \cite{Bra72}. 
In recent years, it has become possible to carry out such studies using 
self-consistent mean fields obtained by solving the constrained HFB equations
\cite{Bar11}. 
The Inglis-Belyaev cranking masses have also been used for 
low-frequency quadrupole collective dynamics
\cite{LGD99,YLQG99,PQSL04,Del10}. 
At present, a systematic investigation on low-lying quadrupole spectra
is underway 
in terms of the five dimensional (5D) collective Hamiltonian
(see Sec.~\ref{sec:5DCH}), 
which is derived from the relativistic (covariant) density functionals 
and by using the Inglis-Belyaev cranking formula 
\cite{Nik09,Li09,LNVM10,Li10,Li11,Nik11,FMXLYM13}.

A problem of the Inglis-Belyaev cranking formula is that
time-odd mean-field effects are ignored, thus,
it underestimates the collective masses (inertial functions) 
\cite{DD95}. 
Moving mean fields induce time-odd components
that change sign under time reversal.
However, the Inglis-Belyaev cranking formula ignores their effects on the
collective masses.
By taking into account such time-odd corrections to the cranking masses,    
one can better reproduce low-lying spectra \cite{HLNNV12}. 
For rotational moments of inertia, we may estimate the time-odd
corrections taking the limit of $\omega_\textrm{rot}\rightarrow 0$ for
the quasi-stationary solution of Eq. (\ref{TDKS_rot}).
Since this provides about $20-40$ \%\ enhancement from the Inglis-Belyaev
formula, the similar enhancement factors of $1.2-1.4$ have been often
utilized for vibrational inertial masses without solid justification.
A better treatment of the time-odd mean-field effects is required
for describing the masses of collective motion and
the effective mass of single-particle motion 
in a self-consistent manner.  
For this purpose, it is highly desirable to apply   
the microscopic theory of LACM in Sec.~\ref{sec:ASCC}
to the TDDFT with realistic EDFs.
At present, however, it remains as a challenge for future. 

\subsubsection{Adiabatic TDHF theory}

Attempts to derive collective Hamiltonian using adiabatic approximation 
to time evolution of mean fields started in 1960's
\cite{Bel65,BK65}. 
In these pioneer works,
the collective quadrupole coordinates $(\beta,\gamma)$ 
were defined in terms of expectation values of the quadrupole operators and 
the 5D collective Hamiltonian was derived 
using the pairing plus quadrupole (P+Q) force model 
\cite{BS69}. 
During 1970's this approach was generalized to a theory applicable
to any effective interaction. 
This advanced approach is called {\it adiabatic TDHF} (ATDHF) theory
\cite{BV78,BGV76,GR78}. 

In the ATDHF theory, the density matrix $\rho(t)$ is written in the following form and 
expanded as a power series with respect to the collective momentum $\chi(t)$. 
\begin{eqnarray*}
\rho(t)&=&e^{i\chi(t)}\rho_0(t)e^{-i\chi(t)}\\
          &=&\rho_0(t)+[i\chi,\rho_0]+\frac{1}{2}[i\chi, [i\chi, \rho_0]]
              + \cdots\\
          &=&\rho_0(t) + \rho_1(t) + \rho_2(t) + \cdots
\end{eqnarray*}
Correspondingly, the time-dependent mean-field Hamiltonian
$h(t)$ is also expanded 
with respect to a power of $\chi(t)$. 
\begin{equation*}
h[\rho(t)]=W_0(t) + W_1(t) + W_2(t) + \cdots
\end{equation*}
Inserting these into the TDHF (TDKS) equation (\ref{TDKS_rho_eq}),
we obtain for the time-odd part
\begin{equation*}
i\dot{\rho}_0=[W_0, \rho_1] + [W_1, \rho_0], 
\end{equation*}
and the time-even part
\begin{equation*}
i\dot{\rho}_1=[W_0, \rho_0] +[W_0, \rho_2] + [W_1, \rho_1] + [W_2, \rho_0] . 
\end{equation*}
These are the basic equations of the ATDHF. 

Let us introduce collective coordinates
$\alpha=(\alpha^1,\cdots,\alpha^f)$ 
as parameters describing the time evolution of the density matrix
$\rho_0(t)$ as
\begin{equation*}
\rho_0(t)=\rho_0(\alpha(t)), \quad
\dot{\rho}_0(t)=\sum_i\frac{\del\rho_0}{\del\alpha^i}\dot{\alpha^i}.  
\end{equation*}
\textcite{BV78}
proposed iterative procedures to 
solve the ATDHF equations for the density matrix parameterized in this way, 
but this idea has not been realized until now. 
The ATDHF does not reduce to the RPA in the small amplitude limit 
if a few collective coordinates are introduced by hand. 
In fact it gives a collective mass different from that of RPA 
\cite{GQ80-1,GQ80-2}.   

The ATDHF theory developed by \textcite{Vil77}
aims at self-consistently determining 
optimum collective coordinates on the basis of
the time-dependent variational principle.  
This approach, however, encountered a difficulty that we cannot get 
a unique solution of its basic equations determining the collective path. 
This non-uniqueness problem was
later solved by treating the second-order terms of the 
momentum expansion in a self-consistent manner
\cite{MP82,KDW91}. 
It was shown that, when the number of collective coordinate is only one,  
a collective path maximally decoupled from non-collective degrees of freedom 
runs along a valley in the multi-dimensional potential-energy surface
associated with the TDHF states.   

In order to describe low-frequency collective motions, 
it is necessary to take into account the pairing correlations. 
Thus, we need to develop the adiabatic TDHFB (ATDHFB) theory. 
This is one of the reasons why applications of the ATDHF theory
have been restricted 
to collective phenomena where the pairing correlations play minor roles,
such as
low-energy collisions between spherical closed-shell nuclei
\cite{Goe83}.
When large-amplitude shape fluctuations take place, 
single-particle level crossings often occur. 
To follow the adiabatic configuration across the level crossing points, 
the pairing correlation plays an essential role. 
Thus, an extension to ATDHFB is indispensable for 
the description of low-frequency collective excitations.  

In the past, \textcite{DS81} 
tried to develop the ATDHFB theory 
assuming the axially symmetric quadrupole deformation parameter
$\beta$ as the collective coordinate.
Quite recently, \textcite{Li12} 
tried to derive
the 5D quadrupole collective Hamiltonian 
on the basis of the ATDHFB. 
However, the extension of ATDHF to ATDHFB is not as straightforward as
we naively expect.
This is because, as we discussed in Sec.~\ref{sec:gauge_invariance},
we need to decouple the pair-rotational degrees of freedom
(number fluctuation) from the LACM of interest.

\subsubsection{Boson expansion method}
\label{sec:BEM}

Boson expansion method is an efficient microscopic method 
of describing anharmonic (non-linear) vibrations going 
beyond the harmonic approximation of QRPA.
In this approach, we first construct a collective subspace 
spanned by many-phonon states of vibrational quanta (determined by the QRPA)  
in the huge-dimensional shell-model space.
These many-phonon states are mapped onto many-boson states
in an ideal boson space. 
Anharmonic effects neglected in the QRPA are treated as higher order terms 
in the power series expansion with respect to the boson creation
and annihilation operators. 
Starting from the QRPA about a spherical shape,
 one can thus derive the 5D quadrupole collective Hamiltonian 
in a fully quantum mechanical manner.  
The boson expansion method has been successfully applied
to low-energy quadrupole excitation spectra 
in a wide range of nuclei including those lying
in transitional regions of quantum phase transitions 
from spherical to deformed shapes
\cite{SK88,KM91}.

In the time-dependent mean-field picture, 
state vectors in the boson expansion method are written 
in terms of the creation and annihilation operators
$(\Gamma_i^\dag, \Gamma_i)$ of the QRPA eigen-modes, 
or, equivalently, in terms of the collective coordinate and momentum operators 
$({\hat Q}^i, {\hat P}_i)$, 
\begin{equation*}
\begin{split}
|\phi(t)\rangle &= e^{i{\hat G}(t)}|\phi_0\rangle,\\
i{\hat G}(t) &= \eta_i(t) \Gamma_i^\dagger - \eta_i^*(t) \Gamma_i
             = ip_i(t) {\hat Q}^i - iq^i(t) {\hat P}_i. 
\end{split}
\end{equation*}
With increasing amplitudes of the quadrupole shape vibration $|\eta_i(t)|$
($|q_i(t)|$), anharmonic (non-linear) effects become stronger. 
Strong non-linear effects may eventually change even the microscopic structure of the collective operators 
$({\hat Q}_i, {\hat P}_i)$ determined by the QRPA. 
In such situations, it is desirable to construct a theory that allows 
variations of microscopic structure of collective operators
as functions of $q_i(t)$. 
The SCC method has accomplished this task
(See Sec.~\ref{sec:derivation_of_H_coll}).

\subsubsection{Generator coordinate method}
\label{sec:GCM}

In the generator coordinate method (GCM), 
quantum eigenstates of collective motion are described as 
superpositions of states $\ket{\phi(\alpha)}$ labeled by the parameters,
$\alpha=(\alpha^1,\cdots,\alpha^f)$,
which are called {\it generator coordinates}.
\begin{equation*}
\ket{\Psi} = \int d\alpha f(\alpha) \ket{\phi(\alpha)} .
\end{equation*}
$\ket{\phi(\alpha)}$ are {\it generating functions},
normally chosen as mean field states (Slater determinants),
which provide non-orthogonal basis for a collective subspace.
The Ritz variational principle then leads to the Hill-Wheeler equation 
\begin{equation*}
\int d\alpha f^*(\alpha) \bra{\phi(\alpha)}\Hhat-E\ket{\phi(\alpha')}=0 
\end{equation*}
determining the weight function $f(\alpha)$. 
Here $\int d\alpha$ denotes multiple integration with respect to the $f$-dimensional 
generator coordinates, and volume elements of integration are 
absorbed in the weight function $f(\alpha)$. 

\begin{figure*}[t]
\includegraphics[clip,width=12cm]{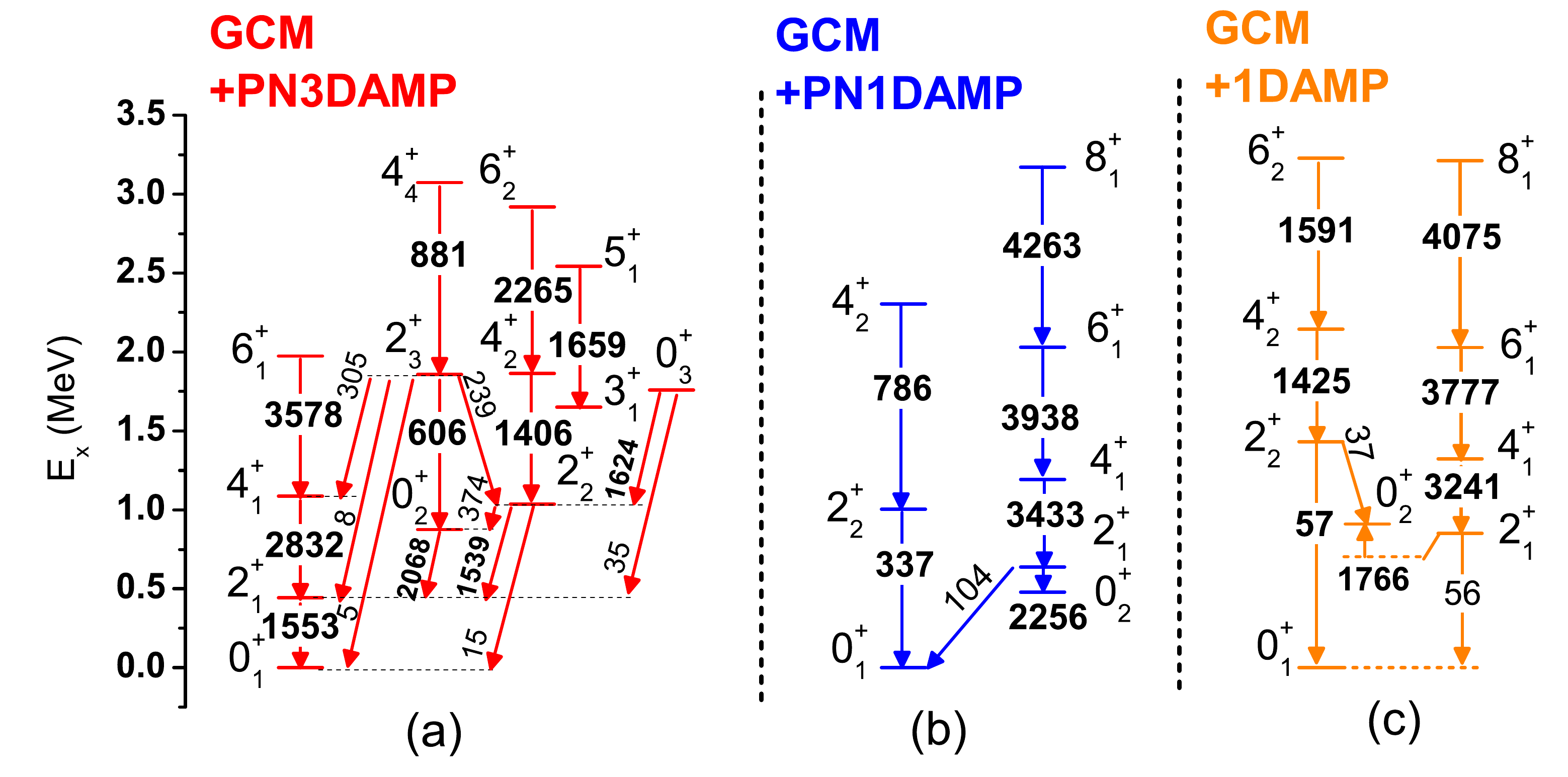}
\caption{
(Color online)
Low-lying spectra and $B(E2)$ values in $e^2$fm$^4$ for $^{76}$Kr;
(a) experimental data from \textcite{Cle07},
(b) the relativistic GCM calculations in the $(\beta,\gamma)$ plane
with the particle number and the angular momentum projections,
(c) the calculation of the 5D collective Hamiltonian
using the cranking inertial masses (Sec.~\ref{sec:CHFB+AP}).
From \textcite{Yao14}.
}
\label{fig:Yao2014}
\end{figure*}

The GCM has been used for a wide variety of nuclear collective phenomena
\cite{RG87,ER04,Ben08,RB11,STS15}.
For low-frequency quadrupole collective motion in superfluid nuclei, 
although the proper generator coordinates are not obvious,
a possible choice
may be the axial and triaxial deformation parameters $(\beta, \gamma$), and
the pairing gaps for neutrons and protons $(\Delta_n,\Delta_p)$.
In addition, to treat the rotational motions associated with spatial
and gauge deformations,
the analytic solution of the angular-momentum eigenstates
and the number eigenstates
are constructed by integration over
the Euler angles of rotation $\Omega=(\vartheta_1,\vartheta_2,\vartheta_3)$, 
and the gauge angles $(\varphi_n,\varphi_p)$, respectively.
In the major applications at the present time, however,   
the pairing gaps $(\Delta_n,\Delta_p)$ are not treated 
as generator coordinates to reduce the dimensionality of integration. 
This leads to the following superpositions
\begin{equation*}
\ket{\Psi^i_{NZIM}} =  \int d\beta d\gamma \sum_K f^i_{NZIK} (\beta,\gamma) 
\hat{P}_N \hat{P}_Z \hat{P}_{IMK}  \ket{\phi(\beta,\gamma)}, 
\end{equation*}
where $\hat{P}_{IMK}$ and $\hat{P}_N$ ($\hat{P}_Z$) denote projection operators 
for angular momentum in three-dimensional space and  
the neutron (proton) number, respectively.  
It has been a great challenge in nuclear structure physics to 
carry out high-dimensional numerical integrations for solving the GCM equation 
using the constrained HFB states.
In recent years, remarkable progress has been taking place, 
which makes it possible to carry out large-scale numerical computations  
\cite{BH08,RE10,RE11,YMRV10,Yao11,Yao14,Rod14}
A recent example is shown in Fig.~\ref{fig:Yao2014}.
As we have seen in Sec.~\ref{sec:basic_formalism},
the HFB calculations using the density-dependent effective interactions 
are better founded by DFT. 
Correspondingly, the modern GCM calculation is often referred to as
{\it multi-reference DFT}
\cite{BH08}.

The GCM is a useful fully quantum approach but the following problems remain
to be solved.
\newline
\noindent
{1. {\it Numerical stability}}\newline
In numerical calculation, one needs to find an optimum discretization
(selection of basis) for the generator coordinates $\alpha$, 
because the continuum limit of integration is not stable in general 
\cite{Bon90}.
It is usually determined semi-empirically but a deeper understanding 
of its physical basis is desirable.  
Another problem is a singular behavior
that may occur during the symmetry projections 
in calculations with use of effective interactions
that depend on non-integer power of density. 
Currently, efforts are underway  to overcome this problem
\cite{AER01,DSNR07,Dug09}.
\newline
\noindent
{2. {\it Necessity of complex coordinates}}\newline
It is well known that one can derive a collective Schr\"odinger equation 
by making Gaussian overlap approximation (GOA) to the GCM equation 
\cite{GW57,OU75,RG87,Roh12}.
There is no guarantee, however, that dynamical effects associated 
with time-odd components of moving mean field are sufficiently
taken into account 
in the collective masses (inertia functions) obtained through this procedure. 
In the case of center of mass motion, 
we need to use complex generator coordinates to obtain the correct mass, 
implying that collective momenta conjugate to collective coordinates
should also be treated as generator coordinates
\cite{PT62,RS80}.
The GOA with respect to the momenta leads to a theory
very similar to ATDHF \cite{GGR83}.
Realistic applications with complex generator coordinates are so far
very few.
\newline
{3. {\it Choice of generator coordinates}}\newline
The most fundamental question is
how to choose the optimum generator coordinates.   
It is desirable to variationally determine the generating functions
$\ket{\phi(\alpha)}$ themselves. 
Let $S$ denotes the space spanned by $\ket{\phi(\alpha)}$. 
The equation determining the space $S$ is then given by 
\begin{equation}
\int_S d\alpha f^*(\alpha)
 \bra{\phi(\alpha)}\Hhat-E\ket{\delta \phi(\alpha')}_{\bot}=0, 
\label{double_projection}
\end{equation}
where $\ket{\delta \phi(\alpha')}_{\bot}$ denotes
 a variation perpendicular to the space $S$. 
Let us add an adjective ``optimum'' to the generator coordinate 
determined by solving the above variational equation. 
It was shown that the mean-field states parameterized
 by a single optimum generator coordinate 
run along a valley of the collective potential energy surface
\cite{HY74}.  
This line of investigation was further developed
\cite{RG79} and greatly stimulated 
the challenge toward constructing microscopic theory of LACM. 
However, direct applications of Eq. (\ref{double_projection})
to realistic EDFs may have a problem.
As we discussed in Sec.~\ref{sec:problems_LACM},
the missing correlations in nuclear EDFs are those in long ranges and
long time scales.
The variation in Eq. (\ref{double_projection}) may take account for
additional short-range correlations, which could lead to
unphysical solutions \cite{SONY06,Fuk13}.

Finally, we note that conventional GCM calculations parameterized
by a few real generator coordinates 
do not reduce to the (Q)RPA in the small-amplitude limit.
It is equivalent to RPA only when all 
the particle-hole degrees of freedom are treated as 
complex generator coordinates \cite{JS64}. 
An extension to the QRPA is not straightforward either.
Thus, systematic comparison of collective inertia masses evaluated
by different approximations 
including the ASCC, the ATDHFB, the GCM+GOA, and the adiabatic cranking methods
is desirable for a better understanding of their physical implications.   

\subsubsection{Time-dependent density matrix theory and higher QRPA}
\label{sec:TDDM}

The TDHF theory describes time-evolution of one-body density matrix 
$\rho_{ij}=\bra{\phi} c_j^\dag c_i \ket{\phi}$ 
on the basis of the time-dependent variational principle. 
To generalize this approach, one may consider, in addition to $\rho_{ij}$, 
time-evolution of two-body correlation matrix 
$C_{ijkl} = \bra{\phi} c_k^\dag c_l^\dag  c_j c_i \ket{\phi}$
$- \rho_{ik}\rho_{jl} + \rho_{il}\rho_{jk}$. 
This approach is called time-dependent density matrix (TDDM) theory
\cite{WC85}.
The extended RPA \cite{TS07}
and the second RPA 
\cite{Dro90,GGC11,Gam12,TN12}
can be derived 
as approximations to 
the small-amplitude limit of the TDDM theory 
\cite{TG89}, 
and have been used to the analysis of damping mechanisms of giant 
resonances and anharmonicities of low-frequency vibrations. 

In the TDDM theory, 
the pairing correlations are taken into account
by the two-body correlations $C_{ijkl}$.
This requires a large computational cost, however. 
The TDDM theory using the HFB quasiparticle representations is not available.
On the other hand,
the higher QRPA  may provide another practical approach to
its small-amplitude approximation. 
In the higher QRPA,
in addition to the two-quasiparticle creation and annihilation operators
in the conventional QRPA,
$
\Gamma_n^{(2)\dag} = \sum_{i,j} \left(
\psi^{n(2)}_{ij} a_i^\dag a_j^\dag + \varphi^{n(2)}_{ij} a_j a_i
\right)
$, 
equations of motion for four quasiparticle creation and annihilation operators, 
\begin{equation*}
\Gamma_n^{(4)\dag} = \sum_{i,j,k,l} 
\Big(
\psi^{n(4)}_{ijkl} a_i^\dag a_j^\dag a_k^\dag a_l^\dag 
 +\phi^{n(4)}_{ijkl} a_i^\dag a_j^\dag a_l a_k \\
 +  \varphi^{n(4)}_{ijkl} a_l a_k a_j a_i \Big) ,
\end{equation*}
are derived.
This approach may be suitable for describing various mode-mode coupling effects 
and anharmonicities arising from Pauli-principle effects in two-phonon states 
where two QRPA vibrational quanta are excited.  
We note that the $a_i^\dag a_j^\dag a_l a_k$ terms in $\Gamma_n^{(4)\dag}$
are often ignored
($\phi^{n(4)}_{ijkl}=0$). 
It is known, however, that collectivities of two-phonon states
cannot be well described without these terms, because 
they are responsible for making the ratio
$B(E2; \textrm{2 phonon} \rightarrow \textrm{1 phonon})/
B(E2; \textrm{1 phonon} \rightarrow \textrm{g.s.}) = 2$
in the harmonic limit \cite{TU64}.
This problem may be overcome by using
the quasiparticle New Tamm-Dancoff method \cite{KMST73-1,KMST73-2,SMT81}.
In the limit of vanishing pairing correlations, 
the quasiparticle-pair scattering terms, $a_i^\dag a_j^\dag a_l a_k$,  
reduce to the particle-hole-pair scattering terms. 
Their effects are taken into account in the extended RPA, 
while they are ignored in the second RPA
\cite{toh01}. 

To our knowledge,
no attempt has been made to introduce collective variables and 
derive collective Hamiltonian on the basis of the TDDM theory.

\subsection{Application to shape coexistence/fluctuation phenomena}

\subsubsection{Five-dimensional quadrupole collective Hamiltonian}
\label{sec:5DCH}

Vibrational and rotational motions of the nucleus can be described as 
time evolution of a self-consistent mean field. 
This is the basic idea underlying the unified model of Bohr and Mottelson
\cite{Boh76,Mot76}.
In this approach, the five-dimensional (5D) collective Hamiltonian
describing the quadrupole vibrational and rotational motions is given by 
\cite{BM75,PR09}
\begin{equation}
H=T_{\rm rot}+T_{\rm vib}+V(\beta,\gamma), 
\label{eq:Hclassical}
\end{equation} 
with $T_{\rm rot}=\frac{1}{2}\sum_{k}\cJ_k\omega_k^2$ and   
$T_{\rm vib}=\frac{1}{2}D_{\beta\beta}\dot \beta^2+D_{\beta\gamma}\dot \beta \dot\gamma
+\frac{1}{2}D_{\gamma\gamma}\dot \gamma^2$,  
where $\omega_k$  and $\cJ_k$ in the rotational energy $T_{\rm rot}$
are the three components of the angular velocities and 
the corresponding moments of inertia, respectively, while 
 ($D_{\beta\beta}, D_{\beta\gamma},D_{\gamma\gamma})$ 
in $T_{\rm vib}$ represent the inertial masses of the vibrational motion. 
Note that $\cJ_{k=1,2,3}$  and ($D_{\beta\beta}, D_{\beta\gamma},D_{\gamma\gamma})$ 
are functions of $\beta$ and $\gamma$.  
The ``deformation parameters" $\beta$ and $\gamma$ are here treated 
as dynamical variables, and 
$\dot \beta$ and $\dot \gamma$ represent their time-derivatives. 
They are related to expectation values of the quadruple operators 
(with respect to the time-dependent mean-field states)
and their variations in time.    
Note also that they are defined with respect to the principal axes of the 
body-fixed (intrinsic) frame that is attached to the instantaneous shape of 
the time-dependent mean-field. 

In the case that the potential energy $V(\beta,\gamma)$ 
has a deep minimum at finite value of $\beta$ and 
$\gamma=0^\circ$ (or $\gamma=60^\circ$), 
a regular rotational spectrum with the $I(I+1)$ pattern may appear. 
In addition to the ground band,
we expect the $\beta$- and $\gamma$-bands to appear, 
where vibrational quanta with respect to
the $\beta$ and $\gamma$ degrees of freedom 
are excited. 
Detailed investigations on the $\gamma$-vibrational bands over many nuclei 
have revealed, however, that they usually exhibit significant
anharmonicities (non-linearities). 
The $\beta$ vibrational bands are even more mysterious, that they couple, 
sometimes very strongly, with the pairing-vibrational modes 
(associated with fluctuations of the pairing gap). 
Recent experimental data indicate the strong need for a radical review
on their characters \cite{HW11}.

\subsubsection{Microscopic derivation of the 5D collective Hamiltonian}
\label{sec:5DCH_derivation}

For collective submanifolds of two dimensions (2D) or higher dimensions, 
an enormous amount of numerical computation is necessary 
to find fully self-consistent solutions of the ASCC equations.  
To handle this problem, a practical approximation scheme, 
called ``local QRPA'' (LQRPA) method, 
has been developed \cite{HSNMM10,SH11,SHYNMM12}. 
This scheme may be regarded as a non-iterative solution of  
Eqs.~(\ref{eq:mfHFB})-(\ref{eq:ASCC2}) 
without the consistency in the generator $\Qhat^i(q)$ between
the moving-frame HFB equation and the moving-frame QRPA equations. 
It may also be regarded as a first-step of the iterative procedure 
for solving the self-consistent equations.   
Further approximation is that, instead of treating the 5D 
collective coordinates simultaneously, 
we first derive the 2D collective Hamiltonian
for vibrational motions corresponding to the $(\beta,\gamma)$ deformations,
and subsequently take into account 
the three-dimensional (3D) rotational motions associated with Euler angles 
at each point of $(\beta,\gamma)$. 
With this procedure, 
we can easily derive the 5D collective Hamiltonian.  

First, we solve the moving-frame HFB equations. 
\begin{eqnarray*}
 &&\delta \bra{\phi(q)} \Hhat_{\rm M}(q) \ket{\phi(q)} = 0,  \\
 &&\Hhat_{\rm M}(q) = \Hhat - \sum_{\tau}\lambda^{(\tau)}(q)\Nt^{(\tau)} 
                                      - \sum_{m = 0, 2} \mu_{m}(q) \Dhatp_{2m}. 
\end{eqnarray*}
This equation corresponds to Eq.~(\ref{eq:mfHFB}) for 
the 2D case with $q=(q^1,q^2)$ 
and with $\Qhat^i(q)$ replaced by the mass quadrupole operators $\Dhatp_{2m}$.
The variables $(\beta,\gamma)$ are defined by 
\begin{eqnarray}
 \beta\cos\gamma &=  \eta D^{(+)}_{20} (q)
= \eta \bra{\phi(q)} \Dhatp_{20} \ket{\phi(q)},
 \label{eq:definition1}  \\
 \frac{1}{\sqrt{2}} \beta\sin\gamma &= \eta D^{(+)}_{22} (q) 
= \eta \bra{\phi(q)} \Dhatp_{22} \ket{\phi(q)}, 
\label{eq:definition2} 
\end{eqnarray}
where $\eta$ is a scaling factor with the dimension of $L^{-2}$.
These equations determine the relation between  $q=(q^1,q^2)$ and $(\beta,\gamma)$. 

Next, we solve the following equations for $i=1$ and 2:
\begin{eqnarray*}
 \delta \bra{\phi(q)} [ \Hhat_{\rm M}(q), \Qhat^i(q) ] 
 - \frac{1}{i} B^i(q) \Phat_i(q) \ket{\phi(q)} &=& 0,
 \\
 \delta \bra{\phi(q)} [ \Hhat_{\rm M}(q), \frac{1}{i} \Phat_i(q)]
 - C_i(q) \Qhat^i(q) \ket{\phi(q)} &=& 0.
\end{eqnarray*}
These are the moving-frame QRPA equations without the curvature terms and 
called local QRPA (LQRPA) equations.

Displacement of the quadrupole deformation
are related to that of $(q_1,q_2)$ by
\begin{equation*}
d\Dp_{2m} = \sum_{i=1,2} \frac{\del \Dp_{2m}}{\del q^i} dq^i,
  \quad\quad m=0,2 .
\end{equation*}
Making a scale transformation such that the inertial masses 
with respect to the collective coordinates $(q_1,q_2)$ become unity 
and using the above relation, 
we can write the kinetic energy of vibrational motions in terms of 
time-derivatives of the quadrupole deformation:   
\begin{eqnarray*}
&& T_{\rm vib} = \frac{1}{2} \sum_{i=1,2} (\dot{q^i})^2 = 
   \frac{1}{2} \sum_{m,m'=0,2} M_{mm'} \Ddotp_{2m}\Ddotp_{2m'}, \\
&& M_{mm'}(\bg) = \sum_{i=1,2} \frac{\del q^i}{\del
 \Dp_{2m}} \frac{\del q^i}{\del \Dp_{2m'}}. 
\end{eqnarray*}
With Eqs. (\ref{eq:definition1}) and (\ref{eq:definition2}),
it is straightforward to rewrite the above expression
using the time-derivatives of $(\bg)$.

Subsequently, we solve the LQRPA equations for 3D rotational motions
at every point of $q$. 
This is given by replacement of $Q^i(q) \rightarrow \hat{\Psi}^k(q)$ and
$B^i(q)\Phat_i(q) \rightarrow \hat{I}_k / \cJ_k(q)$,
where $\hat{\Psi}^k(q)$ represents the local angle operator conjugate to
the angular momentum $\hat{I}_k$.
The solution provides the moments of inertia 
$
\cJ_k(\bg)=4\beta^2D_k(\bg)\sin^2 (\gamma-2\pi k/3) 
$
which determine
the rotational masses $D_k(\beta,\gamma)$
and the rotational energy $T_{\rm rot}$.  

We can quantize the collective Hamiltonian 
(\ref{eq:Hclassical}) 
using the quantization scheme for curvilinear coordinates  
(so-called the Pauli prescription).  
The quantized rotational and vibrational Hamiltonians are given, respectively, by  
$\hat{T}_{\rm rot}=\frac{1}{2}\sum_{k}\hat I_k^2/\cJ_k$   
and 
\begin{eqnarray} 
\hat{T}_{\rm vib}&=&\frac{-1}{2\sqrt{WR}} \left\{ \frac{1}{\beta^4} 
\left[ \dbeta \left(\beta^4\sqrt{\frac{R}{W}}D_{\gamma\gamma}
\dbeta\right)
\right] \right.
\nonumber \\
&-& \left. \dbeta \left(\beta^2\sqrt{\frac{R}{W}}D_{\beta\gamma}\dgamma\right)\right]
\nonumber \\
&+&\frac{1}{\beta^2\sin 3\gamma}\left[
-\dgamma \left(\sqrt{\frac{R}{W}}\sin 3\gamma D_{\beta\gamma}\dbeta\right)
\right.  \nonumber \\
&+& \left. \left.
\dgamma \left(\sqrt{\frac{R}{W}}\sin 3\gamma D_{\beta\beta}\dgamma\right)
\right]
\right\}. 
\end{eqnarray}
with 
$\beta^2 W=D_{\beta\beta}D_{\gamma\gamma}-D_{\beta\gamma}^2$ 
and $R=D_1D_2D_3.$ 

The collective wave functions are written as
\begin{equation*}
 \Psi_{IMk}(\bg,\Omega) = \sum_{K=0}^I \Phi_{IKk}(\bg) 
 \langle\Omega|IMK\rangle,
\end{equation*}
where
$\Phi_{IKk}(\bg)$ and $\langle\Omega|IMK\rangle$ 
represent the vibrational and rotational wave functions, respectively.
Solving the collective Schr\"odinger equations
\begin{equation*}
\left( \hat{T}_{\rm rot}+\hat{T}_{\rm vib}+V(\beta,\gamma) \right)  
\Psi_{IMk}(\bg,\Omega) =E_{IMk}\Psi_{IMk}(\bg,\Omega),  
\end{equation*}
we obtain quantum spectra of quadrupole collective motion. 
Details of the above derivation are given in 
\textcite{HSNMM10} and \textcite{MMNYHS16}.

\subsection{Illustrative examples}

The spherical shell structure gradually changes following the deformation 
of the mean field.  
If we plot single-particle level diagrams as functions of deformation parameters, 
significant gaps, called `deformed magic numbers,'  
appear at the Fermi surface for certain deformations. 
Such deformed shell effects stabilize some deformed shapes of the mean field.
Accordingly, in the HFB calculations, we may encounter 
multiple local minima with different shapes in similar energies.
The LACM 
connecting multiple local minima via
tunneling through potential barriers may take place
to generate the shape fluctuation.
These phenomena may be regarded as a kind of macroscopic quantum tunneling. 
Note that the barriers are not external fields but 
self-consistently generated as 
a consequence of quantum dynamics of the many-body system under consideration.
Quantum spectra of low-energy excitation that involve
dynamics associated with different shapes have been observed 
in almost all regions of the nuclear chart 
\cite{HW11}.
When different kinds of quantum eigenstates associated
with different shapes coexist 
in the same energy region, 
we may call it ``shape coexistence phenomenon''.
This is the case when shape mixing due to tunneling motion is weak and 
collective wave functions retain their localization 
about different equilibrium shapes. 
On the other hand, if the shape mixing is strong, 
large-amplitude shape fluctuations 
extending to different local minima may occur. 
Below, we illustrate these concepts with numerical applications
of the LQRPA method 
to the oblate-prolate shape coexistence/fluctuation phenomena.  

\begin{figure}[t]
\includegraphics[width=6cm]{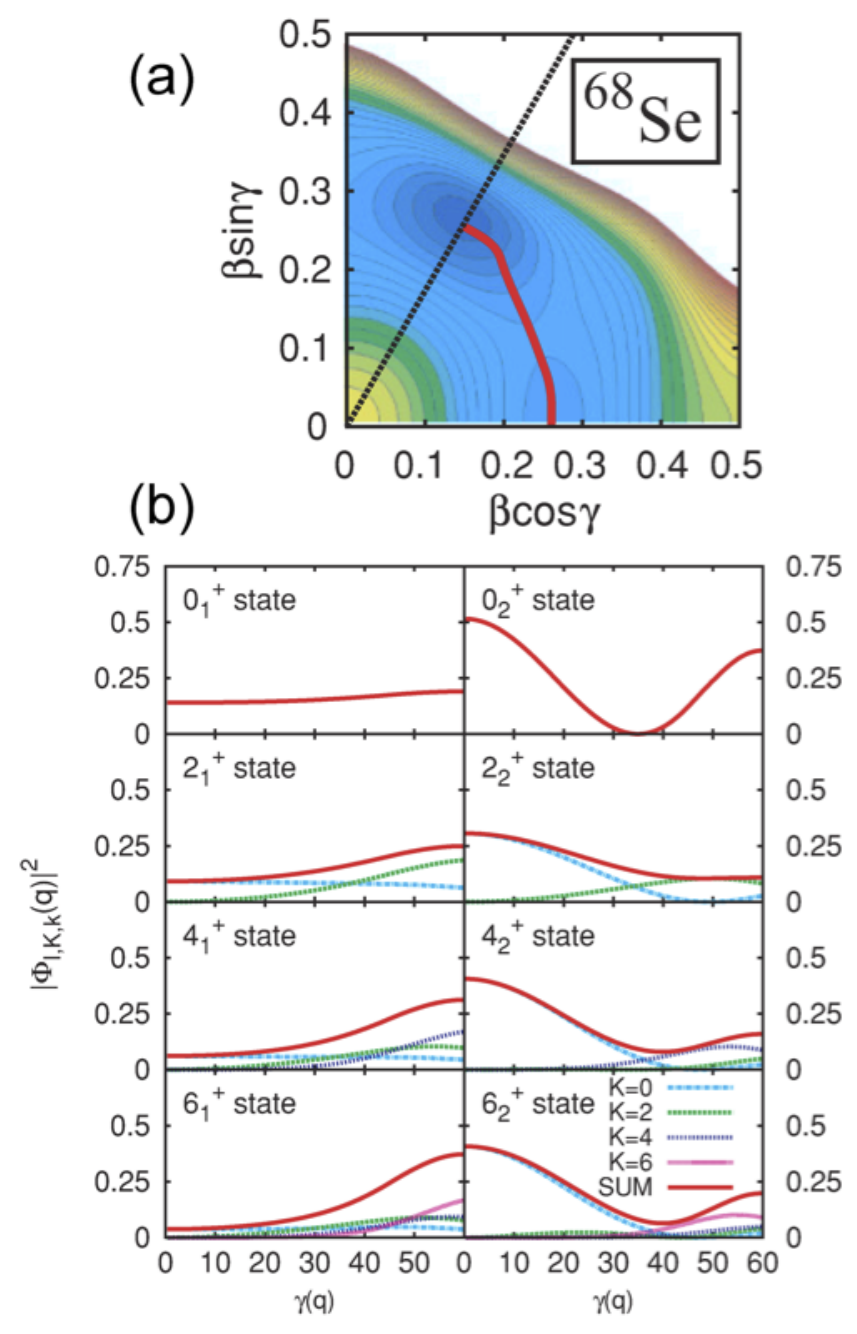}
\caption{
(Color online)
Application of the ASCC method to the oblate-prolate shape coexistence 
phenomenon in $^{68}$Se.
(a)
The collective path for $^{68}$Se obtained by the ASCC method.
The solid (red) line shows the collective path running along the valley
of the potential energy surface 
projected on the $(\bg)$ deformation plane.
(b) 
Vibrational wave functions squared of the lowest (left) and
the second-lowest states (right) 
for each angular momentum. 
In each panel, different $K$-components of the vibrational wave functions 
and the sum of them are plotted as functions of $\gamma(q)$. 
For excitation spectra, see Fig.~\ref{fig:68Se_2D}.
Adapted from \textcite{HNMM09}.
}
\label{fig:68Se_1D}
\end{figure}

\begin{figure}[t]
\includegraphics[width=9cm]{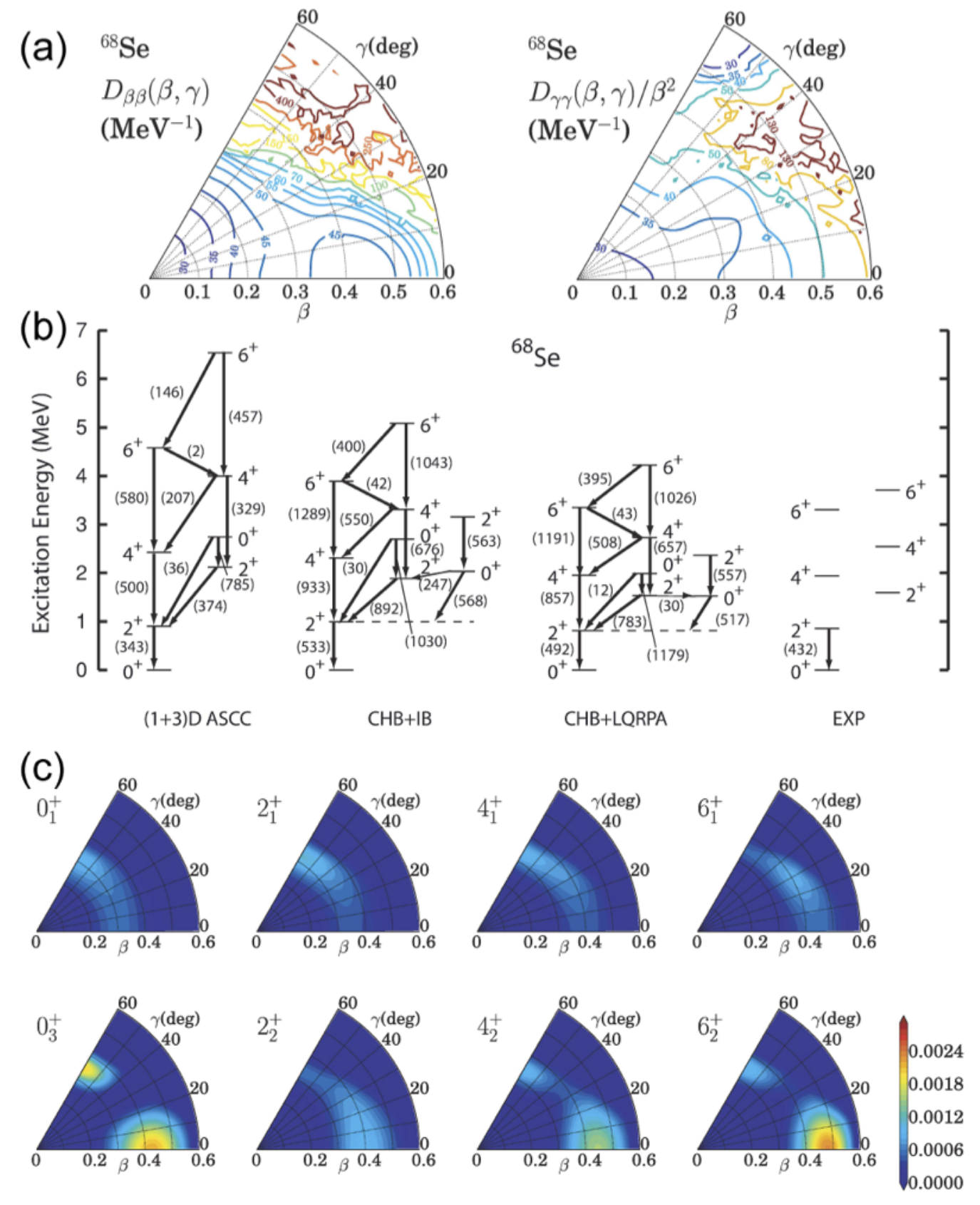}
\caption{
(Color online)
Application of the LQRPA method to the oblate-prolate shape
coexistence/fluctuation phenomenon 
in $^{68}$Se.
(a) Collective inertial masses, $D_{\beta\beta}(\bg)$ and $D_{\gamma\gamma}(\bg)$, 
(b) Excitation spectrum, 
(c) Vibrational wave functions 
$\beta^4 \sum_{K} |\Phi_{IKk}(\bg)|^2$. 
Adapted from \textcite{HSNMM10}. 
}
\label{fig:68Se_2D}
\end{figure}

Figures~\ref{fig:68Se_1D} and \ref{fig:68Se_2D} show
some results of application of the ASCC and QRPA methods to 
the oblate-prolate shape coexistence phenomenon in $^{68}$Se. 
It is clearly seen in Fig.~\ref{fig:68Se_1D} that 
the collective potential exhibits two local minima 
corresponding to the oblate and prolate shapes. 
They are associated with the deformed magic numbers at $N=Z=34$  
appearing for both shapes 
\cite{ham12}. 
The valley runs in the triaxially deformed region and the barrier 
connecting the oblate and prolate minima is low.  
This is an intermediate situation between the oblate-prolate
shape coexistence and the $\gamma$-unstable model of 
\textcite{WJ56}. 
In the former, the barrier is high and the mixing of the oblate 
and prolate shapes is suppressed, 
while the collective potential is flat with respect to the $\gamma$ degree 
of freedom in the latter. 
The theoretical calculation indicates that large-scale quantum shape 
fluctuation occurs along the triaxial valley.  
 
In Fig.~\ref{fig:68Se_1D}, the collective path (one-dimensional 
collective submanifold) self-consistently determined by solving
the ASCC equations, 
(\ref{eq:mfHFB}), (\ref{eq:ASCC1}), and (\ref{eq:ASCC2}), is indicated.
The self-consistent collective path runs along the valley
to connect the prolate and oblate minima. The inertial
mass $B(q)$ is also determined 
by Eqs.~(\ref{eq:ASCC1}) and (\ref{eq:ASCC2}). 
For one dimensional case, 
properly choosing the scale of the collective coordinate $q$,
one can make $B(q)=B$ constant.
The moments of inertia $J_k(q)$ are
calculated by solving the Thouless-Valatin equations at
every point on the collective path. 

The collective wave functions displayed in Fig.~\ref{fig:68Se_1D}(b)
are obtained by solving the collective Schr{\"o}dinger equation 
for the 4D collective Hamiltonian (the 1D collective path 
plus 3D rotational degrees of freedom) microscopically 
derived with the ASCC method
\cite{HNMM09}.  
The ground state shows a $\gamma$-unstable
feature, and accordingly the second $0^+$ state also shows
strong mixing between the prolate and oblate shapes. 
However, increasing the angular momentum, the yrast (yrare)
band becomes more and more oblate (prolate) dominant.
The nuclear shape is localized (stabilized) by the rotation. 
 
In order to confirm that the one-dimensional collective coordinate is 
enough for the low-energy dynamics of $^{68}$Se, it is desirable to 
find the two-dimensional collective submanifold.
This is approximately done according to the LQRPA
(Sec.~\ref{sec:5DCH_derivation}),
in which the self-consistency between the
moving-frame HFB and QRPA equations is ignored, and
no iteration is performed.
Figure~\ref{fig:68Se_2D} shows a result of
the application of the LQRPA method for deriving the
5D collective Hamiltonian (the 2D vibrational and 3D rotational coordinates).
The potential $V(\bg)$ is shown in Fig.~\ref{fig:68Se_2D} (a). 
The vibrational masses $D_{\beta\beta}(\bg)$ and $D_{\gamma\gamma}(\bg)$ 
significantly change as functions of $(\bg)$. 
In addition, considerable variation in the $(\bg)$-plane is also observed 
in the paring gaps (monopole and quadrupole) and 
the rotational moments of inertia.
Due to the time-odd contributions of the moving HFB self-consistent field, 
the collective inertial masses (the vibrational masses and the rotational
moments of inertia) calculated with the LQRPA method
are larger than those evaluated with the Inglis-Belyaev
cranking formula. Their ratios also change as functions of $(\bg)$ 
\cite{HSNMM10}.

A remarkable agreement with experiment is seen in
Fig.~\ref{fig:68Se_2D}(b). An improvement over the 4D calculation
is mostly due to the angular momentum dependence of
the optimal 1D collective path. The calculated collective
wave functions in Fig.~\ref{fig:68Se_2D}(c) clearly indicate the 
importance of the fluctuation with respect to the $\gamma$-degree of
freedom, which is consistent with the 1D collective path
shown in Fig. \ref{fig:68Se_1D}. However, this path should gradually
shifts to larger $\beta$ with increasing angular momentum.
This stretching effect is missing in the 4D calculation.

\section{Relation to TDDFT in electronic systems}
\label{sec:electronic}


DFT and TDDFT have been extensively applied to electronic systems,
matters
composed of electrons and nuclei such as atoms, molecules, nano-materials,
and solids \cite{PY89-B, DG90-B, KH01-B, Mar04-B, SS09-B, Ull12-B, GM12}. 
Electrons in matters always need treatment by quantum mechanics, and 
nuclear motions can be in most cases treated by classical mechanics.
In this chapter, we discuss DFT and TDDFT for electrons in matters, stressing 
similarities with and differences from nuclear DFT.

An apparent difference between electronic and nuclear systems
is the interaction. 
The Hamiltonian of electronic systems is composed of 
the attractive one-body Coulomb potential between  
electrons and nuclei,
and the repulsive Coulomb interaction among electrons.
Besides the difference in the interaction, 
the researchers in the two fields have
different concepts on the DFT and TDDFT.
We first discuss these conceptual differences in Sec.~\ref{sec:conceptual},
and describe the electronic EDFs in practical use in Sec.~\ref{sec:EF}.
We then describe applications of TDDFT in electronic systems.
As in nuclear physics, there are two distinct applications: linear response
TDDFT and TDDFT for large amplitude motion as an initial value problem.
Former applications include electronic excitations and optical responses
in molecules and solids, while the latter applications include electron 
dynamics in matters induced by strong laser pulses.

\subsection{Conceptual difference between electronic and nuclear (TD)DFT}
\label{sec:conceptual}

In electronic systems, DFT and TDDFT are considered as ``self-contained''
theories that can in principle be exact if accurate functionals are obtained.
Improvements of the quality of the calculations should
be achieved through improvement of the EDFs.
There are other theoretical frameworks that can also in principle exactly 
describe properties of electron many-body systems, many-body perturbation 
theory (MBPT) in condensed matter physics and wave function based 
methods in the field of quantum chemistry. These three approaches,
(TD)DFT,
MBPT, and wave function based methods are recognized 
as completely different theories and to constitute independent, self-contained 
theoretical frameworks. In practical applications, DFT and MBPT 
are sometimes used simultaneously: for example, Green's functions that 
appear in the MBPT are approximately constructed from solutions of the 
KS equation. However, in such cases, 
the mixed use of different theories are clearly recognized, 
with some reasons such as computational conveniences. 

The (TD)DFT in nuclear physics is rather different from this: 
for example, DFT and MBPT are often used in a mixed way.
One of the reasons for this difference is probably due to
different roles of the genuine Hartree-Fock (HF) approximation.
In electronic systems,
the HF approximation provides a reasonable starting point for 
the MBPT.
The solutions of the HF and the KS
equations are clearly different.
In nuclear systems, on the other hand,
the HF calculation using
a bare nuclear force does not provide any useful result.
The KS solution is the only appropriate starting point
for the MBPT.

There are also qualitative differences in applications and
interpretations of DFT and TDDFT between two kinds of systems. 
One example is the size of the system that the DFT and TDDFT are
applied to:
In nuclear applications, the DFT and TDDFT are usually
adopted for studies of nuclei
with a few tens of nucleons or more.
In contrast,
for electronic systems, the DFT and TDDFT are applied to 
as small as a few electron systems, even one electron system!
For one electron system,  of course, no potential
originating from the EDF
should appear.
However, due to an 
approximate nature of the EDF in practical use, 
this property is often violated.
The condition of vanishing potential for one electron system
is used to improve the EDF
to remove the self-interaction error, which is 
known as the self-interaction correction \cite{PZ81}.

Another important difference appears in interpretation of linear response 
TDDFT calculations. 
In nuclear TDDFT, we understand
that the linear response TDDFT is accurate only for processes
characterized by small amplitude oscillation around the ground state.
Low-lying excited states are characterized by large amplitude motion
and are considered to need requantization, as described in Sec.
\ref{sec:collective_submanifold}. In electronic TDDFT, on the other
hand, the linear response TDDFT has been applied to any electronic 
excitations no matter how the properties of the states are.
The necessity of requantization has not been recognized
in electronic TDDFT.
The linear response TDDFT for electronic excitations and optical responses
is 
simply called ``TDDFT''.
The {\it linear response} is 
regarded merely as a computational method,
not as an approximation to the TDDFT.

We also find differences in the treatment of collision effects.
In nuclear physics,
theories of the two-body nucleon-nucleon collisions have been
developed, so as to treat these effects in addition to the TDDFT.
In contrast,
efforts been made to incorporate electron-electron collision effects within
the TDKS formalism in electronic TDDFT, introducing correlation potentials 
with retardation. 
One example is an attempt to describe double ionization 
of atoms by strong laser pulse,
which we will discuss in Sec. \ref{sec:applications}.
There are also attempts to treat electron-electron collisions as an
extension of quantum chemistry methods such as multi-configuration 
TDHF and time-dependent configuration
interaction theories \cite{CZKKKS05}.

In electronic systems, DFT and TDDFT have been widely applied 
to extended systems. In describing electronic motions 
in infinitely periodic systems (crystalline solids), the KS equation
is solved in a unit cell of the solid, which is called ``first-principles 
band calculations''.
Extended systems are classified into metallic 
and insulating systems, depending on presence or absence of 
the band gap. Applying an external field to insulators, there appears
a dielectric polarization and a surface charge. The surface charge
has an influence on electrons inside the solid.  Since it is
the long-range effect,  it cannot be incorporated in the LDA.
To include the polarization effect in the DFT, density polarization
functional theory \cite{GL97} has been developed in which the
polarization is treated as an independent degree of freedom.
A similar argument is applicable to electron dynamics in the TDDFT.
Consider a current flowing in an extended system, or in a finite
system, for example, a circular current flowing a nano material 
of ring shape. It is difficult to incorporate effects of the current 
on electron dynamics by local approximation. 
For such cases, time-dependent current density functional theory 
(TDCDFT) treating current and vector potential as basic variables 
has been developed \cite{Ull12-B}. The TDCDFT also attracts 
interests to incorporate retardation effects. It has been realized
that the retardation effects cannot be introduced consistently in
TDDFT, if one assumes the 
LDA \cite{Dob94}. 
In the TDCDFT, it is possible to include the retardation effect 
in the local approximation scheme \cite{VK96}.

\subsection{Energy density functionals}
\label{sec:EF}

In this section, we describe properties of EDFs
of electronic systems in practical use,
with some emphasis on differences from
those in nuclear systems. As in nuclear TDDFT, the adiabatic 
approximation of Eq. (\ref{adiabatic_potential})
is usually adopted for most applications of electronic TDDFT;
one employs the same EDF
as that in the static calculation, replacing a static density with 
a time-dependent density without retardation.
Therefore, here, we mainly describe EDF for the static (ground state) 
calculations.
At the end of this section, we briefly mention progresses 
beyond the adiabatic approximation.

In nuclear DFT, a general form of the EDF
as a functional of density, density gradient, kinetic energy density, 
current density, spin density, pair density, and so on has been 
considered since early stage of its progress \cite{Eng75}.
In contrast, electronic DFT started with an EDF of 
density only in the LDA and gradually developed to include more 
complex elements. 

Energy density of a uniform system as a function of density is
the most fundamental information for the EDF.
Accurate energy density of an electron gas system in the ground
state has been obtained around 1980 \cite{CA80}.
It has been obtained by the MBPT at high density and by numerical
calculations using quantum Monte Carlo method at medium and
low density, connecting to the energy density of the Wigner crystal
at very low density. Since then, a number of LDA calculations have
been carried out for various systems, utilizing analytic forms
of functional which are obtained by fitting the numerical energy density. 
When treating systems with spin polarization 
such as isolated atoms and ferromagnetic materials, local spin density
approximation treating densities of spin up and spin down as basic
variables has been developed.

As a step toward higher accuracy from the LDA, EDFs
including a gradient of electron density have been developed.
A group of EDFs with density gradient
that are widely used today is called the generalized
gradient approximation (GGA).
They are constructed
around 1990 and succeeded to increase the accuracy
substantially from the LDA \cite{SFR07}. 
To further improve the accuracy, EDFs including a kinetic
energy density have been developed.
They are called the meta-GGA  \cite{TPSS03}. 
In developing these new EDFs,
exact analytical properties, that should be satisfied
by EDF, are respected.
These attempts to increase
the accuracy of the EDFs employing more and more
elements are named the Jacob's ladder of the DFT by \textcite{Per05}.

At present, most successful EDFs in the sense of
accurate description of measured properties
are those called ``hybrid functional'' \cite{KH01-B}.
They use a mixture of semi-local and nonlocal forms
for the exchange energy.
The ratio of the mixture,
which is determined empirically, is chosen to be about 3:1. 
In molecules, the functional named B3LYP \cite{SDCF94}
is known to give good results for many systems and has been
quite often used \cite{LJ13}.
In infinitely periodic systems, hybrid functionals have
also been proposed \cite{HSE03}.
However, the use is somewhat limited 
because calculation of the nonlocal exchange terms is computationally
expensive in plane wave basis method
that is popular in the solid-state calculations.

In electronic systems, computational methods to solve the KS equation 
is classified into two. 
One is the local basis 
expansion method in which the basis functions are given with respect to
atomic positions.
This is adopted in most quantum chemistry 
codes for molecules.
The other is the grid representation either 
in the coordinate or in the momentum spaces. The grid representation 
in momentum, which is often called the plane wave basis method, 
has been widely adopted in computational codes of crystalline solids.
Recently, the real-space grid representation becomes more and more 
popular,
since it is superior for calculations 
with massively parallel computers \cite{GPAW10,octopus12}. 
In the grid approach, it is difficult to describe inner orbitals that are 
strongly bound to nuclei. The pseudo-potential methods have been 
developed to avoid the difficulty.
In the local basis expansion methods, nonlocal exchange terms can be 
managed with a reasonable computational cost. However, in the
grid representation methods, the computational cost becomes 
extremely high. This situation is similar to the nuclear DFT calculations. 
In Skyrme HF calculations in which no nonlocal term 
appears, the real-space grid representation is a popular computational 
method, while in the HF calculations with
Gogny interaction, the basis expansion method such as the harmonic
oscillator basis is used to handle the nonlocal Fock terms.

Even with hybrid functionals, it is not possible to incorporate
long-range electron correlations that are responsible for the van der
Waals forces which are important between two neutral molecules.
For this problem, one practical and successful approach is to add 
a long-range potential energy, $-C/R^6$, to every pair of atoms,
on top of the DFT \cite{Gr06}.
Microscopic approaches to construct EDFs incorporating 
the long-range electron correlations have also been actively 
pursued \cite{BCLS15}. 

While accurate calculation of the ground-state energy is
the principal goal of the DFT calculations, orbital energies, in particular
the energy gap between occupied and unoccupied orbitals, are
important to describe electronic excitations and dynamics in TDDFT.
Comparing energy gaps of insulators obtained from eigenvalues of the 
KS equation with measured energy gaps,
the KS energy gaps are systematically too small. 
For a better description of energy gaps, potentials as 
functionals of the density gradient and of the kinetic energy density 
have been developed. 
For atoms and
molecules, a potential named LB94 \cite{LB94}, which includes
the density gradient, has been successfully used for optical response
calculations. The potential is so constructed that it has the correct 
asymptotic form, $-e^2/r$ , which should be satisfied in electrically 
neutral systems. 
For extended systems, the meta-GGA potential 
that includes kinetic energy density was proposed by
\textcite{TB09}, which attracts recent interests.
These potentials are directly given as a functional of density, gradient
of the density, and kinetic energy density.
The EDFs that provide these potentials are not constructed.
We do not know even whether such EDFs exist or not. 

Beyond the adiabatic approximation is certainly an important issue.
In the linear response TDDFT, the number of excited states is
equal to the number of $1p-1h$ configurations. If one would
hope to describe many-particle-many-hole-like configurations 
within the linear response TDDFT, frequency dependence of the 
exchange correlation kernel, the second derivative of the 
energy density functional with respect to densities, should be crucial. 
Inclusion of electron-electron collision effects through energy density
functional will also require the frequency dependencies.
Although extensive efforts have been made to construct nonadiabatic
functionals, the functionals which are useful for wide purposes have
not yet been obtained. A nonadiabatic energy functional in TDCDFT 
proposed by \textcite{VK96} has been tested for 
several problems. In that functional, the nonadiabaticity has been 
discussed making relations to the viscoelastic stresses of 
electronic quantum liquid.

\subsection{Applications}
\label{sec:applications}

\subsubsection{Linear Response}

Among applications of electronic TDDFT, the linear response TDDFT 
in the adiabatic approximation has been widely used and highly successful
to describe electronic excitations and optical responses of molecules.
As in nuclear TDDFT, the basic idea is to extract excitation energies
and response functions from the density change induced by a weak 
external field applied to molecules.

Historically, optical responses of spherical systems have been
investigated first. Using a similar approach to that in nuclear theory
employing the continuum Green's function, optical responses of rare gas
atoms have been investigated by \textcite{ZS80}
and of metallic clusters by  \textcite{Ek84}, respectively
(Sec.~\ref{sec:Green's_function_method}).

In middle 1990's and later, efficient computational methods have 
been developed for linear response TDDFT calculations of molecules 
without any spatial symmetries. 
A matrix diagonalization method preparing occupied and 
unoccupied orbitals has been developed by \textcite{CJCC98}
and has been named ``Casida method''
(Sec.~\ref{sec:normal_modes}). 
A method solving linear Schr\"odinger-like equation for a given
external field with a fixed frequency is known as the Sternheimer
method \cite{NY01}. 
Real-time method has also been 
developed \cite{YB96, YNIB06}, 
solving the TDKS equation in real time after an impulsive external 
field applied to the system
(Sec.~\ref{sec:linear_real_time}).
The matrix diagonalization method is
the most widely used in practical purposes. The real time method is
superior to calculate collective excitations
 to which a large number 
of electron-hole pairs contribute.
After middle 1990's, linear response TDDFT has been implemented 
in many quantum chemistry codes as a tool to calculate
electronically excited states of molecules with reasonable accuracy and
cost. Using these codes, researchers who do not have much knowledge 
and experience on TDDFT, including experimentalists, 
can easily perform the linear response TDDFT calculations of molecules. 
After 2011, the number of papers that include TDDFT as keywords
exceeds 1,000 per year.

As the method has been applied to a wide variety of molecules,
it has been realized that linear response TDDFT with local
or semilocal approximation fails systematically \cite{Ull12-B}. 
For example, electronic excitation energies of long-chain molecules are 
systematically underestimated. Excitation energies of charge-transfer 
excitations, in which the electron and the hole are spatially remote,
are also underestimated.
These failures are attributed to
the incomplete cancellation of the electron self-energy.

Linear responses of extended systems are characterized by
dielectric functions, $\epsilon(\vec q,\omega)$. 
The dielectric functions
of metallic systems that are dominated
by plasmon are reasonably 
described by the adiabatic TDDFT. In contrast, it does not give 
satisfactory results for semiconductors and insulators.
In these solids, optical responses around the band gap energy
are characterized by excitons, bound excited states of electrons 
and holes. It has been realized that the excitons cannot be described 
in the adiabatic TDDFT with local approximations \cite{ORR02}. 
For optical responses in semiconductors and insulators, 
the GW-plus-Bethe-Salpeter approach,
solving the Bethe-Salpeter equation with the Green 
functions containing self-energy given by GW approximation, 
has been quite successful \cite{RL00}. 

\subsubsection{Electron dynamics under strong field}

In nuclear physics, TDDFT calculations as initial value problems 
have been developed in the studies of heavy ion collisions.
In electronic systems, similar initial-value approaches have been 
widely applied to interactions of a strong laser pulse with matters.

One of active frontiers of laser science is to produce strong 
and ultra-short light pulses and to explore their interaction with 
matters. At extremely intense limit, high energy phenomena
such as vacuum breakdown and nuclear reactions induced by
strong laser pulses are actively investigated \cite{DMHK12}. 
In material sciences, interactions of light pulses 
whose scales are approaching to atomic units have been attracting 
significant interests. 
When the magnitude of the laser electric field approaches
to those of binding electrons to ions, the electron
dynamics induced by the laser pulse will become extremely nonlinear \cite{BK00}.
The shortest light pulse available today is comparable to the 
period of hydrogen atom. 
Using such a ultra-short
laser pulse as a flash light, 
there have been intense attempts to take snapshots of electron
dynamics in atoms, molecules, and solids \cite{KI09}.
To theoretically investigate extremely nonlinear and 
ultrafast electron dynamics in matters, computational approaches 
solving time-dependent Schr\"dinger equation for one-electron 
systems and TDKS equation for many-electron systems have 
been extensively developed.

In strong laser pulse irradiations on atoms and molecules,
various phenomena like tunnel and multiphoton ionizations,
above threshold ionization,
high harmonic generation, and Coulomb explosion have been described
by the real-time TDDFT \cite{MG04-B, CT04, Ull12-B}. 
In the interaction of strong laser pulses with
metallic clusters, nonlinear interactions between strong
laser pulse and the plasmon, collective electronic excitation,
play an important role
\cite{CRSU00,WDRS15}.
In the multiple ionizations of atoms at relatively low laser
intensities, it is known that the secondary ionizations proceed mainly 
through the rescattering process: an ionized electron is 
accelerated by the applied laser pulse and collides with 
the atom from which the electron was first emitted.
This collision process has been regarded as a test case to
develop EDFs that could describe collision effects.
However, it turned out that finding such
functional is, as anticipated, not an easy task \cite{Ull12-B}.

Recently, interactions of strong laser pulses with solids
have been attracting interests, aiming at exploring new phenomena
that could bring innovative optical devices. 
The TDDFT calculations have been
carried out to analyze nonlinear electron dynamics in solids,
including ultrafast current generation in transparent material \cite{WLB14}, 
and coupled dynamics of electrons and macroscopic electromagnetic 
fields \cite{YSS12}

Real-time TDDFT calculations have been applied to fields other than 
laser sciences. One example is electron transfer dynamics in ion collisions.
Electronic TDHF calculations have been also applied to nuclear fusion 
reactions in astrophysical environments to investigate electronic 
screening effects \cite{SKLS93}. Collision of energetic ions impinging on
graphene sheet has been explored \cite{BWPV12,ZMR12}.
Collisions between 
multiply ionized and neutral atoms have been investigated \cite{NYTA00}.

\subsubsection{Coupled dynamics of electrons and atoms}

Before ending this section,
we present a simultaneous description of electronic
and atomic motions. In nuclear physics, there is no degrees of freedom
corresponding to atomic motion. However, coupling of a slow collective
motion with fast internal motions as in nuclear fusion and fission dynamics
may have some similarities.

If the material 
has an energy gap
and electrons always stay in their ground state,
we may assume the adiabatic, 
Born-Oppenheimer approximation. 
In such cases, we may separate
the problem into two steps: For a given atomic configuration, we first 
solve the static KS equation to obtain the electronic ground state. 
Then the forces acting on atoms are calculated using the 
Feynman-Hellman theorem. Finally the atomic motions are 
calculated solving the Newton's equation. This is the so-called 
ab-initio molecular dynamics calculation, initiated with 
a slightly different implementation by \textcite{CP85}.

Simultaneous descriptions of electronic excitation and atomic motion,
which are often termed nonadiabatic molecular 
dynamics, are much more involved. 
We first consider a simple molecule 
where one or at most a few electronic 
states are important. 
When the electronic levels are well separated,
we may assume the Newtonian 
motion for atoms on the adiabatic potential energy surface.
When the two electronic states come 
close in energy at a certain atomic configuration, quantum transitions 
between different potential energy surfaces need to be treated. 
The potential energy surfaces may be efficiently calculated by 
the linear response TDDFT.
Such simulations have been widely 
applied to photo-molecule interactions \cite{PG14}. 
We note that, in such simulations, the TDKS equation needs not
to be solved in real time.

How can we treat cases in which a number of electronic levels
are close in energy and transitions frequently take place?
The electronic excitation spectra can even form the
continuum in solids.
There is an alternative method called the
Ehrenfest dynamics. In this method, the TDKS equations for electrons 
and Newtonian equations for atoms are solved simultaneously in real
time, as coupled equations. 
At each time, the force acting on each atom
is calculated from the electron density \cite{Tav15, SYK10}. 
 
These two methods are conceptually very different. The former
method utilizes the linear response TDDFT to prepare potential energy surfaces,
while the latter utilizes solution of real time TDKS equation as
an initial value problem. At present, it is empirically decided which 
method to use for a given problem. 
Accumulation
of results will eventually make it possible to assess the quality of
approximation of the two approaches.

\section{Summary and future outlook}
\label{sec:sumary}

The TDDFT using modern nuclear EDFs provides a unified,
systematic, and quantitative description of nuclear structure and reaction.
Thanks to its non-trivial density dependence,
these EDFs are capable of simultaneously reproducing the bulk
properties of nuclei (saturation, EOS, etc.)
and properties of individual nucleus (shell effects, deformation, etc.).
The nuclear EDF also shows various kinds of 
spontaneous breaking of the symmetry (SSB).
Especially, the translational symmetry is always
violated for finite nuclei.
The SSB can be incorporated in a stringent manner
by the DFT theorems for the wave-packet states.
Nevertheless,
there remain several open questions for rigorous justification of
the DFT in nuclear physics \cite{Gir10}.
Because of significant increase in computational resources and 
development in parallelized computer programs, 
the TDDFT serves as 
modern approaches to a variety of nuclear
phenomena which were addressed only with phenomenological models.
Since all the parameters in nuclear EDFs are basically fixed,
it can provide non-empirical predictions.
In the present review, we summarize recent developments in the
three categories: Linear density response, real-time method,
and requantization of TDDFT collective submanifold.

The linear density response around the ground state
is known as (Q)RPA
in nuclear physics.
Recent calculations treat all the residual fields
induced by the density variations in the EDF.
This is particularly important for the separation of ANG modes
associated with the SSB.
The program coding and numerical computation have been facilitated by
the finite amplitude method and other iterative methods to
the linear response.
These developments significantly reduce the computational costs
and necessary memory capacity for heavy deformed nuclei. 

The treatment of the continuum is another issue which has been
extensively studied in recent years 
to explore unique properties of weakly bound nuclei near the drip lines. 
The most complete formalism is the continuum QRPA simultaneously
treating the continuum in the particle-hole and particle-particle (hole-hole)
channels with the Green's function method.
However, so far, the numerical calculation has been achieved only for
spherical systems.

The real-time TDDFT calculation provides useful insights into nuclear
many-body dynamics, such as microscopic understanding of nuclear
reaction and energy dissipation.
One of the recent major achievements is the large-scale 3D calculation
in the TDBdGKS (TDHFB) scheme (Sec~\ref{sec:real_time}).
Although the full calculations for nuclear dynamics
in this scheme are so far limited to the linear response,
one can expect further applications to large amplitude dynamics in
near future.
Meantime, the approximate treatment of the BCS-like pairing may
provide a useful guidance for that (Sec.~\ref{sec:Cb-TDHFB}).

A microscopic derivation of the 
internucleus 
potential and the dissipation
has been developed by several authors recently,
and applied to many systems (Sec.~\ref{sec:NN_pot_friction}).
This provides a connection between the microscopic TDDFT simulation
and the phenomenological potential approaches to nuclear fusion.
The method even quantitatively describes the sub-barrier fusion reaction
for some cases,
by extracting the potential from the TDDFT calculation \cite{UO07,UO08}.
These methods may be justifiable before two nuclei
overlaps substantially in the fusion process.
However, it requires further developments and studies in clarifying 
the entire dynamics in the fusion process.
The real-time TDDFT studies of quasifission are in progress too
(Sec.~\ref{sec:real_time}).

Recent studies on the multi-nucleon transfer reaction show a reasonable
agreement with experimental mass distribution (Sec.~\ref{sec:transfer}).
The fluctuations in major channels seem to be taken into account by the
TDDFT simulation with the particle-number projection.
However, some discrepancies were also identified, especially in minor
channels.
For the improvement, the stochastic mean-field and Baranger-V\'en\'eroni
variational approaches may provide a tool to correct
these missing fluctuations and correlations (\ref{sec:fluctuations}).
It has been partially successful but further studies are desired.

At present, all the available nuclear EDFs seem not to
be able to express,in the KS scheme,
correlations associated with low-energy modes
of (slow) collective motion.
They have been addressed by additional correlations beyond the KS
scheme,
which includes the particle-vibration coupling,
the higher random-phase approximation,
the time-dependent density-matrix (TDDM) method,
the generator coordinate method (GCM), and so on.
In this review, we put some emphasis on the requantization of the
TDDFT collective submanifold to take into account the missing correlations
(Sec~\ref{sec:collective_submanifold}).
The self-consistent derivation of a collective Hamiltonian (submanifold)
suitable for description of low-energy large amplitude motion
can be achieved by solving the adiabatic self-consistent collective
coordinate (ASCC) equations.
The inertial masses include time-odd effects and
are guaranteed to produce the correct total mass for the translation.
The method also overcomes known difficulties in the adiabatic TDHF method.
It has been applied to studies of nuclear quadrupole dynamics in the
pairing-plus-quadrupole model.
For the aim of deriving collective Hamiltonian for various kinds of
large amplitude collective motion (LACM)
on the basis of the modern EDFs,
the finite-amplitude method and new iterative solvers in
Sec.~\ref{sec:iterative_methods}
may be utilized to numerically solve the moving-frame QRPA equations
in an efficient way. 

The collective inertial masses should be studied furthermore.
The collective inertial mass, which is locally defined,
represents the inertia of the many-body system 
against an infinitesimal change of the collective coordinate.
As the single-particle-energy spectrum in the mean field changes
during the LACM,
the level crossing at the Fermi energy successively occurs.
We expect that the configuration rearrangement at the level crossing
is essential to keep the system at low energy.
Thus, for low-energy nuclear dynamics, the pairing correlation plays
an essential role in determination of the collective mass parameters
\cite{BBBV90}.
It remains as an interesting subject to investigate
how the self-consistent determination of 
the optimal directions of collective motion and
 the finite frequency $\omega(q)$ 
of the moving-frame QRPA modes 
affect the level crossing dynamics of the superfluid nuclear systems.

In addition to the quadrupole collective motions,
large-amplitude collective phenomena associated with instability
toward octupole deformations 
of the mean field as well as interplay of the quadrupole and octupole
modes of excitations have 
been widely observed in low-lying states of nuclei
\cite{BN96}.
In the high-spin yrast region where the nucleus is highly excited
but cold (zero-temperature), 
new types of rotations and vibrations may emerge 
\cite{SW05},
such as wobbling motions 
\cite{HH03,SS09,FD14}
and superdeformed shape vibrations 
\cite{NMMS96}.
It is quite interesting to apply the microscopic theory of LACM to 
these new collective phenomena 
\cite{Mat10}.
Macroscopic quantum tunnelings through self-consistently generated barriers,  
like spontaneous fissions and deep sub-barrier fusions, are, needless to say, 
great challenges of nuclear structure physics.

In electronic TDDFT, the linear response is considered to be exact,
and the anharmonic large amplitude nature should not matter
(Sec.~\ref{sec:conceptual}).
The failures in describing a certain class of excited states are
due to incomplete EDFs,
not to the limited applicability of the linear response.
This makes a striking contrast to the concept of nuclear DFT/TDDFT.
Because of these conceptual differences,
major efforts in the electronic DFT/TDDFT are devoted to
improvement in quality of EDFs.
Construction of a practical and accurate EDF
including the retardation effects beyond the
adiabatic local density approximation is currently under investigation.
This is a challenging subject in the electronic TDDFT.
Nevertheless, using the adiabatic EDFs,
there have been numerous successful applications both
in the linear response and the initial-value TDDFT
for molecules and solids (Sec.~\ref{sec:applications}).

The nuclear many-body dynamics in the large-amplitude collective motion
is still a big challenge for nuclear physics.
This review has described theoretical and computational progress
in the nuclear TDDFT studies,
which we think significant in last decades.
We hope it provides stimulus to researchers in the field.

\begin{acknowledgments}
We are grateful to many collaborators and colleagues,
including P. Avogadro, S. Ebata, N. Hinohara, T. Inakura, H. Z. Liang,
K. Mizuyama, K. Sato, K. Sekizawa, K. Washiyama. and K. Yoshida.
This work was supported in part by JSPS KAKENHI Grants No.
24105006, No. 25287065, No. 26400268, and No. 15H03674.
\end{acknowledgments}

\appendix
\section{Krylov reduction of the RPA space}
\label{sec:appendix_krylov}

It is easy to see that the 2$d$ Krylov subspace (\ref{Krylov_subspace})
contains the RPA-conjugate partners.
In fact,
$({\cal NH})^m F_v$ and $({\cal NH})^m\tilde{F}_v$
are RPA conjugate to each other
($({\cal NH})^m\tilde{F}_v =(-1)^m {\cal I} \{({\cal NH})^m {F}_v\}^*$).
We can show this using ${\cal HI}={\cal IH}^*$
and $\left\{ {\cal N},{\cal I} \right\} = 0$.

Next, let us map the RPA equation in the $2D$ space to
that in the $2d$ space.
Suppose that we construct
the ${\cal N}$-orthonormalized basis,
$\{ {Q}_1,\cdots,{Q}_d;
\tilde{Q}_1,\cdots,\tilde{Q}_d \}$
from Eq. (\ref{Krylov_subspace}).
Let us define the $2D\times 2d$ rectangular matrix,
${\cal Q}\equiv ({Q},\tilde{Q})$,
which is a projection from the $2D$ full space into
the $2d$ subspace.
For instance, the Hamiltonian and the norm matrix 
in Eq. (\ref{H_and_N}) are transformed into $2d\times 2d$
Hermitian matrices, as
\begin{equation*}
h\equiv {\cal Q}^\dagger {\cal H Q} =
\begin{pmatrix}
a & b\\
b^* & a^* 
\end{pmatrix}
,\quad
n\equiv {\cal Q}^\dagger {\cal N Q} =
\begin{pmatrix}
1 & 0 \\
0 & -1
\end{pmatrix}
.
\end{equation*}
Here, $a$ and $b$ are $d\times d$ matrices, given by
$a_{ij}\equiv {Q}_i^\dagger {\cal NH} {Q}_j
=-(\tilde{Q}_i^\dagger {\cal NH} \tilde{Q}_j)^*$
and
$b_{ij}\equiv {Q}_i^\dagger {\cal NH} \tilde{Q}_j
= -(\tilde{Q}_i^\dagger {\cal NH} {Q}_j)^*
$.
The eigenvectors
\begin{equation*}
{z}_n\equiv 
\begin{pmatrix}
x_n \\
y_n
\end{pmatrix}
,\quad
{\tilde{z}}_n\equiv 
\begin{pmatrix}
y_n^* \\
x_n^*
\end{pmatrix}
\end{equation*}
are obtained by the diagonalizing
the $2d\times 2d$ matrix, $nh$.
In analogy to Eq. (\ref{Z_and_Omega}), we define the matrix notation
\begin{equation*}
z \equiv ({z},\tilde{z}) =
\begin{pmatrix}
x & y^* \\
y & x^*
\end{pmatrix}
,\quad
\omega\equiv
\begin{pmatrix}
\omega_d & 0 \\
0 & \omega_d
\end{pmatrix}
.
\end{equation*}
The eigenvalue equation (\ref{NHZ=ZOmegaN}) is mapped to
\begin{equation}
nhz = z \omega n .
\label{nhz=zomegan}
\end{equation}

It is easy to show that the reduction (\ref{nhz=zomegan}) preserves
the sum rules $m_L$ with odd $L$.
Since the subspace (\ref{Krylov_subspace}) is complete for
intermediate states ($L<2d$) in Eq. (\ref{EWSR_odd_L}),
we can replace the norm matrix ${\cal N}$ by ${\cal Q}n{\cal Q}^\dagger$.
Then, Eq. (\ref{EWSR_odd_L}) can be rewritten as
\begin{eqnarray*}
&&m_L
=\frac{1}{2}
({F}_v+\tilde{F}_v)^\dagger
{\cal Q} (nh)^L n {\cal Q}^\dagger
({F}_v+\tilde{F}_v) \\
&&=\frac{1}{2}
({F}_v+\tilde{F}_v)^\dagger
{\cal Q} z \omega^L z^\dagger {\cal Q}^\dagger
({F}_v+\tilde{F}_v) 
= \sum_{n=1}^d \omega_n^L |\bra{n} F \ket{0}|^2
,
\end{eqnarray*}
where we used the relation $z^\dagger h z = \omega$ which is derived from
Eq. (\ref{nhz=zomegan}).
Here, $z^\dagger {\cal Q}^\dagger ({F}_v+\tilde{F}_v)$
is nothing but the transition strength $\bra{n}F\ket{0}$
calculated with the approximate eigenvectors, ${Z}' = {\cal Q} z
=\begin{pmatrix} X' & Y'^*\\ Y'& X'^* \end{pmatrix}$.

\section{Response function with the Green's function}
\label{sec:appendix_response}

In this appendix, we show the derivation of Eq. (\ref{Pi_0}).
The unperturbed (independent-particle) density response
$\delta R^0(\omega)$
is defined by the limit of the vanishing residual kernels, $w=w'=0$.
Since the response function
$\Pi_0(\omega)$ is diagonal in the quasiparticle basis,
it can be easily obtained from Eq. (\ref{QRPA_2}) as
\begin{eqnarray}
\delta R^0(\omega) &=&
\sum_{i,j} \left\{
\frac{ \Psi_i^0 V^{(+-)}_{ij} \tilde{\Psi}_j^{0\dagger}}{\omega-E_i-E_j}
+ \frac{\tilde\Psi_i^0 V^{(-+)}_{ij} \Psi_j^{0\dagger}}{-\omega-E_i-E_j}
\right\} \nonumber \\
&=&\sum_{i,j} \left\{
\frac{ \Psi_i^0 \Psi_i^{0\dagger} V \tilde{\Psi}_j^0 \tilde{\Psi}_j^{0\dagger}}{\omega-E_i-E_j}
+ \frac{\tilde\Psi_i^0 \tilde\Psi_i^{0\dagger} V \Psi_j^0  \Psi_j^{0\dagger}}{-\omega-E_i-E_j}
\right\}
 , \quad\quad
\label{delta_R_0_spectrum}
\end{eqnarray}
which has poles at the two-quasiparticle energies, $\omega=\pm (E_i+E_j)$.
Note that we have converted the quasiparticle representation
to a general form
(cf. the transition densities of Eq. (\ref{delta_R_omega})).
Adding the following zero in the right hand side,
\begin{equation*}
\sum_{i,j} \left\{
\frac{ \tilde\Psi_i^0 \tilde\Psi_i^{0\dagger} {V}(\omega) \tilde{\Psi}_j^0 \tilde{\Psi}_j^{0\dagger}}{\omega+E_i-E_j}
+ \frac{\tilde\Psi_i^0 \tilde\Psi_i^{0\dagger} {V}(\omega) \tilde\Psi_j^0  \tilde\Psi_j^{0\dagger}}{-\omega-E_i+E_j}
\right\}
=0 ,
\end{equation*}
leads to an expression
\begin{equation}
\label{delta_R_0}
\begin{split}
\delta R^0(\omega)=\sum_i &\left\{
{\cal G}_0(\omega-E_i) {V}(\omega) \tilde\Psi_i^0\tilde\Psi_i^{0\dagger}
\right.\\
&\left.
+\tilde\Psi_i^0\tilde\Psi_i^{0\dagger} {V}(\omega) {\cal G}_0(-\omega-E_i)
\right\}
.
\end{split}
\end{equation}
Here, the Green's function ${\cal G}_0(E)$ is given by
\begin{equation}
\label{G_0}
{\cal G}_0(E) = 
\left(E-H_s[R_0]\right)^{-1}
=\sum_i \left\{
\frac{\Psi_i^0\Psi_i^{0\dagger}}{E-E_i}
+\frac{\tilde\Psi_i^0\tilde\Psi_i^{0\dagger}}{E+E_i}
\right\}
.
\end{equation}
This Green's function contains both normal and abnormal
Green's function, $G_0(E)$ and $F_0(E)$, in the $2\times 2$ matrix form.
Equation (\ref{delta_R_0}) can also be derived by the Fourier
transform of Eq. (\ref{delta_R_t})
\begin{equation*}
\delta R^0(\omega)= \sum_i \left\{
\delta \tilde{\Psi}_i(\omega) \tilde\Psi_i^{0\dagger}
+\tilde\Psi_i^0 \delta \tilde{\Psi}_i^\dagger(-\omega) \right\}
\end{equation*}
and
$
\delta\tilde\Psi_i(\omega)={\cal G}_0(\omega-E_i) V(\omega) \tilde\Psi_i^0
$.

Now, let us adopt a single-particle representation,
$\{ \alpha \}$.
It should be noted that,
since the quasiparticle state has upper and lower components,
$U(\alpha)$ and $V(\alpha)$,
the quantities with two single-particle indices, such as
$V(\omega)$, $\delta R^0(\omega)$, and
${\cal G}_0(E)$, 
are expressed in $2\times 2$ matrix form.
The response function, $\Pi_0(\omega)$ and $\Pi(\omega)$,
with four indices should be expressed in
the $(2\times 2) \otimes (2\times 2)$ form.
In order to avoid these complications,
we adopt the primed indices, $\alpha',\beta',\cdots$,
which 
are given after Eq. (\ref{H_s}).

Equation (\ref{delta_R_0}) is represented as
\begin{eqnarray}
&\delta R^0(\alpha'\beta';\omega) = \sum_{\mu'\nu'}
\Pi_0(\alpha'\beta',\mu'\nu';\omega) {V}(\mu'\nu';\omega) , & \nonumber\\
\label{Pi_0_4_rep}
&\Pi_0(\alpha'\beta',\mu'\nu';\omega)= \sum_{i} \left\{
{\cal G}_0(\alpha'\mu';\omega-E_i) \tilde\Phi_i(\nu')\tilde\Phi_i^\dagger(\beta')
\right. &\nonumber \\
&\left. +\tilde\Phi_i(\alpha')\tilde\Phi_i^\dagger(\mu') {\cal G}_0(\nu'\beta';-\omega-E_i)
\right\}
. &
\end{eqnarray}
Similarly, the residual kernel ${\cal W}$ is represented by four indices.
In principle, according to Eq. (\ref{Pi}), we may obtain the QRPA
response function $\Pi(\omega)$ and the density response $\delta R(\omega)$.

Here, we distinguish the upper $\Psi_i^{0(1)}=U_i$ 
and lower components $\Psi_i^{0(2)}=V_i$ of the quasiparticle $\Psi_i^0$,
and introduce the $2\times 2$ matrix form for
the density $\delta R^{(mn)}$ and the external potential $V^{(mn)}$,
and the $(2\times 2)\otimes(2\times 2)$ form for the response function
$\Pi^{(mn,pq)}$ and the residual kernels ${\cal W}^{(mn,pq)}$,
with the indices $m,n,p,q=1$ and 2.
If the potential and residual kernels have the diagonal character,
$V^{(mn)}(\alpha\beta)=V^{(mn)}(\alpha)\delta_{\alpha\beta}$,
${\cal W}^{(mn,pq)}(\alpha\beta,\mu\nu)={\cal W}^{(mn,pq)}(\alpha,\mu)
\delta_{\alpha\beta}\delta_{\mu\nu}$,
we may simplify Eq. (\ref{Pi_0_4_rep}) to its diagonal
representation
\begin{equation}
\label{Pi_0_2_rep}
\begin{split}
\Pi_0^{(mn,pq)}&(\alpha,\beta;\omega)= \\
&\sum_{i} \left\{
{\cal G}_0^{(mp)}(\alpha\beta;\omega-E_i)
\tilde\Phi_i^{(q)}(\beta)
\tilde\Phi_i^{(n)\dagger}(\alpha) 
\right.\\
&\left. + \tilde\Phi_i^{(m)}(\alpha)\tilde\Phi_i^{(p)\dagger}(\beta)
 {\cal G}_0^{(qn)}(\beta\alpha;-\omega-E_i)
\right\}
.
\end{split}
\end{equation}
and the unperturbed density response is given by
\begin{equation*}
\delta R_0^{(mn)}(\alpha\alpha)=\sum_{p,q=1,2} \sum_\beta
\Pi_0^{(mn,pq)}(\alpha,\beta) V^{(pq)}(\beta) .
\end{equation*}

\bibliography{nuclear_physics,chemical_physics,myself,current}

\end{document}